\documentclass{article}

\usepackage[numbers]{natbib}

\usepackage[preprint]{neurips_2025}

\usepackage[utf8]{inputenc} 
\usepackage[T1]{fontenc}    
\usepackage{hyperref}       
\usepackage{url}            
\usepackage{booktabs}       
\usepackage{amsfonts}       
\usepackage{nicefrac}       
\usepackage{microtype}      
\usepackage{xcolor}         
\usepackage{wrapfig}
\usepackage{graphicx}
\usepackage{amsmath}
\newcommand{\equalcontrib}{\textsuperscript{*}}
\newcommand{\corrauth}{\textsuperscript{\dag}}
\title{Sparse Autoencoders Bridge The Deep Learning Model and The Brain}

\author{%
  Ziming Mao\equalcontrib$^{1}$ \quad
  Jia Xu\equalcontrib$^{1}$ \quad
  Zeqi Zheng\equalcontrib$^{2,3}$ \quad
  Haofang Zheng\equalcontrib$^{1}$ \quad
  Dabing Sheng\equalcontrib$^{1}$ \\
  \And
  Mufan Xue\equalcontrib$^{1}$ \quad
  Yaochu Jin\corrauth$^{3}$ \quad
  Guoyuan Yang\corrauth$^{1}$ \\
  \\
  $^{1}$Beijing Institute of Technology \quad
  $^{2}$Zhejiang University \quad
  $^{3}$Westlake University \\[0.5em]
  \texttt{mzimo@bit.edu.cn} \quad
  \texttt{jinyaochu@westlake.edu.cn} \quad
  \texttt{yanggy@bit.edu.cn}
}

\begin{document}

\maketitle
\begingroup
\renewcommand{\thefootnote}{\fnsymbol{footnote}}
\footnotetext[1]{Equal contribution}
\footnotetext[2]{Corresponding authors}
\endgroup

\begin{abstract}
    We present SAE-BrainMap, a novel framework that directly aligns deep learning visual model representations with voxel-level fMRI responses using sparse autoencoders (SAEs). First, we train layer-wise SAEs on model activations and compute the correlations between SAE unit activations and cortical fMRI signals elicited by the same natural image stimuli with cosine similarity, revealing strong activation correspondence (maximum similarity up to 0.76). Depending on this alignment, we construct a voxel dictionary by optimally assigning the most similar SAE feature to each voxel, demonstrating that SAE units preserve the functional structure of predefined regions of interest (ROIs) and exhibit ROI-consistent selectivity. Finally, we establish fine-grained hierarchical mapping between model layers and the human ventral visual pathway, also by projecting voxel dictionary activations onto individual cortical surfaces, we visualize the dynamic transformation of the visual information in deep learning models. It is found that ViT-B/16\textsubscript{CLIP} tends to utilize low-level information to generate high-level semantic information in the early layers and reconstructs the low-dimension information later. Our results establish a direct, downstream-task-free bridge between deep neural networks and human visual cortex, offering new insights into model interpretability. 

\end{abstract}

\section{Introduction}
\label{introduction}
Deep neural networks (DNNs) share several fundamental similarities to the brain: both comprise vast populations of interconnected units (neurons), are  well hierarchically organized, and demonstrate remarkable capabilities in solving complex visual tasks~\cite{dicarloUntanglingInvariantObject2007, yangBrainDecodesDeep2024, DifferencesSelectivityNatural}.
Despite decades of research, we understand the brain much better than DNNs, which are often treated as "black-boxes" due to their complex internal dynamics~\cite{oikarinenCLIPDissectAutomaticDescription2023, zouRepresentationEngineeringTopDown2023}.
Forging explicit links between these systems and leveraging our biological understanding to interpret DNNs holds promise for improving model transparency and deepening our insights into the computational principles underlying deep learning~\cite{cerdasBrainACTIVIdentifyingVisuosemantic2024, conwellLargescaleExaminationInductive2024a, xuLimitsVisualRepresentational2021}.
However, building such connections remains a major challenge due to the fundamental differences in structural and computational procedures between artificial and biological neural systems.

To bridge this gap, researchers have developed two complementary strategies:
(1) \textit{Representation-based analysis} methods compare activation patterns in DNNs and the brain in response to the same visual stimuli~\cite{conwellLargescaleExaminationInductive2024a, xuLimitsVisualRepresentational2021, khaligh-razaviFixedMixedRSA2017, kriegeskorteNeuralTuningRepresentational2021a}. A widely used technique, Representational Similarity Analysis (RSA)~\cite{kriegeskorteRepresentationalSimilarityAnalysis2008}, identifies hierarchical correspondences between DNN layers and visual cortex regions. Its extension, voxel-encoding RSA (veRSA)~\cite{conwellLargescaleExaminationInductive2024a, khaligh-razaviFixedMixedRSA2017}, further improves alignment by linearly reweighting DNN features to predict voxel-level responses. 
(2) \textit{Prediction-based modeling} approaches aim to predict brain activity from DNN features elicited by visual input~\cite{luoBrainSCUBAFineGrainedNatural2024, luoBrainMappingDense2024}. FactorTopy~\cite{wangBetterModelsHuman2023} introduces topological constraints and factorized feature selection to enhance the stability and interpretability of brain-model alignment. Conversely, BrainDiVE~\cite{luoBrainDiffusionVisual} reverses this mapping, generating images that maximize activation in target brain regions, thereby enabling data-driven cortical exploration. 

Despite revealing intriguing correspondences between DNNs and brain activity, both strategies face significant limitations. \textit{Representation-based analysis} methods is constrainted to revealing correlations at the level of representational geometry and heavily rely on the manually selected ROIs, failing to support additional insights and fine-grained variations in the representations~\cite{conwellLargescaleExaminationInductive2024a, raghuVisionTransformersSee2021}. \textit{Prediction-based modeling} approaches, while effective, typically rely on supervised learning paradigms that are susceptible to dataset scale, task bias, and neural noise~\cite{luoBrainMappingDense2024, yangCLIPMSMMultiSemanticMapping2025}. Critically, \textit{prediction-based modeling} methods construct only indirect mappings and depend heavily on large-scale datasets and the performance of the downsteam task. This prompts a fundamental question: \textit{can we design a more direct and downstream-task-free alignment between model and brain representations?}

To address this, we introduce SAE-BrainMap, a framework that leverages sparse autoencoders 
(SAEs)~\cite{cunninghamSparseAutoencodersFind2023} to directly bridge visual model representations and voxel-wise functional magnetic resonance imaging (fMRI) responses based on shared visual stimuli. Our motivation for this idea is: SAEs decompose internal model activations into interpretable, disentangled features whose activations reflect the model's functional performance~\cite{marksSparseFeatureCircuits2025}, while fMRI measures cortex activities by capturing changes in blood oxygenation, reflecting the functional role of voxels in visual processing~\cite{allenMassive7TFMRI2022}.

Our findings can be divided into three main components. (1) Notably, we find a strong activation similarity between SAE units and cortical voxels (with a maximum cosine similarity of up to 0.76), enabling a direct, fine-grained hierarchical alignment between model layers and brain cortex. (2) We furthermore construct a voxel dictionary by optimally assigning units exhibiting the highest activation pattern similarity to each voxel. We figure out that SAEs units could keep the functional structure of brain ROIs, also voxels and units with similar activation patterns exhibit functional similarity. (3) By projecting voxel dictionary activations onto the cortical surface, we use the brain as a functional template to trace the visual processing dynamics during the model visual procedure. Specifically in CLIP ViT-B/16, we observe a semantic information transformation by utilizing low-level information to generate high-level abstract representation across the early layers, while the later layers simultaneously engage both levels of information, with an increasing emphasis on abstract semantics.

\begin{figure}[!t]
    \centering
    \includegraphics[width=1\linewidth]{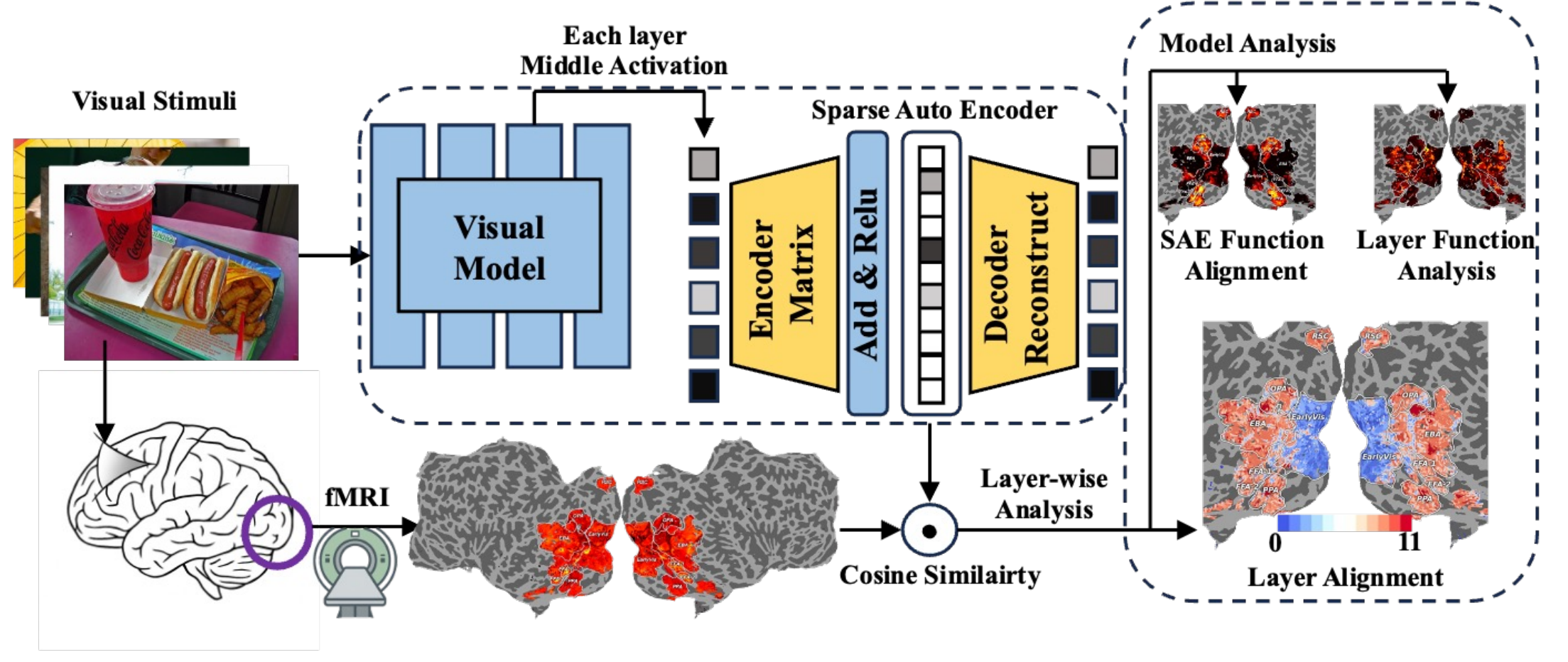}
    \caption{\textbf{Architecture of SAE-BrainMap.} First, our framework first trains Sparse Autoencoders for \textit{each model layer}. Second, it computes \textit{cosine similarity matrix} between SAE units activations and cortical fMRI responses. Finally, it interprets deep learning model with brain based on the \textit{cosine similarity matrix}.}
    \label{fig1}
\end{figure}

\section{Relate Work}
\label{section 2}

\paragraph{Sparse Autoencoders (SAEs)}
Sparse Autoencoders (SAEs) are a mechanistic interpretability method 
that reverse-engineers deep learning models by decomposing their activation 
vectors into linear combinations of interpretable, monosemantic sparse features~\cite{zouRepresentationEngineeringTopDown2023, cunninghamSparseAutoencodersFind2023, leaskSparseAutoencodersNot2025}. 
Currently, SAEs have been widely applied in interpretability studies 
of both Large Language Models~\cite{cunninghamSparseAutoencodersFind2023, leaskSparseAutoencodersNot2025, gaoScalingEvaluatingSparse2024, rajamanoharanJumpingAheadImproving2024} and Vision Models~\cite{felArchetypalSAEAdaptive2025, thasarathanUniversalSparseAutoencoders2025, zaigrajewInterpretingCLIPHierarchical2025, tianSparseAutoencoderZeroShot2025, gortonMissingCurveDetectors2024}, mainly because its ability to tackle the 
superposition phenomenon, individual neurons is often polysemantic~\cite{elhageToyModelsSuperposition2022, sharkeyInterimResearchReport2022, yunTransformerVisualizationDictionary2023}.
It could uncover several model functional circuits~\cite{cunninghamSparseAutoencodersFind2023, marksSparseFeatureCircuits2025}, characterize the representational roles of individual functional 
vectors within the feature space~\cite{leaskSparseAutoencodersNot2025, gaoScalingEvaluatingSparse2024, felArchetypalSAEAdaptive2025}, and illuminate shared information 
across diverse deep learning models~\cite{thasarathanUniversalSparseAutoencoders2025}. The sparse vector activations extracted by SAEs accurately capture model's function, revealing what the model 
does and to what extent~\cite{cunninghamSparseAutoencodersFind2023, shuSurveySparseAutoencoders2025}, whereas fMRI measures blood-oxygen consumption 
to indicate neural activity under different tasks or stimuli, thereby 
uncovering functional specialization across brain areas~\cite{yangCLIPMSMMultiSemanticMapping2025, wangBetterModelsHuman2023, allenMassive7TFMRI2022, khoslaHighlySelectiveResponse2022, xueConvolutionalNeuralNetwork2024}. Consequently, 
we aim to leverage SAEs as a bridge to explore the alignment between 
deep learning models and human brain.

\begin{wrapfigure}{r}{0.4\textwidth}
    \centering
    \includegraphics[width=0.38\textwidth]{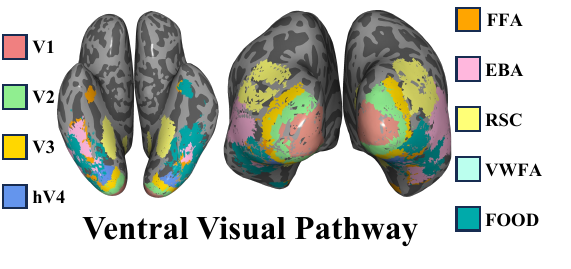}
    \caption{Visualizing S5 selective ROIs}
    \label{fig2}
  \end{wrapfigure}
\paragraph{Ventral Visual Pathway} \label{ventral visual pathway} 
Brain processes visual information in a hierarchical manner~\cite{dicarloUntanglingInvariantObject2007, goodaleSeparateNeuralPathways1994, grill-spectorFunctionalArchitectureVentral2014, SelectivityShapeSizea}. 
Visual information starts from the early visual cortex V1, and progresses through V2, V3, 
and hV4. Voxels along this pathway exhibit increasingly 
larger receptive fields~\cite{dumoulinPopulationReceptiveField2008} and transit from 
encoding low-level visual features such as edges and curvature~\cite{hubelReceptiveFieldsBinocular1962, WeKnowWhat} to processing more complex attributes, including shape, color, texture, and depth~\cite{DifferencesSelectivityNatural, ReceptiveFieldsFunctional, freemanFunctionalPerceptualSignature2013}.
Subsequently, high-level semantic understanding is accomplished in the high-level visual cortex.
Notable examples include the fusiform face area (FFA) for face selectivity~\cite{Kanwisher4302}, 
the retrosplenial cortex (RSC) for scene selectivity~\cite{Epstein1998}, 
the extrastriate body area (EBA) for body selectivity~\cite{Downing2001}, 
the visual word form area (VWFA) for word selectivity~\cite{Cohen2000}, 
and a region showing selectivity for food-related stimuli~\cite{KHOSLA20224159}. 
Our study centers on the nine ROIs mentioned above, as illustrated in Figure \ref{fig2}.

\paragraph{Model-Brain Alignment}
Model-brain alignment aims to investigate the correspondence between deep neural networks (DNNs) and the human brain~\cite{yangBrainDecodesDeep2024, conwellLargescaleExaminationInductive2024}.
Current approaches fall into two broad families: Representational Similarity Analysis (RSA) ~\cite{kriegeskorteRepresentationalSimilarityAnalysis2008a} and brain-encoder method  ~\cite{wangBetterModelsHuman2023}.
Using veRSA, it is shown in~\cite{khaligh-razaviFixedMixedRSA2017} that lower network layers 
align best with early visual cortex, whereas higher layers align with higher-level 
visual areas. Xu et al. ~\cite{xuLimitsVisualRepresentational2021} reported  
representational divergence between high-level visual cortex and the upper layers of CNNs. 
Brain-encoder method work such as FactorTopy ~\cite{yangBrainDecodesDeep2024} maps 
DNN layers to voxels, effectively “painting” the network onto the cortex at fine spatial resolution. 
The work in ~\cite{luoBrainSCUBAFineGrainedNatural2024, cerdasBrainACTIVIdentifyingVisuosemantic2024, luoBrainDiffusionVisual} reverse the alignment and describe the cortical function with the generation model. Each line of work has limitations: RSA depends on predefined ROIs and therefore cannot 
deliver voxel-wise insight, while voxel-wise encoding models provide only indirect 
correspondence because they hinge on a downstream prediction task and are sensitive to its performance. Our approach 
leverages Sparse Autoencoders to achieve direct and training-free voxel-wise model-brain alignment, eliminating the dependence on downstream prediction performance like ~\cite{yangBrainDecodesDeep2024} and without the need for predefined ROIs as in ~\cite{conwellLargescaleExaminationInductive2024a}.

\section{Method}

We aim to establish a voxel-wise directly correspondence between deep learning models
and the ventral visual pathway using paired natural image stimulis and fMRI signals. First,
we introduce the background of SAEs. Then we compare the similarity between the activation 
patterns of SAEs' units and the fMRI signals. Finally, we describe how to utilize the 
similarity matrix to establish a model-brain mapping by visualizing each layer's function in visual process.
\subsection{SAEs Background and Training}
Sparse autoencoders aim to reconstruct the model activation through an encoder-decoder architecture.
Given a dataset of $n$ images $\mathcal{I}$ and a DNN model to embed the images $f : \mathcal{I} \to A$.
Define the image embeddings as $A \in \mathbb{R}^{n \times d}$, where $d$ denotes the embedding dimension, and $A$ refers to the "CLS" token in most models, except in the SAM model, where it is computed as the average embedding of all patch tokens.
SAEs learn a linear encoder $\Psi_{\theta}(\cdot)$ that maps embeddings $A$ into the activation of $k$ sparse features $Z \in \mathbb{R}^{n \times k}$.
The linear decoder $\Phi_{\theta}(\cdot)$, which shares the same linear weight $W^T \in \mathbb{R}^{k \times d}$ with the encoder, 
reconstructs the original embeddings $\hat{A}$ from the sparse features activation $Z$. We setup SAEs as ~\cite{cunninghamSparseAutoencodersFind2023}. It can be written as Equation \ref{q1},\ref{q2}:
\begin{align}
    Z = \Psi_{\theta}(A) &= ReLU(WA + b) \label{q1} \\ 
    \hat{A} = \Phi_{\theta}(Z) &= W^TZ = \sum_{i=0}^{k-1} W^T_{i}Z_{i} \label{q2},
\end{align}
where $W \in \mathbb{R}^{d \times k}$ and $b \in \mathbb{R}^{k}$ are the learned parameters
Each units $W^T_{i}$ is a sparse feature vector capturing the core components of the embeddings, which make up the SAEs feature dictionary~\cite{cunninghamSparseAutoencodersFind2023}. 
Reconstructed embeddings $\hat{A}$ is the linear combination of the units with positive activations.
The training process aims to minimize the loss function $\mathcal{L}$ in Equation \ref{q3}. $\alpha$ is a hyperparameter controlling the sparsity of the reconstruction.
\begin{align}
    \mathcal{L}(A, \hat{A}, Z) = \| \hat{A} - A \|_2^2 + \alpha \| Z \|_1 \label{q3}
\end{align}
We train SAEs for each layer of a deep learning model. Indicate $A^{(j)}$ is the output after the $j$-th layer process, 
$\Psi_{\theta}^{(j)}$ is the encoder of the j-th layer and $Z^{(j)}$ is the SAEs activation of the $j$-th layer. The 
extraction of $A^{(j)}$ is mentioned in the Appendix.

\subsection{SAE-BrainMap}

SAE-BrainMap aims to interpret model's visual procedures by establishing a voxel-wise model-brain alignment.
For the given image stimuli $\mathcal{I} \in \mathbb{R}^{n \times 3 \times H \times W}$, we can extract the brain fMRI z-score values as a $v$ element vector $B \in \mathbb{R}^{n \times v}$,
and the SAEs' units activation sets $[Z_{zscore}^{(0)}, Z_{zscore}^{(1)}, ..., Z_{zscore}^{(L)}]$, where $Z_{zscore}^{j}$ is Z-Scored SAE unit activations (without-ReLU), matching the z-score preprocessed of fMRI signals~\cite{allenMassive7TFMRI2022}, $v$ is the number of voxels in our selected ROIs, as referred to in Section \ref{ventral visual pathway}, and $L$ is the all layer number in the deep learning model.
We use the cosine similarity to measure the correspondence between two activation patterns as Equation \ref{q4}.
\begin{align}
    M^{(j)}_{\text{simi}} = \cos\left( \frac{B^\top}{\|B\|_2}, \frac{Z_{zscore}^{(j)}}{\|Z_{zscore}^{(j)} \|_2} \right), \quad &Z_{zscore}^{(j)} = Zscore(WA^{(j)} + b) \label{q4} \\ 
    \mathcal{C}[x] = \operatorname*{argmax}_{{j \in \{0, 1, \dots, L\}}}& \left ( \max_{y} M^{j}_{x, y} \right ) \label{q5}.
\end{align}
We performed normalization along the dimension $n$ corresponding to the number of images, and get an activation similarity map $M^{(j)}_{Simi} \in \mathbb{R}^{v \times n}$ for each layer $j$, where $M^{(j)}_{simi}[x, y]$ indicate the activation similarity between the $x$-th voxel and the $y$-th SAEs units. To investigate the model's overall information processing trend, we identify the most corresponding layer for each voxel $x$ and get a voxel-wise layer alignment matrix $\mathcal{C} \in \mathbb{R}^{v}$, as described Equation \ref{q5}.

By establishing a voxel dictionary $\mathcal{D}^{(j)} \in \mathbb{R}^{v \times d}$ for $j$-th layer, we can map SAEs' units activation onto the template of brain cortex. Specifically, for each voxel $x$, the model selects the most similar SAE unit by identifying the index $y$ that maximizes the similarity $M_{x,y}$ between voxel $x$ and unit $y$. The corresponding representation $W[y]$ of that unit is then assigned to $\mathcal{D}^{(j)}[x]$, effectively associating each voxel with its most representative model feature.
This mapping allows us to visualize the model layer's corresponding SAEs activation patterns $\mathcal{S}^{(j)} \in \mathbb{R}^{v}$ with human visual cortex, as Equation \ref{q6}.
\begin{align}
    \mathcal{D}^{(j)}[x] &= W[ \operatorname*{argmax}_y M_{x, y}], \quad x \in \{0, 1, \dots, v\} \label{q6} \\
    \mathcal{S}^{(j)} &= A^{(j)}D^{(j)} \quad j \in \{0, 1, \dots, L\} \label{q7}.
\end{align}

\section{Experiment}

In this section, we first introduce the similarities between the SAE's units activation 
and voxels' fMRI signal. We then explore the functional alignment between units and ROIs with high activation correspondence. Based on the alignment, we paint model layers onto the cortex and visualize image information procedure with the help of the brain template.
\subsection{Setup}
\begin{figure}[!t]
    \centering
    \includegraphics[width=1\linewidth]{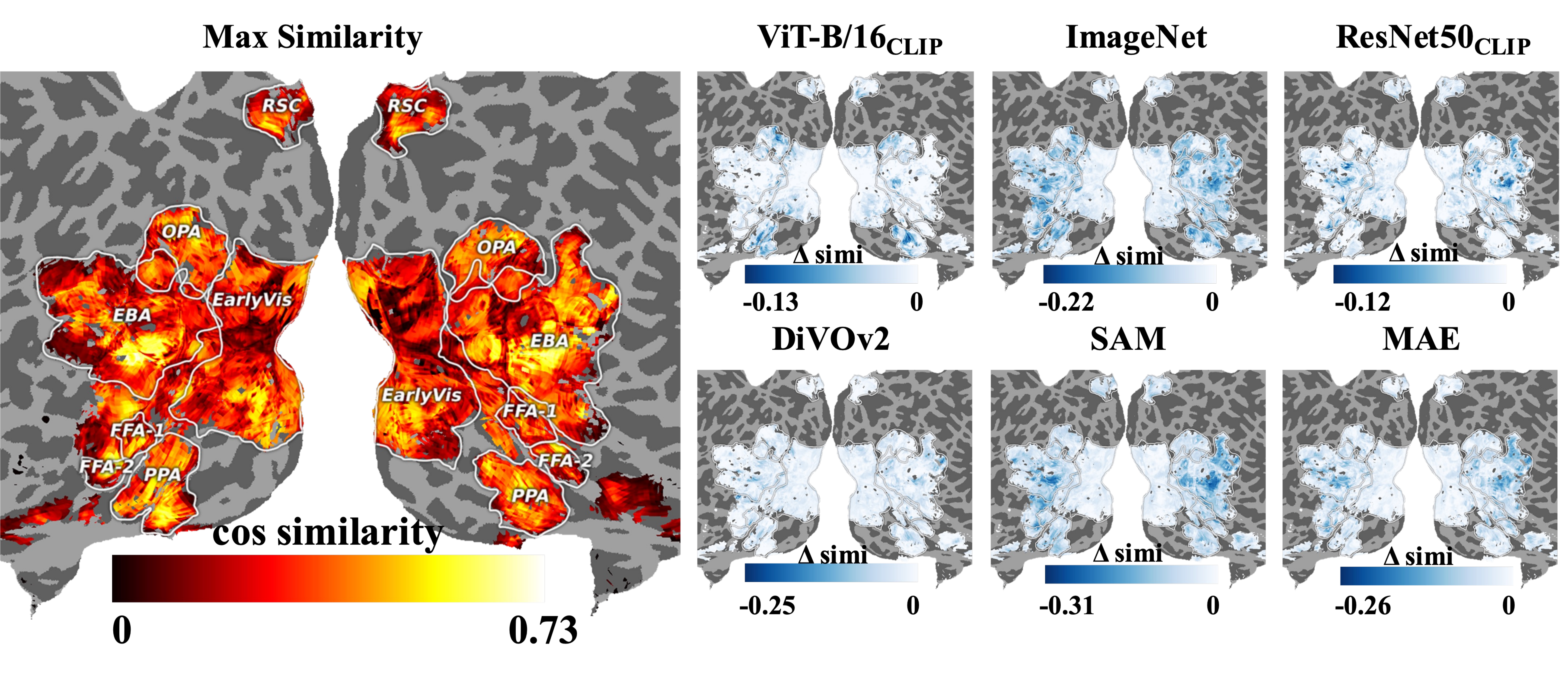}
    \caption{\textbf{Brain-SAEs Activation Similarity on S5.} Left: voxel-wise maximum similarity score across models. Right: difference in similarity scores between models and the voxel-wise maximum}
    \label{fig3}
\end{figure}
    
\begin{table}[!t]
    \centering
    
    \resizebox{\textwidth}{!}{
      \begin{tabular}{c 
                      cc cc cc cc cc cc cc}
        \toprule
        Subject 
        & \multicolumn{2}{c}{All Model} 
        & \multicolumn{2}{c}{ViT-B/16\textsubscript{CLIP}} 
        & \multicolumn{2}{c}{ImageNet} 
        & \multicolumn{2}{c}{MAE} 
        & \multicolumn{2}{c}{DiNOv2} 
        & \multicolumn{2}{c}{SAM} 
        & \multicolumn{2}{c}{ResNet50\textsubscript{CLIP}} 
        \\
        & Max & Avg
        & Max & Avg 
        & Max & Avg 
        & Max & Avg 
        & Max & Avg 
        & Max & Avg 
        & Max & Avg 
        \\
        \midrule
        S1 & \textbf{0.762} & 0.259 & 0.618 & 0.249 & 0.503 & 0.220 & 0.605 & 0.239 & 0.484 & 0.231 & 0.421 & 0.212 & \textbf{0.644} & 0.244 \\
        S2 & 0.716 & 0.265 & 0.573 & 0.254 & 0.470 & 0.228 & 0.523 & 0.245 & \textbf{0.517} & 0.238 & \textbf{0.489} & 0.215 & 0.622 & 0.252 \\
        S3 & 0.705 & 0.236 & 0.561 & 0.225 & \textbf{0.519} & 0.200 & \textbf{0.513} & 0.218 & 0.507 & 0.210 & 0.444 & 0.193 & 0.615 & 0.227 \\
        S4 & 0.666 & 0.233 & 0.547 & 0.224 & 0.430 & 0.198 & 0.509 & 0.215 & 0.492 & 0.206 & 0.431 & 0.189 & 0.542 & 0.220 \\
        S5 & 0.725 & \textbf{0.293} & \textbf{0.591} & \textbf{0.281} & 0.502 & \textbf{0.250} & 0.478 & \textbf{0.270} & 0.470 & \textbf{0.258} & 0.413 & \textbf{0.235} & 0.604 & \textbf{0.280} \\
        S6 & 0.687 & 0.238 & 0.550 & 0.229 & 0.484 & 0.203 & 0.512 & 0.220 & 0.492 & 0.208 & 0.440 & 0.193 & 0.593 & 0.226 \\
        S7 & 0.657 & 0.227 & 0.545 & 0.217 & 0.425 & 0.193 & 0.506 & 0.210 & 0.468 & 0.202 & 0.401 & 0.184 & 0.544 & 0.216 \\
        S8 & 0.639 & 0.196 & 0.533 & 0.188 & 0.451 & 0.169 & 0.461 & 0.182 & 0.456 & 0.175 & 0.401 & 0.166 & 0.538 & 0.186 \\
        \midrule
        \textbf{Cos Avg} 
          & 0.695 & 0.243 & 0.565 & 0.233 & 0.473 & 0.207 & 0.513 & 0.225 & 0.486 & 0.216 & 0.430 & 0.198 & 0.588 & 0.231 \\
         \textbf{Predict Cos Avg}& 
         0.850 & 0.337 & 0.846 & 0.275 & 0.809 & 0.259 & 0.740 & 0.223 & 0.838 & 0.264 & 0.746 & 0.242 & 0.848 & 0.190 \\
        \bottomrule \\
        \end{tabular}
    }

    \caption{\textbf{Brain-SAEs activation similarity}. All models SAEs units shares a similar activation pattern with voxels fMRI response. The final row presents the \textit{Brain Encoder} predict avg cosine similarity with the fMRI response. The predict performance likely due to their inner similarity.}
    \label{table1}
\end{table}

\label{setup}
The fMRI dataset we use in the experiment is Nature Scenes Dataset (NSD)~\cite{allenMassive7TFMRI2022}, 
which includes high-density fMRI data from eight participants. Each of the subject 
was expected to view 10,000 natural scene images selected from COCO~\cite{linMicrosoftCOCOCommon2015} for three times, while all the subject share the same set of 1000 images and an independent set of 9000 images. 
fMRI beta values were z-scored across runs and averaged across up to three repetitions per image, yielding one fMRI response per voxel per image. Four of the subjects (S1,S2,S5,S7) finished all the experiment, while another four did not. We performed the same preprocessing for subjects who did not finish all the tasks as~\cite{yangCLIPMSMMultiSemanticMapping2025, allenMassive7TFMRI2022, xueConvolutionalNeuralNetwork2024}.
The cortical visualization results were rendered in each subject's native space using Pycortex~\cite{gaoPycortexInteractiveSurface2015}. 

We train SAEs for each layer of each model in our analyses. SAEs' hidden feature dimension $k = R \times d$, 
where $d$ denotes target layer's activation dimension, and $R$ is a hyperparameter which is kept
$R=16$ for all of the models. The hyperparameter $\alpha=0.00086$, as ~\cite{cunninghamSparseAutoencodersFind2023}. 
We train all SAEs with the ImageNet-1k (2012) training set~\cite{dengImageNetLargeScaleHierarchical} for 3 epochs and learning rate set to $5 \times 10^{-5}$. 
The pretrained vision models used in our experiment are OpenAI CLIP visual backbone ViT-16/B and ResNet50~\cite{radfordLearningTransferableVisual2021, heDeepResidualLearning2015}, 
MAE ViT-B/16~\cite{heMaskedAutoencodersAre2021}, DiNOv2 ViT-B/14~\cite{oquabDINOv2LearningRobust2024}, SAM ViT-B/14~\cite{kirillovSegmentAnything2023} 
and ImageNet trained ViT-16/B~\cite{dosovitskiyImageWorth16x162021a}(the model is named ImageNet in the paper) loaded from pytorch~\cite{paszkePyTorchImperativeStyle2019}. More training information could be checked in Appendix~\ref{appendix3}.

\subsection{Brain-SAEs Activation Similarity}

We extract the SAE units' activations on each subject's independent images and directly compare them with the subject's fMRI responses using cosine similarity as Equation~\ref{q4}. 
The voxel activations and SAE unit activations exhibit high similarity, indicating a shared representational structure between the brain and the model. More importantly, this phenomenon
widely exists within models with different architectures, different training datasets and different training method. 
We visualize the maximum similarity for each voxel across all models and layers, 
along with the difference between each model's max similarity within their layers and the maximum for subject 5 in Figure \ref{fig3}. We also find that the CLIP 
visual backbone achieves the highest average similarity to brain responses among all visual models, 
while SAM exhibits the lowest similarity. This discovery is in consistent with the recent work showing that CLIP perform better on brain activation prediction downstream tasks and voxel function analysis~\cite{cerdasBrainACTIVIdentifyingVisuosemantic2024, luoBrainSCUBAFineGrainedNatural2024, luoBrainMappingDense2024, wangBetterModelsHuman2023, luoBrainDiffusionVisual}. 
By training a brain encoder for each model~\cite{wangBetterModelsHuman2023}, we found that brain encoder models high predict performance is likely due to their inner correspondences as the prediction performance is comparable to their intrinsic similarity. as shown in Table~\ref{table1}.

\begin{figure}[!t]
    \centering
    \includegraphics[width=1\linewidth]{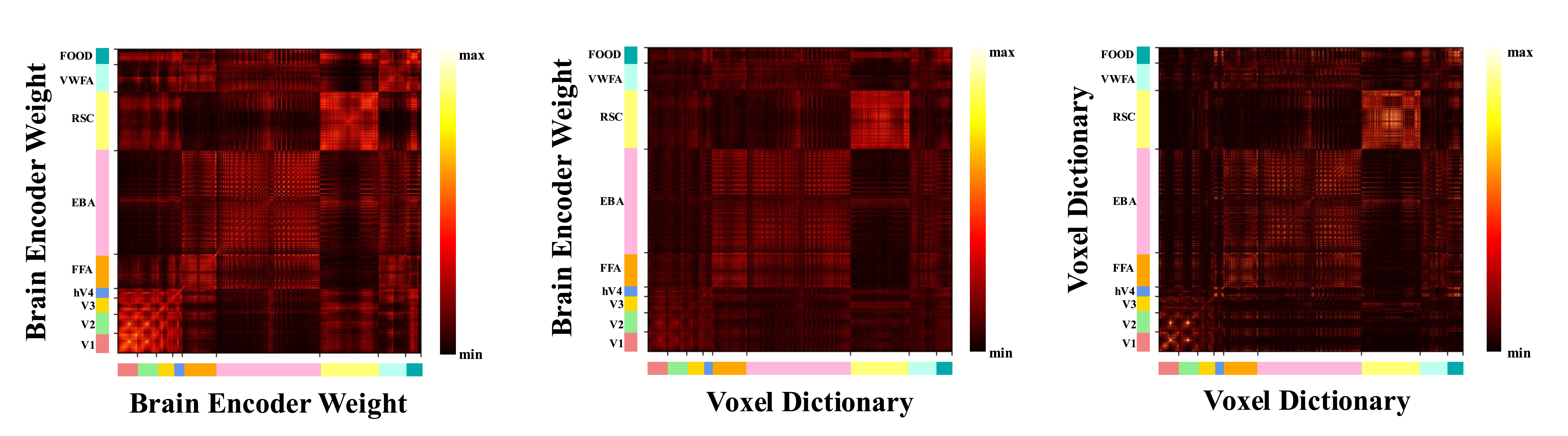}
    \caption{\textbf{Representation Similarity Matrix} between the brain encoder weights of the last layer of ViT-B/16\textsubscript{CLIP}, the voxel dictionary, and their mutual similarity. The voxel dictionary preserves the brain's functional structure and share a high RSA score with brain encoder weight.}

    \label{fig4}
\end{figure}
\begin{table}[!t]
    \centering

    \begin{tabular}{cccccccccc}
    \toprule
    \textbf{Model} & \textbf{S1} & \textbf{S2} & \textbf{S3} & \textbf{S4} & \textbf{S5} & \textbf{S6} & \textbf{S7} & \textbf{S8} & \textbf{Avg}\\
    \midrule
    ViT-B/16\textsubscript{CLIP} & 0.048 & 0.134 & 0.223 & 0.219 & \textbf{0.252} & 0.207 & 0.208 & 0.214 & 0.188 \\
    Imagenet & 0.222 & 0.247 & 0.214 & 0.221 & \textbf{0.258} & 0.204 & 0.183 & 0.176 & 0.216 \\
    MAE & 0.484 & 0.523 & 0.492 & 0.484 & 0.509 & \textbf{0.516} & 0.502 & 0.498 & 0.501 \\
    DiNOv2 & 0.210 & 0.211 & 0.198 & 0.198 & \textbf{0.230} & 0.179 & 0.185 & 0.143 & 0.194 \\
    SAM & 0.368 & 0.379 & 0.388 & 0.403 & \textbf{0.407} & 0.358 & 0.336 & 0.328 & 0.371 \\
    ResNet50\textsubscript{CLIP} & 0.342 & \textbf{0.413} & 0.400 & 0.361 & 0.376 & 0.335 & 0.300 & 0.331 & 0.357 \\
    \bottomrule \\
    \end{tabular}

    \caption{\textbf{RSA Similarity Score} between the brain encoder weights of the last layer and the voxel dictionary. This indicates that SAE units maintain a strong structural correspondence with the brain encoder weights, which are regarded as the ground truth representation of cortical functional structure.}
    \label{table2}
\end{table}

\subsection{Brain-Model Function Correspondence}
\label{section 4.3}

In this section, we indicate that SAE units not only exhibit high activation similarity with brain voxels but also align functionally with the brain's prior distribution. We first establish a voxel dictionary as defined in Equation~\ref{q6} and organize it according to the ROIs. We also train a brain encoder for each layer and regard the encoder weights as the ground truth of brain function, due to their outstanding performance in brain functional analysis ~\cite{cerdasBrainACTIVIdentifyingVisuosemantic2024, luoBrainSCUBAFineGrainedNatural2024, luoBrainDiffusionVisual}.

By computing the representation correlation matrix based on pairwise cosine similarities, we find that the voxel dictionary preserves functional separability across ROIs and shares a high RSA score with the ground truth (highest: 0.516, see Table~\ref{table2}). We visualize the similarity matrix between the brain encoder weights and the voxel dictionary for the last layer of ViT-B/16\textsubscript{CLIP} in Figure~\ref{fig4}, which indicates that the voxel dictionary retains the ROI prior.

In addition, we select units that show the highest similarity to voxels within specific ROIs. We find that these units' functions are consistent with the functional priors of those ROIs. Specifically, we select SAE units from the last layer of ViT-B/16\textsubscript{CLIP} that are most similar to the high-level visual cortex, FFA, EBA, RSC, VWFA, and FOOD regions. We then estimate the activation of these units on the COCO validation dataset ~\cite{linMicrosoftCOCOCommon2015} and Broden~\cite{bauNetworkDissectionQuantifying2017}, selecting the top-5 activated images. Next, we estimate the unit activations on the patches of these selected images, and apply bilinear interpolation to produce pixel-wise activation heatmaps, following the methods in ~\cite{felArchetypalSAEAdaptive2025, thasarathanUniversalSparseAutoencoders2025} (see Figure~\ref{fig4}).

We find that the selected units perform functions corresponding to their associated ROIs—such as face, body, space, word, and food selectivity. This demonstrates that SAE units not only show high similarity to brain voxel activations but also exhibit similar functional properties. For each unit, we obtain a vector $M_{\text{simi}}[:, y] \in \mathbb{R}^{v}$ representing its similarity to all voxels. By visualizing $M_{\text{simi}}[:, y]$ on the cortex, we can describe each unit's function in terms of brain regions.

We visualize the selected units' unit-to-voxel similarity in Figure~\ref{fig4}, and observe that the unit most similar to the FFA and EBA shares similar pattern due to their selectivity for human faces and bodies. Additionally, the unit most similar to the FOOD area also shows strong similarity to the early visual cortex, suggesting that the perception of food requires substantial low-level semantic information.

\begin{figure}[!t]
    \centering
    \label{fig5}
    \includegraphics[width=1\linewidth]{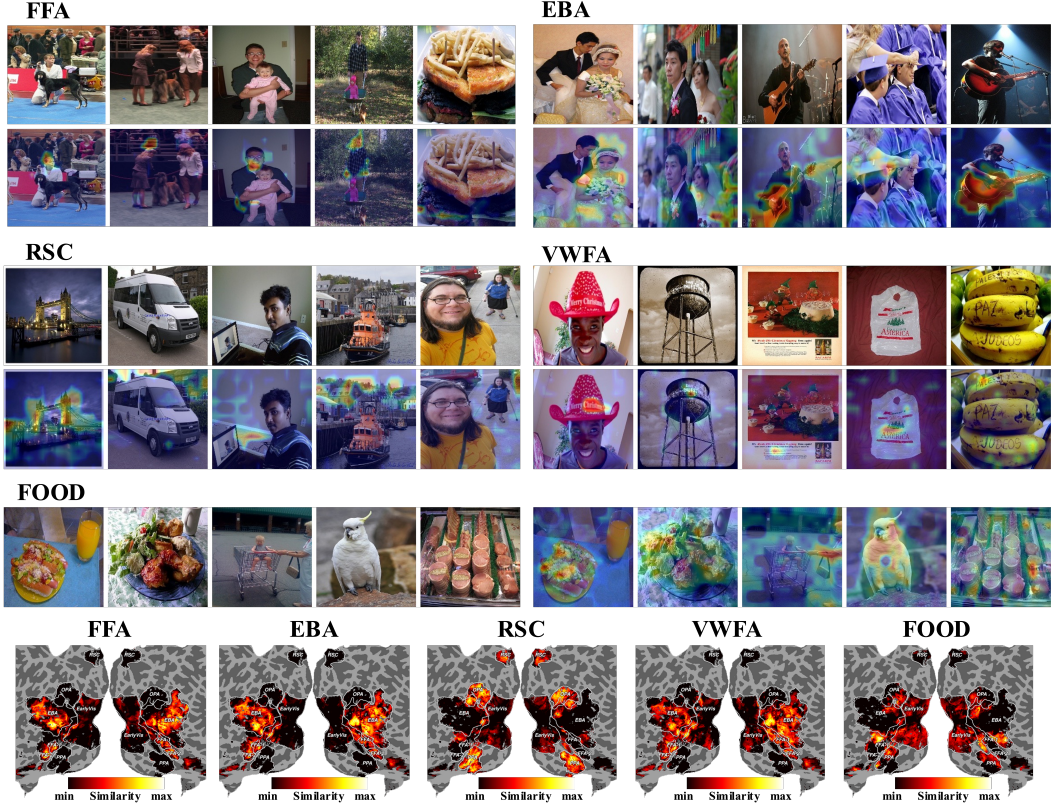}

    \caption{\textbf{Visualization of Unit Selectivity and Functionality.} \textbf{Top}: Top-5 activated images for units with highest similarity to voxels in each ROI on ViT-B/16\textsubscript{CLIP}, demonstrating consistent selectivity for faces, bodies, scenes, words, and food. \textbf{Bottom}: Cortical visualization of unit-to-voxel similarity}
    
\end{figure}
\subsection{Brain-Model Layer Alignment}

\begin{figure}[!t]
    \centering
    \includegraphics[width=1\linewidth]{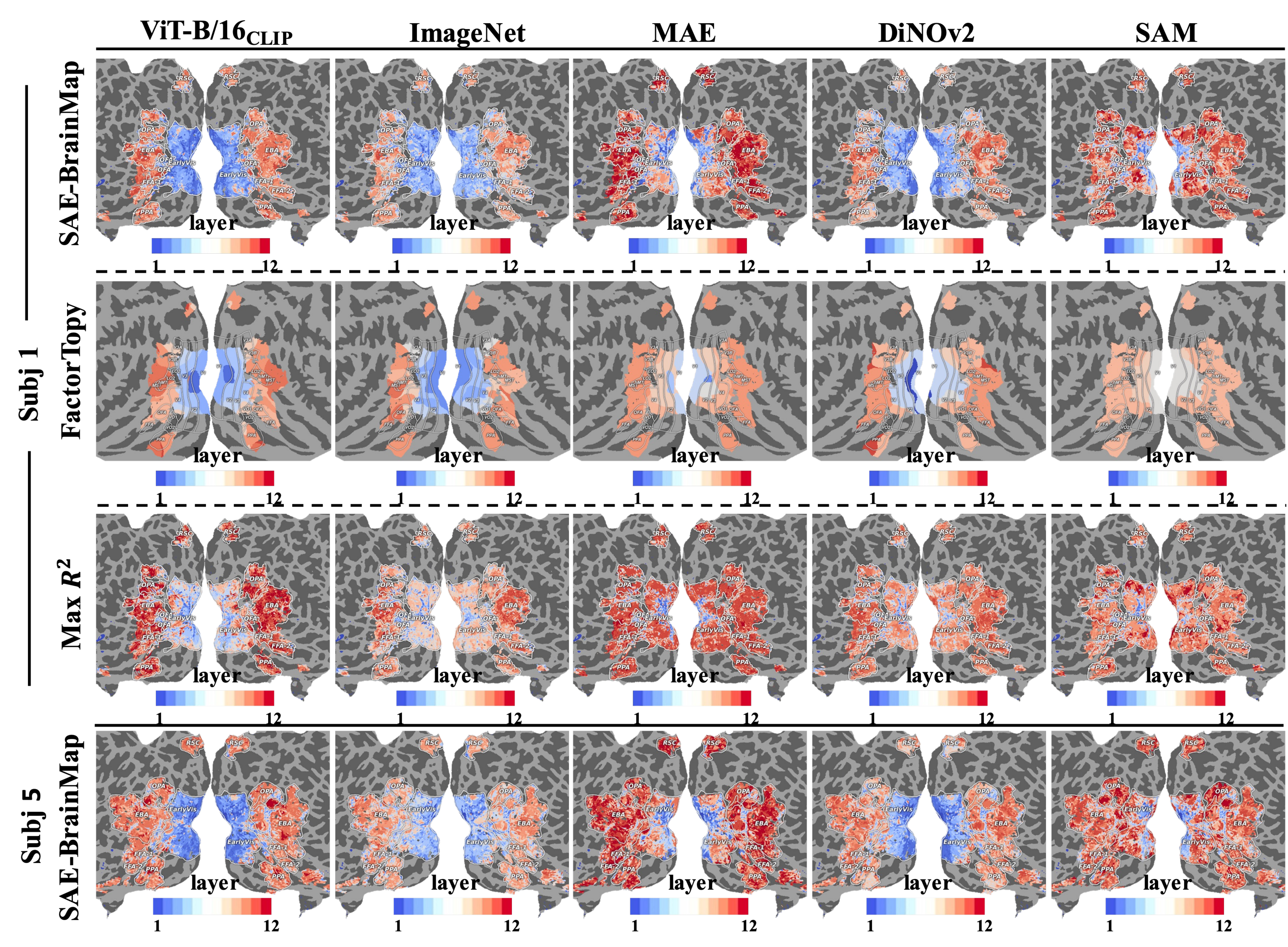}
    \caption{\textbf{Brain-Model Layer Alignment}. Top: Layer alignment for S1 based on SAE-BrainMap, FactorTopy~\cite{yangBrainDecodesDeep2024} and Max $R^2$~\cite{wangBetterModelsHuman2023}, we visualize five models with attention structure and have 12 layers. Bottom: SAE-BrainMap layer alignment on S5.}
    \label{fig6}
\end{figure}

As shown in Figure~\ref{fig6}, we establish a brain-model layer alignment by matching each voxel to its most similar model layer (Equation.~\ref{q5}), thereby capturing the model's characteristic information processing tendency ~\cite{yangBrainDecodesDeep2024}. All models exhibit a clear and consistent hierarchical correspondence with the human visual cortex: lower layers align with the early visual cortex, while higher layers correspond to the high-level visual cortex. This result is well correlated with the previous methods ~\cite{yangBrainDecodesDeep2024, xueConvolutionalNeuralNetwork2024}.

Table~\ref{table2} presents the cosine similarity between our alignment results and two voxel-wise mapping methods, Max $R^2$ ~\cite{wangBetterModelsHuman2023} and FactorTopy ~\cite{yangBrainDecodesDeep2024}. For supervised models, ViT-B/16\textsubscript{CLIP} and ImageNet-trained ViT-B/16 exhibit highly similar alignment trends. Notably, high-level visual cortex show the strongest similarity to layers 10 and 11, but not layer 12, probably because the last layer is optimized for downstream tasks. In contrast, the SAM model's final layer aligns with the early visual cortex and RSC, suggesting that segmentation tasks rely more heavily on low-level visual features. For self-supervised models, MAE's last layer corresponds to the high-level visual cortex, while DiNOv2 resembles supervised pretrained models, possibly due to differences in training strategy.

\begin{table}[!t]
    \centering
    
    \resizebox{\textwidth}{!}{
      \begin{tabular}{c 
                      ccc ccc ccc ccc ccc}
                      \toprule
                      & \multicolumn{3}{c}{ViT-B/16\textsubscript{CLIP}} 
                      & \multicolumn{3}{c}{ImageNet} 
                      & \multicolumn{3}{c}{MAE} 
                      & \multicolumn{3}{c}{DiNOv2} 
                      & \multicolumn{3}{c}{SAM} 
                      \\
                      & S1 & S5 & Avg 
                      & S1 & S5 & Avg 
                      & S1 & S5 & Avg 
                      & S1 & S5 & Avg 
                      & S1 & S5 & Avg 
                      \\
                      \midrule
                      Similarity with ~\cite{yangBrainDecodesDeep2024} 
                      & 0.79 & 0.54 & 0.69 
                      & 0.91 & 0.84 & 0.70
                      & 0.25 & 0.20 & 0.22
                      & 0.54 & 0.55 & 0.61
                      & 0.18 & 0.25 & 0.22
                      \\
                      Similarity with ~\cite{wangBetterModelsHuman2023}
                      & 0.59 & 0.65 & 0.73
                      & 0.61 & 0.73 & 0.68 
                      & 0.69 & 0.68 & 0.70
                      & 0.82 & 0.88 & 0.84
                      & 0.71 & 0.64 & 0.73
                      \\
                      \textbf{slop} $\uparrow$
                      & 0.93 & 0.95 & 0.94
                      & 0.91 & 0.92 & 0.92
                      & 0.89 & 0.92 & 0.91
                      & 0.92 & 0.94 & 0.93
                      & 0.84 & 0.88 & 0.87 \\
                      \bottomrule

        \end{tabular}
    }

    \caption{\textbf{Similarity and Slope}. The similarity of brain-model layer alignment between SAE-BrainMap and FactorTopy ~\cite{yangBrainDecodesDeep2024}, as well as with the Max $R^2$~\cite{wangBetterModelsHuman2023}. The slope estimates the hierarchical similarity between model and brain, and a higher slope indicates a better alignment.}
    \label{table3}
\end{table}
Table~\ref{table3} reports the \textbf{hierarchy slope} values, quantifying the alignment between each model's layer hierarchy and the brain's ventral visual pathway using the method in~\cite{yangBrainDecodesDeep2024}. To compute this metric, we first separate ROIs into a two-level structure 1.[V1, V2, V3, hV4], 2.[FFA, EBA, RSC, VWFA, FOOD]. Voxels inside these ROIs are assigned with a rough value 
$l_x \in \{0, 1\}$ to represent voxel's prior hierarchical structure $l\in \mathbb{R}^{v}$. Next, we compute slope $b = \langle \frac{f}{\|f\|_2}, \frac{C}{\|C\|_2} \rangle $ to get the slope score. Higher
slope value indicate that model's layer structure shares more similarities with the brain hierarchical structure. We demonstrate that all models have a high slope value, indicate that models process information hierarchically and share a high similarities with brain hierarchical.

\subsection{Visual Information Procedure Analysis}

\begin{figure}[!t]
    \centering
    \includegraphics[width=1\linewidth]{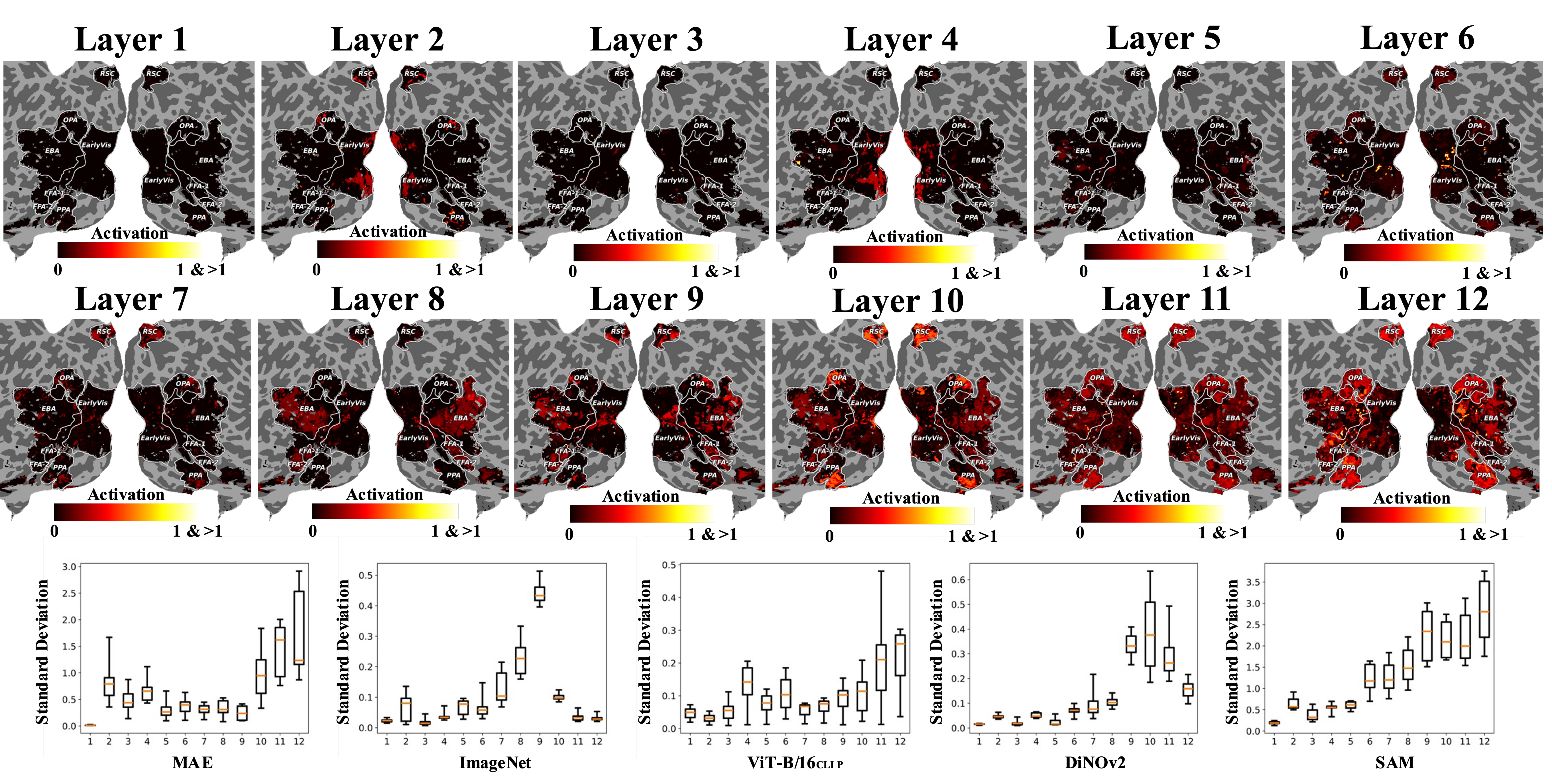}
    \caption{\textbf{Top}: Layer wise Voxel Dictionary average activation for ViT-B/16\textsubscript{CLIP}, S5. This reveals 
    the model forms low-level information first, then generates high-level information while several low-level information disappeared 
    in the middle layers and was reconstructed in the last layers. \textbf{Bottom}: The standard deviation of the Voxel Dictionary 
    activities on test images across subjects. The x-axis denotes the model layers (from 1 to 12), while the y-axis represents the standard deviation, quantifying how much the activations vary across images.}
    \label{fig7}
\end{figure}

We constructed a voxel dictionary \(\mathcal{D}^{(l)}\) for each layer \(l\) and calculated the activation pattern \(\mathcal{S}^{(j)}\) of the SAE units using shared images from each subject as in Equation~\ref{q7}. The average pattern \( Avg(\mathcal{S}^{(j)}) \) across all images is visualized on the brain template, and the layer-wise standard deviation box plots for each subject are illustrate in Figure~\ref{fig7}. 
For ViT-B/16\textsubscript{CLIP}, the layer activation pattern provides insights into the model's hierarchical information processing mechanism: (i) in the early layers, the model primarily processes low-level information; (ii) subsequently, low-level information temporarily diminishes and is transformed into high-level abstract representations such as scenes and bodies; (iii) in the later stages, the model reconstructs the low-level information simultaneously integrating high-level information to generate the image embedding. This procedure illustrates that as image representations propagate through the DNNs, different levels of information are represented in an intermingled manner. Specifically, low-level information can be transformed into high-level semantic information, while high-level information can also be restructured to incorporate low-level features. This bidirectional transformation is likely facilitated by the residual stream, which retains information from the initial stages of processing and conveys it through to the highest layers.

The boxplot in Figure~\ref{fig7} illustrates the variance distribution of the model's activation patterns $\mathcal{S}^{(j)}$. For all models, the variance of SAEs units' activities gets higher as with layers. This indicates that the information processed in early layers is similar and basic, while in higher layers, models deal with more complex and diverse semantic information. A higher variance also suggests that higher layers in the model learn and understand the various and distinct information hidden in the images.

\section{Discussion}
\label{section 5}
\subparagraph{Limitations:} The training config of SAEs kept the same during all the experiments, we do not consider the effect of SAEs' dimension, training methods, training epochs and training dataset on the Brain-Model alignment, which may affect the results.

\subparagraph{Conclusion:} We bridge the gap between deep learning visual models and the human ventral visual pathway with the help of sparse autoencoders. By comparing the activation of SAEs units and voxel fMRI signals, we find that two systems exhibit a high degree of similarity. Furthermore, voxels and units with similar activation patterns exhibit functional similarity, and units could retain the ROIs prior functional structure. Depending on these similarities, we demonstrate that the models share hierarchical alignment with the brain, and reveal the visual information procedure of ViT-B/16\textsubscript{CLIP} in the cortex.

{
    \small
    \bibliography{ref} 

\begin{thebibliography}{61}
\providecommand{\natexlab}[1]{#1}
\providecommand{\url}[1]{\texttt{#1}}
\expandafter\ifx\csname urlstyle\endcsname\relax
  \providecommand{\doi}[1]{doi: #1}\else
  \providecommand{\doi}{doi: \begingroup \urlstyle{rm}\Url}\fi

\bibitem[DiCarlo and Cox(2007)]{dicarloUntanglingInvariantObject2007}
James~J. DiCarlo and David~D. Cox.
\newblock Untangling invariant object recognition.
\newblock \emph{Trends in Cognitive Sciences}, 11\penalty0 (8):\penalty0 333--341, August 2007.
\newblock ISSN 13646613.
\newblock \doi{10.1016/j.tics.2007.06.010}.

\bibitem[Yang et~al.(2024)Yang, Gee, and Shi]{yangBrainDecodesDeep2024}
Huzheng Yang, James Gee, and Jianbo Shi.
\newblock Brain {{Decodes Deep Nets}}, March 2024.

\bibitem[Coggan et~al.(2017)Coggan, Allen, Farrar, et~al.]{DifferencesSelectivityNatural}
David~D. Coggan, Luke~A. Allen, Oliver R.~H. Farrar, et~al.
\newblock Differences in selectivity to natural images in early visual areas (v1--v3).
\newblock \emph{Scientific Reports}, 7:\penalty0 2444, 2017.
\newblock \doi{10.1038/s41598-017-02569-4}.

\bibitem[Oikarinen and Weng(2023)]{oikarinenCLIPDissectAutomaticDescription2023}
Tuomas Oikarinen and Tsui-Wei Weng.
\newblock {{CLIP-Dissect}}: {{Automatic Description}} of {{Neuron Representations}} in {{Deep Vision Networks}}, June 2023.

\bibitem[Zou et~al.(2023)Zou, Phan, Chen, Campbell, Guo, Ren, Pan, Yin, Mazeika, Dombrowski, Goel, Li, Byun, Wang, Mallen, Basart, Koyejo, Song, Fredrikson, Kolter, and Hendrycks]{zouRepresentationEngineeringTopDown2023}
Andy Zou, Long Phan, Sarah Chen, James Campbell, Phillip Guo, Richard Ren, Alexander Pan, Xuwang Yin, Mantas Mazeika, Ann-Kathrin Dombrowski, Shashwat Goel, Nathaniel Li, Michael~J. Byun, Zifan Wang, Alex Mallen, Steven Basart, Sanmi Koyejo, Dawn Song, Matt Fredrikson, J.~Zico Kolter, and Dan Hendrycks.
\newblock Representation {{Engineering}}: {{A Top-Down Approach}} to {{AI Transparency}}, October 2023.

\bibitem[Cerdas et~al.(2024)Cerdas, Sartzetaki, Petersen, Roig, Mettes, and Groen]{cerdasBrainACTIVIdentifyingVisuosemantic2024}
Diego~Garc{\'i}a Cerdas, Christina Sartzetaki, Magnus Petersen, Gemma Roig, Pascal Mettes, and Iris Groen.
\newblock {{BrainACTIV}}: {{Identifying}} visuo-semantic properties driving cortical selectivity using diffusion-based image manipulation, October 2024.

\bibitem[Conwell et~al.(2024{\natexlab{a}})Conwell, Prince, Kay, Alvarez, and Konkle]{conwellLargescaleExaminationInductive2024a}
Colin Conwell, Jacob~S. Prince, Kendrick~N. Kay, George~A. Alvarez, and Talia Konkle.
\newblock A large-scale examination of inductive biases shaping high-level visual representation in brains and machines.
\newblock \emph{Nature Communications}, 15\penalty0 (1):\penalty0 9383, October 2024{\natexlab{a}}.
\newblock ISSN 2041-1723.
\newblock \doi{10.1038/s41467-024-53147-y}.

\bibitem[Xu and {Vaziri-Pashkam}(2021)]{xuLimitsVisualRepresentational2021}
Yaoda Xu and Maryam {Vaziri-Pashkam}.
\newblock Limits to visual representational correspondence between convolutional neural networks and the human brain.
\newblock \emph{Nature Communications}, 12\penalty0 (1):\penalty0 2065, April 2021.
\newblock ISSN 2041-1723.
\newblock \doi{10.1038/s41467-021-22244-7}.

\bibitem[{Khaligh-Razavi} et~al.(2017){Khaligh-Razavi}, Henriksson, Kay, and Kriegeskorte]{khaligh-razaviFixedMixedRSA2017}
Seyed-Mahdi {Khaligh-Razavi}, Linda Henriksson, Kendrick Kay, and Nikolaus Kriegeskorte.
\newblock Fixed versus mixed {{RSA}}:~{{Explaining}} visual representations by fixed and mixed feature sets from shallow and deep computational models.
\newblock \emph{Journal of Mathematical Psychology}, 76:\penalty0 184--197, February 2017.
\newblock ISSN 0022-2496.
\newblock \doi{10.1016/j.jmp.2016.10.007}.

\bibitem[Kriegeskorte and Wei(2021)]{kriegeskorteNeuralTuningRepresentational2021a}
Nikolaus Kriegeskorte and Xue-Xin Wei.
\newblock Neural tuning and representational geometry.
\newblock \emph{Nature Reviews Neuroscience}, 22\penalty0 (11):\penalty0 703--718, November 2021.
\newblock ISSN 1471-003X, 1471-0048.
\newblock \doi{10.1038/s41583-021-00502-3}.

\bibitem[Kriegeskorte et~al.(2008{\natexlab{a}})Kriegeskorte, Mur, and Bandettini]{kriegeskorteRepresentationalSimilarityAnalysis2008}
Nikolaus Kriegeskorte, Marieke Mur, and Peter Bandettini.
\newblock Representational {{Similarity Analysis}} -- {{Connecting}} the {{Branches}} of {{Systems Neuroscience}}.
\newblock \emph{Frontiers in Systems Neuroscience}, 2:\penalty0 4, November 2008{\natexlab{a}}.
\newblock ISSN 1662-5137.
\newblock \doi{10.3389/neuro.06.004.2008}.

\bibitem[Luo et~al.(2024{\natexlab{a}})Luo, Henderson, Tarr, and Wehbe]{luoBrainSCUBAFineGrainedNatural2024}
Andrew~F. Luo, Margaret~M. Henderson, Michael~J. Tarr, and Leila Wehbe.
\newblock {{BrainSCUBA}}: {{Fine-Grained Natural Language Captions}} of {{Visual Cortex Selectivity}}, May 2024{\natexlab{a}}.

\bibitem[Luo et~al.(2024{\natexlab{b}})Luo, Yeung, Zawar, Dewan, Henderson, Wehbe, and Tarr]{luoBrainMappingDense2024}
Andrew~F. Luo, Jacob Yeung, Rushikesh Zawar, Shaurya Dewan, Margaret~M. Henderson, Leila Wehbe, and Michael~J. Tarr.
\newblock Brain {{Mapping}} with {{Dense Features}}: {{Grounding Cortical Semantic Selectivity}} in {{Natural Images With Vision Transformers}}, October 2024{\natexlab{b}}.

\bibitem[Wang et~al.(2023)Wang, Kay, Naselaris, Tarr, and Wehbe]{wangBetterModelsHuman2023}
Aria~Y. Wang, Kendrick Kay, Thomas Naselaris, Michael~J. Tarr, and Leila Wehbe.
\newblock Better models of human high-level visual cortex emerge from natural language supervision with a large and diverse dataset.
\newblock \emph{Nature Machine Intelligence}, 5\penalty0 (12):\penalty0 1415--1426, November 2023.
\newblock ISSN 2522-5839.
\newblock \doi{10.1038/s42256-023-00753-y}.

\bibitem[Luo et~al.()Luo, Henderson, Wehbe, and Tarr]{luoBrainDiffusionVisual}
Andrew~F Luo, Margaret~M Henderson, Leila Wehbe, and Michael~J Tarr.
\newblock Brain {{Diffusion}} for {{Visual Exploration}}: {{Cortical Discovery}} using {{Large Scale Generative Models}}.

\bibitem[Raghu et~al.(2021)Raghu, Unterthiner, Kornblith, Zhang, and Dosovitskiy]{raghuVisionTransformersSee2021}
Maithra Raghu, Thomas Unterthiner, Simon Kornblith, Chiyuan Zhang, and Alexey Dosovitskiy.
\newblock Do {{Vision Transformers See Like Convolutional Neural Networks}}?
\newblock In \emph{Advances in {{Neural Information Processing Systems}}}, volume~34, pages 12116--12128. Curran Associates, Inc., 2021.

\bibitem[Yang et~al.(2025)Yang, Xue, Mao, Zheng, Xu, Sheng, Sun, Yang, and Li]{yangCLIPMSMMultiSemanticMapping2025}
Guoyuan Yang, Mufan Xue, Ziming Mao, Haofang Zheng, Jia Xu, Dabin Sheng, Ruotian Sun, Ruoqi Yang, and Xuesong Li.
\newblock {{CLIP-MSM}}: {{A Multi-Semantic Mapping Brain Representation}} for {{Human High-Level Visual Cortex}}.
\newblock \emph{Proceedings of the AAAI Conference on Artificial Intelligence}, 39\penalty0 (9):\penalty0 9184--9192, April 2025.
\newblock ISSN 2374-3468.
\newblock \doi{10.1609/aaai.v39i9.32994}.

\bibitem[Cunningham et~al.(2023)Cunningham, Ewart, Riggs, Huben, and Sharkey]{cunninghamSparseAutoencodersFind2023}
Hoagy Cunningham, Aidan Ewart, Logan Riggs, Robert Huben, and Lee Sharkey.
\newblock Sparse {{Autoencoders Find Highly Interpretable Features}} in {{Language Models}}, October 2023.

\bibitem[Marks et~al.(2025)Marks, Rager, Michaud, Belinkov, Bau, and Mueller]{marksSparseFeatureCircuits2025}
Samuel Marks, Can Rager, Eric~J. Michaud, Yonatan Belinkov, David Bau, and Aaron Mueller.
\newblock Sparse {{Feature Circuits}}: {{Discovering}} and {{Editing Interpretable Causal Graphs}} in {{Language Models}}, March 2025.

\bibitem[Allen et~al.(2022)Allen, {St-Yves}, Wu, Breedlove, Prince, Dowdle, Nau, Caron, Pestilli, Charest, Hutchinson, Naselaris, and Kay]{allenMassive7TFMRI2022}
Emily~J. Allen, Ghislain {St-Yves}, Yihan Wu, Jesse~L. Breedlove, Jacob~S. Prince, Logan~T. Dowdle, Matthias Nau, Brad Caron, Franco Pestilli, Ian Charest, J.~Benjamin Hutchinson, Thomas Naselaris, and Kendrick Kay.
\newblock A massive {{7T fMRI}} dataset to bridge cognitive neuroscience and artificial intelligence.
\newblock \emph{Nature Neuroscience}, 25\penalty0 (1):\penalty0 116--126, January 2022.
\newblock ISSN 1097-6256, 1546-1726.
\newblock \doi{10.1038/s41593-021-00962-x}.

\bibitem[Leask et~al.(2025)Leask, Bussmann, Pearce, Bloom, Tigges, Moubayed, Sharkey, and Nanda]{leaskSparseAutoencodersNot2025}
Patrick Leask, Bart Bussmann, Michael Pearce, Joseph Bloom, Curt Tigges, Noura~Al Moubayed, Lee Sharkey, and Neel Nanda.
\newblock Sparse {{Autoencoders Do Not Find Canonical Units}} of {{Analysis}}, February 2025.

\bibitem[Gao et~al.(2024)Gao, la~Tour, Tillman, Goh, Troll, Radford, Sutskever, Leike, and Wu]{gaoScalingEvaluatingSparse2024}
Leo Gao, Tom~Dupr{\'e} la~Tour, Henk Tillman, Gabriel Goh, Rajan Troll, Alec Radford, Ilya Sutskever, Jan Leike, and Jeffrey Wu.
\newblock Scaling and evaluating sparse autoencoders, June 2024.

\bibitem[Rajamanoharan et~al.(2024)Rajamanoharan, Lieberum, Sonnerat, Conmy, Varma, Kram{\'a}r, and Nanda]{rajamanoharanJumpingAheadImproving2024}
Senthooran Rajamanoharan, Tom Lieberum, Nicolas Sonnerat, Arthur Conmy, Vikrant Varma, J{\'a}nos Kram{\'a}r, and Neel Nanda.
\newblock Jumping {{Ahead}}: {{Improving Reconstruction Fidelity}} with {{JumpReLU Sparse Autoencoders}}, August 2024.

\bibitem[Fel et~al.(2025)Fel, Lubana, Prince, Kowal, Boutin, Papadimitriou, Wang, Wattenberg, Ba, and Konkle]{felArchetypalSAEAdaptive2025}
Thomas Fel, Ekdeep~Singh Lubana, Jacob~S. Prince, Matthew Kowal, Victor Boutin, Isabel Papadimitriou, Binxu Wang, Martin Wattenberg, Demba Ba, and Talia Konkle.
\newblock Archetypal {{SAE}}: {{Adaptive}} and {{Stable Dictionary Learning}} for {{Concept Extraction}} in {{Large Vision Models}}, February 2025.

\bibitem[Thasarathan et~al.(2025)Thasarathan, Forsyth, Fel, Kowal, and Derpanis]{thasarathanUniversalSparseAutoencoders2025}
Harrish Thasarathan, Julian Forsyth, Thomas Fel, Matthew Kowal, and Konstantinos Derpanis.
\newblock Universal {{Sparse Autoencoders}}: {{Interpretable Cross-Model Concept Alignment}}, February 2025.

\bibitem[Zaigrajew et~al.(2025)Zaigrajew, Baniecki, and Biecek]{zaigrajewInterpretingCLIPHierarchical2025}
Vladimir Zaigrajew, Hubert Baniecki, and Przemyslaw Biecek.
\newblock Interpreting {{CLIP}} with {{Hierarchical Sparse Autoencoders}}, February 2025.

\bibitem[Tian et~al.(2025)Tian, Nan, Xu, Zhai, Qu, Liu, Ren, Jia, and Zhang]{tianSparseAutoencoderZeroShot2025}
Zhihua Tian, Sirun Nan, Ming Xu, Shengfang Zhai, Wenjie Qu, Jian Liu, Kui Ren, Ruoxi Jia, and Jiaheng Zhang.
\newblock Sparse {{Autoencoder}} as a {{Zero-Shot Classifier}} for {{Concept Erasing}} in {{Text-to-Image Diffusion Models}}, March 2025.

\bibitem[Gorton(2024)]{gortonMissingCurveDetectors2024}
Liv Gorton.
\newblock The {{Missing Curve Detectors}} of {{InceptionV1}}: {{Applying Sparse Autoencoders}} to {{InceptionV1 Early Vision}}, September 2024.

\bibitem[Elhage et~al.(2022)Elhage, Hume, Olsson, Schiefer, Henighan, Kravec, {Hatfield-Dodds}, Lasenby, Drain, Chen, Grosse, McCandlish, Kaplan, Amodei, Wattenberg, and Olah]{elhageToyModelsSuperposition2022}
Nelson Elhage, Tristan Hume, Catherine Olsson, Nicholas Schiefer, Tom Henighan, Shauna Kravec, Zac {Hatfield-Dodds}, Robert Lasenby, Dawn Drain, Carol Chen, Roger Grosse, Sam McCandlish, Jared Kaplan, Dario Amodei, Martin Wattenberg, and Christopher Olah.
\newblock Toy {{Models}} of {{Superposition}}, September 2022.

\bibitem[Sharkey et~al.(2022)Sharkey, Braun, and {beren}]{sharkeyInterimResearchReport2022}
Lee Sharkey, Dan Braun, and {beren}.
\newblock [{{Interim}} research report] {{Taking}} features out of superposition with sparse autoencoders.
\newblock December 2022.

\bibitem[Yun et~al.(2023)Yun, Chen, Olshausen, and LeCun]{yunTransformerVisualizationDictionary2023}
Zeyu Yun, Yubei Chen, Bruno~A. Olshausen, and Yann LeCun.
\newblock Transformer visualization via dictionary learning: Contextualized embedding as a linear superposition of transformer factors, April 2023.

\bibitem[Shu et~al.(2025)Shu, Wu, Zhao, Rai, Yao, Liu, and Du]{shuSurveySparseAutoencoders2025}
Dong Shu, Xuansheng Wu, Haiyan Zhao, Daking Rai, Ziyu Yao, Ninghao Liu, and Mengnan Du.
\newblock A {{Survey}} on {{Sparse Autoencoders}}: {{Interpreting}} the {{Internal Mechanisms}} of {{Large Language Models}}, March 2025.

\bibitem[Khosla et~al.(2022{\natexlab{a}})Khosla, Ratan~Murty, and Kanwisher]{khoslaHighlySelectiveResponse2022}
Meenakshi Khosla, N.~Apurva Ratan~Murty, and Nancy Kanwisher.
\newblock A highly selective response to food in human visual cortex revealed by hypothesis-free voxel decomposition.
\newblock \emph{Current Biology}, 32\penalty0 (19):\penalty0 4159--4171.e9, October 2022{\natexlab{a}}.
\newblock ISSN 09609822.
\newblock \doi{10.1016/j.cub.2022.08.009}.

\bibitem[Xue et~al.(2024)Xue, Wu, Li, Li, and Yang]{xueConvolutionalNeuralNetwork2024}
Mufan Xue, Xinyu Wu, Jinlong Li, Xuesong Li, and Guoyuan Yang.
\newblock A {{Convolutional Neural Network Interpretable Framework}} for {{Human Ventral Visual Pathway Representation}}.
\newblock \emph{Proceedings of the AAAI Conference on Artificial Intelligence}, 38\penalty0 (6):\penalty0 6413--6421, March 2024.
\newblock ISSN 2374-3468.
\newblock \doi{10.1609/aaai.v38i6.28461}.

\bibitem[Goodale et~al.(1994)Goodale, Meenan, B{\"u}lthoff, Nicolle, Murphy, and Racicot]{goodaleSeparateNeuralPathways1994}
Melvyn~A. Goodale, John~Paul Meenan, Heinrich~H. B{\"u}lthoff, David~A. Nicolle, Kelly~J. Murphy, and Carolynn~I. Racicot.
\newblock Separate neural pathways for the visual analysis of object shape in perception and prehension.
\newblock \emph{Current Biology}, 4\penalty0 (7):\penalty0 604--610, July 1994.
\newblock ISSN 0960-9822.
\newblock \doi{10.1016/S0960-9822(00)00132-9}.

\bibitem[{Grill-Spector} and Weiner(2014)]{grill-spectorFunctionalArchitectureVentral2014}
Kalanit {Grill-Spector} and Kevin~S. Weiner.
\newblock The functional architecture of the ventral temporal cortex and its role in categorization.
\newblock \emph{Nature Reviews Neuroscience}, 15\penalty0 (8):\penalty0 536--548, August 2014.
\newblock ISSN 1471-0048.
\newblock \doi{10.1038/nrn3747}.

\bibitem[Murata et~al.(2000)Murata, Gallese, Luppino, Kaseda, and Sakata]{SelectivityShapeSizea}
Akira Murata, Vittorio Gallese, Giuseppe Luppino, Masakazu Kaseda, and Hideo Sakata.
\newblock Selectivity for the shape, size, and orientation of objects for grasping in neurons of monkey parietal area aip.
\newblock \emph{Journal of Neurophysiology}, 83\penalty0 (5):\penalty0 2580--2601, 2000.
\newblock \doi{10.1152/jn.2000.83.5.2580}.

\bibitem[Dumoulin and Wandell(2008)]{dumoulinPopulationReceptiveField2008}
Serge~O. Dumoulin and Brian~A. Wandell.
\newblock Population receptive field estimates in human visual cortex.
\newblock \emph{NeuroImage}, 39\penalty0 (2):\penalty0 647--660, January 2008.
\newblock ISSN 1053-8119.
\newblock \doi{10.1016/j.neuroimage.2007.09.034}.

\bibitem[Hubel and Wiesel(1962)]{hubelReceptiveFieldsBinocular1962}
D.~H. Hubel and T.~N. Wiesel.
\newblock Receptive fields, binocular interaction and functional architecture in the cat's visual cortex.
\newblock \emph{The Journal of Physiology}, 160\penalty0 (1):\penalty0 106--154.2, January 1962.
\newblock ISSN 0022-3751.

\bibitem[Carandini et~al.(2005)Carandini, Demb, Mante, Tolhurst, Dan, Olshausen, Gallant, and Rust]{WeKnowWhat}
Matteo Carandini, Jonathan~B. Demb, Valerio Mante, David~J. Tolhurst, Yang Dan, Bruno~A. Olshausen, Jack~L. Gallant, and Nicole~C. Rust.
\newblock Do we know what the early visual system does?
\newblock \emph{Journal of Neuroscience}, 25\penalty0 (46):\penalty0 10577--10597, 2005.
\newblock ISSN 0270-6474.
\newblock \doi{10.1523/JNEUROSCI.3726-05.2005}.

\bibitem[Levitt et~al.(1994)Levitt, Kiper, and Movshon]{ReceptiveFieldsFunctional}
Jonathan~B. Levitt, David~C. Kiper, and J.~Anthony Movshon.
\newblock Receptive fields and functional architecture of macaque v2.
\newblock \emph{Journal of Neurophysiology}, 71\penalty0 (6):\penalty0 2517--2542, June 1994.
\newblock \doi{10.1152/jn.1994.71.6.2517}.

\bibitem[Freeman et~al.(2013)Freeman, Ziemba, Heeger, Simoncelli, and Movshon]{freemanFunctionalPerceptualSignature2013}
Jeremy Freeman, Corey~M. Ziemba, David~J. Heeger, Eero~P. Simoncelli, and J.~Anthony Movshon.
\newblock A functional and perceptual signature of the second visual area in primates.
\newblock \emph{Nature Neuroscience}, 16\penalty0 (7):\penalty0 974--981, July 2013.
\newblock ISSN 1546-1726.
\newblock \doi{10.1038/nn.3402}.

\bibitem[Kanwisher et~al.(1997)Kanwisher, McDermott, and Chun]{Kanwisher4302}
Nancy Kanwisher, Josh McDermott, and Marvin~M. Chun.
\newblock The fusiform face area: A module in human extrastriate cortex specialized for face perception.
\newblock \emph{Journal of Neuroscience}, 17\penalty0 (11):\penalty0 4302--4311, 1997.
\newblock ISSN 0270-6474.
\newblock \doi{10.1523/JNEUROSCI.17-11-04302.1997}.

\bibitem[Epstein and Kanwisher(1998)]{Epstein1998}
R.~Epstein and N.~Kanwisher.
\newblock A cortical representation of the local visual environment.
\newblock \emph{Nature}, 392:\penalty0 598--601, 1998.
\newblock \doi{10.1038/33402}.

\bibitem[Downing et~al.(2001)Downing, Jiang, Shuman, and Kanwisher]{Downing2001}
P.~E. Downing, Y.~Jiang, M.~Shuman, and N.~Kanwisher.
\newblock A cortical area selective for visual processing of the human body.
\newblock \emph{Science}, 293\penalty0 (5539):\penalty0 2470--2473, September 2001.
\newblock \doi{10.1126/science.1063414}.

\bibitem[Cohen et~al.(2000)Cohen, Dehaene, Naccache, Leh{\'e}ricy, Dehaene-Lambertz, H{\'e}naff, and Michel]{Cohen2000}
Laurent Cohen, Stanislas Dehaene, Lionel Naccache, Stéphane Leh{\'e}ricy, Ghislaine Dehaene-Lambertz, Marie-Anne H{\'e}naff, and Franck Michel.
\newblock The visual word form area: spatial and temporal characterization of an initial stage of reading in normal subjects and posterior split-brain patients.
\newblock \emph{Brain}, 123\penalty0 (2):\penalty0 291--307, February 2000.
\newblock \doi{10.1093/brain/123.2.291}.

\bibitem[Khosla et~al.(2022{\natexlab{b}})Khosla, {Ratan Murty}, and Kanwisher]{KHOSLA20224159}
Meenakshi Khosla, N.~Apurva {Ratan Murty}, and Nancy Kanwisher.
\newblock A highly selective response to food in human visual cortex revealed by hypothesis-free voxel decomposition.
\newblock \emph{Current Biology}, 32\penalty0 (19):\penalty0 4159--4171.e9, 2022{\natexlab{b}}.
\newblock ISSN 0960-9822.
\newblock \doi{10.1016/j.cub.2022.08.009}.

\bibitem[Conwell et~al.(2024{\natexlab{b}})Conwell, Prince, Kay, Alvarez, and Konkle]{conwellLargescaleExaminationInductive2024}
Colin Conwell, Jacob~S. Prince, Kendrick~N. Kay, George~A. Alvarez, and Talia Konkle.
\newblock A large-scale examination of inductive biases shaping high-level visual representation in brains and machines.
\newblock \emph{Nature Communications}, 15\penalty0 (1):\penalty0 9383, October 2024{\natexlab{b}}.
\newblock ISSN 2041-1723.
\newblock \doi{10.1038/s41467-024-53147-y}.

\bibitem[Kriegeskorte et~al.(2008{\natexlab{b}})Kriegeskorte, Mur, and Bandettini]{kriegeskorteRepresentationalSimilarityAnalysis2008a}
Nikolaus Kriegeskorte, Marieke Mur, and Peter~A. Bandettini.
\newblock Representational similarity analysis - connecting the branches of systems neuroscience.
\newblock \emph{Frontiers in Systems Neuroscience}, 2, November 2008{\natexlab{b}}.
\newblock ISSN 1662-5137.
\newblock \doi{10.3389/neuro.06.004.2008}.

\bibitem[Lin et~al.(2015)Lin, Maire, Belongie, Bourdev, Girshick, Hays, Perona, Ramanan, Zitnick, and Doll{\'a}r]{linMicrosoftCOCOCommon2015}
Tsung-Yi Lin, Michael Maire, Serge Belongie, Lubomir Bourdev, Ross Girshick, James Hays, Pietro Perona, Deva Ramanan, C.~Lawrence Zitnick, and Piotr Doll{\'a}r.
\newblock Microsoft {{COCO}}: {{Common Objects}} in {{Context}}, February 2015.

\bibitem[Gao et~al.(2015)Gao, Huth, Lescroart, and Gallant]{gaoPycortexInteractiveSurface2015}
James~S. Gao, Alexander~G. Huth, Mark~D. Lescroart, and Jack~L. Gallant.
\newblock Pycortex: An interactive surface visualizer for {{fMRI}}.
\newblock \emph{Frontiers in Neuroinformatics}, 9, September 2015.
\newblock ISSN 1662-5196.
\newblock \doi{10.3389/fninf.2015.00023}.

\bibitem[Deng et~al.()Deng, Dong, Socher, Li, Li, and {Fei-Fei}]{dengImageNetLargeScaleHierarchical}
Jia Deng, Wei Dong, Richard Socher, Li-Jia Li, Kai Li, and Li~{Fei-Fei}.
\newblock {{ImageNet}}: {{A Large-Scale Hierarchical Image Database}}.

\bibitem[Radford et~al.(2021)Radford, Kim, Hallacy, Ramesh, Goh, Agarwal, Sastry, Askell, Mishkin, Clark, Krueger, and Sutskever]{radfordLearningTransferableVisual2021}
Alec Radford, Jong~Wook Kim, Chris Hallacy, Aditya Ramesh, Gabriel Goh, Sandhini Agarwal, Girish Sastry, Amanda Askell, Pamela Mishkin, Jack Clark, Gretchen Krueger, and Ilya Sutskever.
\newblock Learning {{Transferable Visual Models From Natural Language Supervision}}.
\newblock In \emph{Proceedings of the 38th {{International Conference}} on {{Machine Learning}}}, pages 8748--8763. PMLR, July 2021.

\bibitem[He et~al.(2015)He, Zhang, Ren, and Sun]{heDeepResidualLearning2015}
Kaiming He, Xiangyu Zhang, Shaoqing Ren, and Jian Sun.
\newblock Deep {{Residual Learning}} for {{Image Recognition}}, December 2015.

\bibitem[He et~al.(2021)He, Chen, Xie, Li, Doll{\'a}r, and Girshick]{heMaskedAutoencodersAre2021}
Kaiming He, Xinlei Chen, Saining Xie, Yanghao Li, Piotr Doll{\'a}r, and Ross Girshick.
\newblock Masked {{Autoencoders Are Scalable Vision Learners}}, December 2021.

\bibitem[Oquab et~al.(2024)Oquab, Darcet, Moutakanni, Vo, Szafraniec, Khalidov, Fernandez, Haziza, Massa, {El-Nouby}, Assran, Ballas, Galuba, Howes, Huang, Li, Misra, Rabbat, Sharma, Synnaeve, Xu, Jegou, Mairal, Labatut, Joulin, and Bojanowski]{oquabDINOv2LearningRobust2024}
Maxime Oquab, Timoth{\'e}e Darcet, Th{\'e}o Moutakanni, Huy Vo, Marc Szafraniec, Vasil Khalidov, Pierre Fernandez, Daniel Haziza, Francisco Massa, Alaaeldin {El-Nouby}, Mahmoud Assran, Nicolas Ballas, Wojciech Galuba, Russell Howes, Po-Yao Huang, Shang-Wen Li, Ishan Misra, Michael Rabbat, Vasu Sharma, Gabriel Synnaeve, Hu~Xu, Herv{\'e} Jegou, Julien Mairal, Patrick Labatut, Armand Joulin, and Piotr Bojanowski.
\newblock {{DINOv2}}: {{Learning Robust Visual Features}} without {{Supervision}}, February 2024.

\bibitem[Kirillov et~al.(2023)Kirillov, Mintun, Ravi, Mao, Rolland, Gustafson, Xiao, Whitehead, Berg, Lo, Doll{\'a}r, and Girshick]{kirillovSegmentAnything2023}
Alexander Kirillov, Eric Mintun, Nikhila Ravi, Hanzi Mao, Chloe Rolland, Laura Gustafson, Tete Xiao, Spencer Whitehead, Alexander~C. Berg, Wan-Yen Lo, Piotr Doll{\'a}r, and Ross Girshick.
\newblock Segment {{Anything}}, April 2023.

\bibitem[Dosovitskiy et~al.(2021)Dosovitskiy, Beyer, Kolesnikov, Weissenborn, Zhai, Unterthiner, Dehghani, Minderer, Heigold, Gelly, Uszkoreit, and Houlsby]{dosovitskiyImageWorth16x162021a}
Alexey Dosovitskiy, Lucas Beyer, Alexander Kolesnikov, Dirk Weissenborn, Xiaohua Zhai, Thomas Unterthiner, Mostafa Dehghani, Matthias Minderer, Georg Heigold, Sylvain Gelly, Jakob Uszkoreit, and Neil Houlsby.
\newblock An {{Image}} is {{Worth}} 16x16 {{Words}}: {{Transformers}} for {{Image Recognition}} at {{Scale}}, June 2021.

\bibitem[Paszke et~al.(2019)Paszke, Gross, Massa, Lerer, Bradbury, Chanan, Killeen, Lin, Gimelshein, Antiga, Desmaison, K{\"o}pf, Yang, DeVito, Raison, Tejani, Chilamkurthy, Steiner, Fang, Bai, and Chintala]{paszkePyTorchImperativeStyle2019}
Adam Paszke, Sam Gross, Francisco Massa, Adam Lerer, James Bradbury, Gregory Chanan, Trevor Killeen, Zeming Lin, Natalia Gimelshein, Luca Antiga, Alban Desmaison, Andreas K{\"o}pf, Edward Yang, Zach DeVito, Martin Raison, Alykhan Tejani, Sasank Chilamkurthy, Benoit Steiner, Lu~Fang, Junjie Bai, and Soumith Chintala.
\newblock {{PyTorch}}: {{An Imperative Style}}, {{High-Performance Deep Learning Library}}, December 2019.

\bibitem[Bau et~al.(2017)Bau, Zhou, Khosla, Oliva, and Torralba]{bauNetworkDissectionQuantifying2017}
David Bau, Bolei Zhou, Aditya Khosla, Aude Oliva, and Antonio Torralba.
\newblock Network {{Dissection}}: {{Quantifying Interpretability}} of {{Deep Visual Representations}}, April 2017.

\bibitem[Kornblith et~al.(2019)Kornblith, Norouzi, Lee, and Hinton]{kornblithSimilarityNeuralNetwork2019}
Simon Kornblith, Mohammad Norouzi, Honglak Lee, and Geoffrey Hinton.
\newblock Similarity of {{Neural Network Representations Revisited}}.
\newblock In \emph{Proceedings of the 36th {{International Conference}} on {{Machine Learning}}}, pages 3519--3529. PMLR, May 2019.

\end{thebibliography}
}

\newpage
\appendix

\section{Broader impact}
\label{appendix1}
Our work introduces the similarity between deep learning models and ventral visual pathway. We can utilize these alignment both explore the function of deep learning model and understand the function of human visual cortex. We can understand deep learning models information procedure for \textbf{any} ANN structure, also, we can explore voxel's selectivity on the image as what ~\cite{luoBrainMappingDense2024} have down. Also we can utilize the SAEs units to intervene the generation process of Diffusion models to explore the voxel's function as ~\cite{luoBrainDiffusionVisual}. Moreover, our method could bring more insight into the deep learning procedure, and explain the activation pattern discovered by CKA~\cite{kornblithSimilarityNeuralNetwork2019, raghuVisionTransformersSee2021}, see more detail in Appendix~\ref{appendix2}

\section{CKA result discussion:}
\label{appendix2}
~\cite{kornblithSimilarityNeuralNetwork2019} introduces CKA method, a kind of representation analysis method similar to Representation Similarities to explore two deep learning models' activation similarity. For model $A$ and $B$, $\text{Let } K_{ij} = k(\mathbf{x}_i, \mathbf{x}_j), \quad L_{ij} = l(\mathbf{y}_i, \mathbf{y}_j), \text{ where } k, l \text{ are kernel functions}$ and $\mathbf{x}_i$ and $\mathbf{y}_i$ are the i-th image activation of model A and model B. $\text{Define the centering matrix: } H_n = I_n - \frac{1}{n} \mathbf{1} \mathbf{1}^\top. $ The CKA calculate method is discribed as follow.

\begin{align}
    \quad \mathrm{HSIC}(K, L) &= \frac{1}{(n - 1)^2} \operatorname{tr}(KHLH). \\
    \quad \mathrm{CKA}(K, L) &= \frac{\mathrm{HSIC}(K, L)}{\sqrt{\mathrm{HSIC}(K, K) \cdot \mathrm{HSIC}(L, L)}}.
\end{align}
\begin{wrapfigure}{r}{0.45\textwidth}
    \centering
    \includegraphics[width=0.38\textwidth]{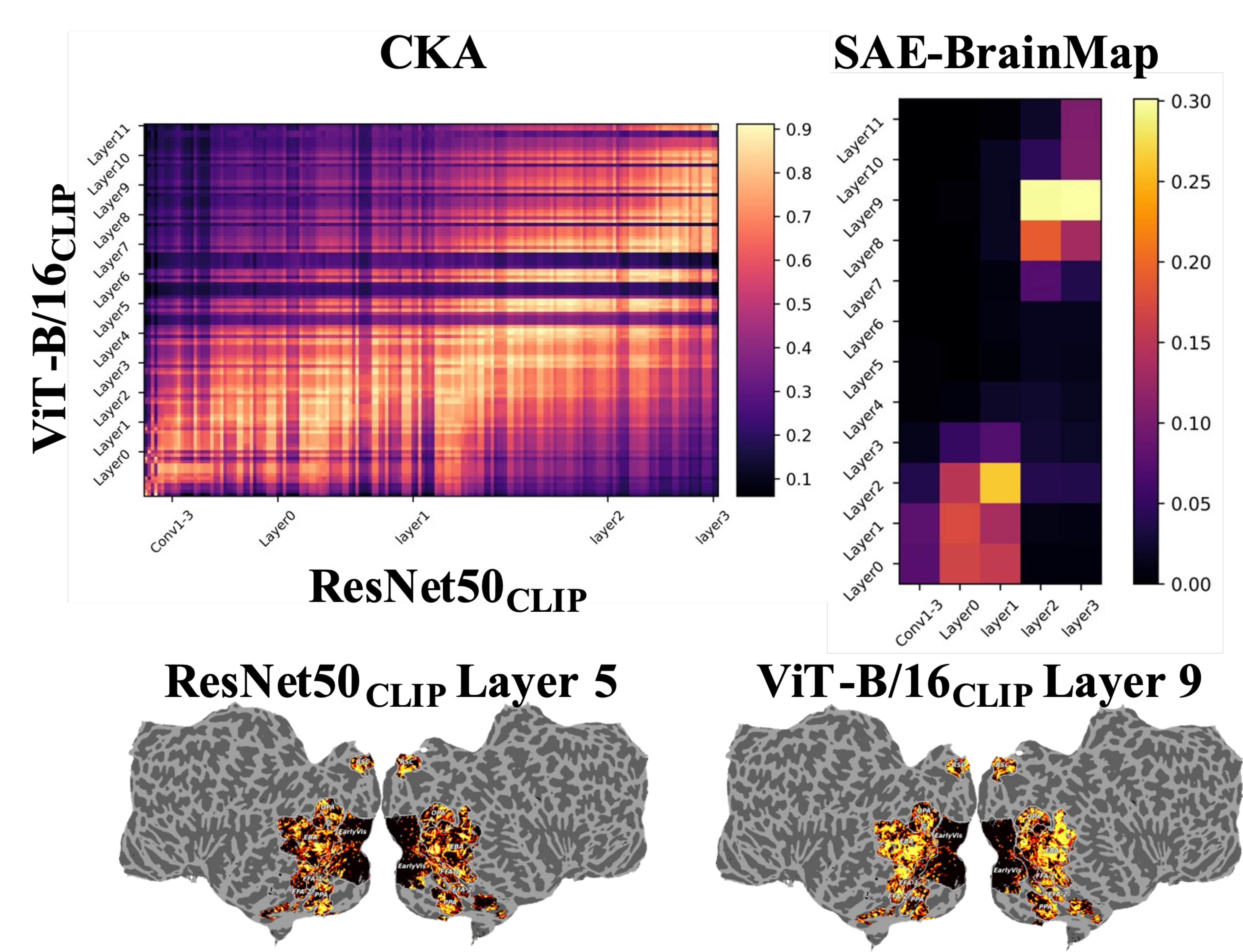}
    \caption{The CKA result for ViT-B/16\textsubscript{CLIP} and ResNet50\textsubscript{CLIP} on top left corner. We compare two models' Model-Brain Layer alignment similarity on the top right corner, and visualize the Layer 5 of ResNet50\textsubscript{CLIP} and Layer 9 of  ViT-B/16\textsubscript{CLIP} on the bottom.}
    \label{fig8}
  \end{wrapfigure}

We compare the CKA matrix between ViT-B/16\textsubscript{CLIP} and ResNet50\textsubscript{CLIP}, and calculate the similarity between each model each layer's Brain-Model Layer Alignment Result, as illustrate in Figure~\ref{fig8}. The similarity is calculated through a method similar to Intersection over Union (IoU). The similarity score is calculated through Equation: $ \mathrm{IoU}_{i,j} = \frac{|\{v \mid l_{\text{ViT}}(v) = i \land l_{\text{ResNet}}(v) = j\}|}{|\{v \mid l_{\text{ViT}}(v) = i \lor l_{\text{ResNet}}(v) = j\}|}$ Here, $l_{\mathrm{ViT}}(v)$ denotes the index of the layer in ViT that best matches voxel $v$, and $l_{\mathrm{ResNet}}(v)$ denotes the corresponding best-matching layer in ResNet. The numerator represents the number of voxels for which both models assign the same respective layer (i.e., the intersection), while the denominator represents the number of voxels assigned to that layer by at least one of the two models (i.e., the union). 
We found that our layer alignment IoU could reveal the activation similarity pattern between two models activation similarity, we visualize the voxel that similar to layer 9 of ViT-B/16\textsubscript{CLIP} and layer 5 of ResNet50\textsubscript{CLIP} in Figure~\ref{fig8}.

\section{Training Environment and Details}
\label{appendix3}
\textbf{Environment:} We perform all the experiment on device with 512G RAM, 2 Nvidia L20 GPU with 48G memory, with CUDA version of 12.4.

\textbf{SAEs training details:} We train the Sparse Autoencoders as ~\cite{cunninghamSparseAutoencodersFind2023}. We discribe several superparameters in section~\ref{setup}. We use AdamW Optimizer through the SAEs training procedures. The image preprocess during training follows ~\cite{yangBrainDecodesDeep2024}. The image is resized to $224 \times 224$, but for SAM, to $1024 \times 1024$ and is normalized channel-wise using the mean $[0.485, 0.456, 0.406]$ and standard deviation $[0.229, 0.224, 0.225]$, which are standard values derived from the ImageNet dataset~\cite{heDeepResidualLearning2015}.

\textbf{Brain Encoder training details:} We train brain encoders for each layer and each model. For each model, we extract j-th layer's activation $A^{(j)}$ and training a linear layer with equation: $A^{(j)}W + b = B$, $B$ is the voxel activation for a certain image. We use AdamW Optimizer through the training process. We use the independent 9000 image for each subject as ~\cite{luoBrainDiffusionVisual}. We performed image preprocess as ~\cite{luoBrainSCUBAFineGrainedNatural2024},  input images are resized to $224 \times 224$ pixels and $1024 \times 1024$ for SAM model. Data augmentation includes randomly scaling pixel intensities within the range [0.95, 1.05], followed by normalization using the mean and standard deviation of CLIP-preprocessed images. Prior to being fed into the network, each image undergoes a random spatial offset of up to 4 pixels along both the horizontal and vertical axes, with edge padding used to fill any resulting empty regions. Additionally, independent Gaussian noise with mean $\mu = 0$ and variance $\sigma^2 = 0.05$ is added to each pixel. We measure the predict performance with \textit{cosine similarity} and $R^2 score$, illustrating in Table~\ref{AppendixTable1}.

\begin{table}[h]
    \centering
    
    \resizebox{\textwidth}{!}{
      \begin{tabular}{c 
                      cc cc cc cc cc cc cc}
        \toprule
        Subject 
        & \multicolumn{2}{c}{All Model} 
        & \multicolumn{2}{c}{ViT-B/16\textsubscript{CLIP}} 
        & \multicolumn{2}{c}{ImageNet} 
        & \multicolumn{2}{c}{MAE} 
        & \multicolumn{2}{c}{DiNOv2} 
        & \multicolumn{2}{c}{SAM} 
        & \multicolumn{2}{c}{ResNet50\textsubscript{CLIP}} 
        \\
        \textit{Cos Similarity}
        & Max & Avg
        & Max & Avg 
        & Max & Avg 
        & Max & Avg 
        & Max & Avg 
        & Max & Avg 
        & Max & Avg 
        \\
        \midrule
        S1 & 0.868& 0.372& 0.868& 0.306& 0.838& 0.287& 0.751& 0.245& 0.859 & 0.289& 0.756 & 0.267 & 0.865& 0.209  \\
        
        S2 & 0.868& 0.368 & 0.868 & 0.307 & 0.827& 0.289& 0.799& 0.250& 0.862& 0.296 & 0.797& 0.270 & 0.854& 0.208  \\

        S3 & 0.848& 0.323& 0.842 & 0.265 & 0.798& 0.249& 0.706 & 0.214& 0.825 & 0.254& 0.748& 0.235 & 0.848 & 0.185 \\

        S4 & 0.856& 0.319 & 0.856& 0.258 & 0.819 & 0.242 & 0.721 & 0.208 & 0.846& 0.248 & 0.723 & 0.226& 0.856 & 0.174 \\

        S5 & 0.889 & 0.414& 0.889 & 0.338 & 0.848& 0.318 & 0.799 & 0.274 & 0.880 & 0.324 & 0.787 & 0.296 & 0.888 & 0.240 \\

        S6 & 0.812& 0.324& 0.802 & 0.259 & 0.778 & 0.244 & 0.700 & 0.209 & 0.798 & 0.248 & 0.715 & 0.227 & 0.812 & 0.175 \\

        S7 & 0.870& 0.327&0.862 & 0.264 & 0.840 & 0.246 & 0.759 & 0.212 & 0.864 & 0.253 & 0.756 & 0.230 & 0.870 & 0.182 \\

        S8 & 0.791 & 0.252 & 0.783 & 0.206 & 0.723 & 0.196 & 0.685 & 0.171 & 0.769 & 0.199 & 0.683 & 0.184 & 0.790 & 0.146 \\
        \midrule
        \textbf{Cos Avg} & 
          0.850 & 0.337 & 0.846 & 0.275 & 0.809 & 0.259 & 0.740 & 0.223 & 0.838 & 0.264 & 0.746 & 0.242 & 0.848 & 0.190 \\	
        \bottomrule \\
        \end{tabular}
    }
    \caption{\textbf{Brain Encoder Cos Similarity}. The cosine similarity between Brain Encoder Predictions and the voxel fMRI response. We select the max voxel's cosine similarity across all voxels and the average cosine similarity across all voxels.}
    \label{AppendixTable1}
    \end{table}

\section{Similarity Visualize for SAE-Brain Similarity on Subject 1-8}
\begin{figure}[!t]
    \centering
    \includegraphics[scale=0.8]{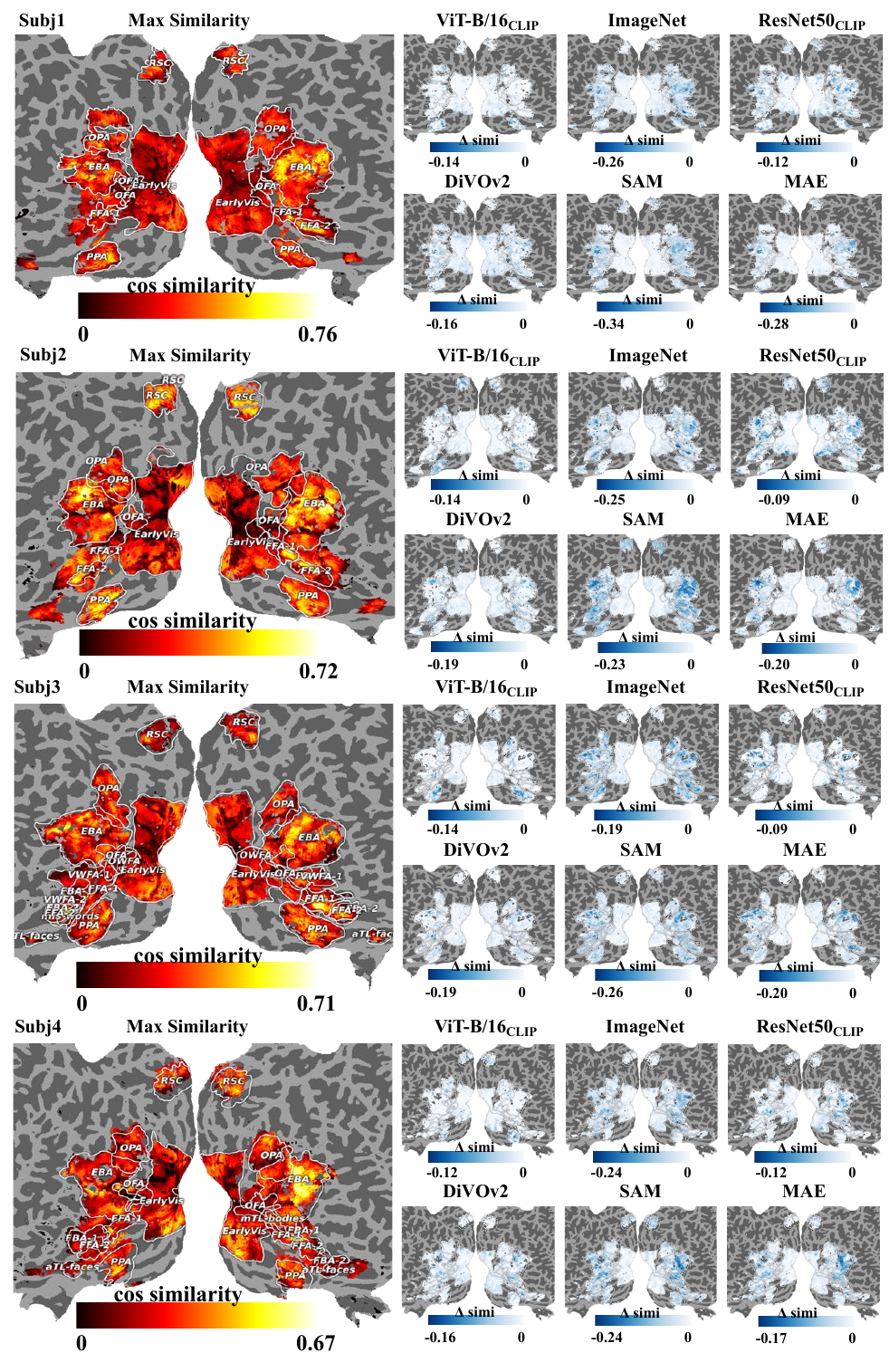}
    \caption{\textbf{Brain-SAEs Activation Similarity on S1-4.} Left: voxel-wise maximum similarity score across models. Right: difference in similarity scores between models and the voxel-wise maximum.}
    \label{AppendixFig1}
\end{figure}

\begin{figure}[!t]
    \centering
    \includegraphics[scale=0.8]{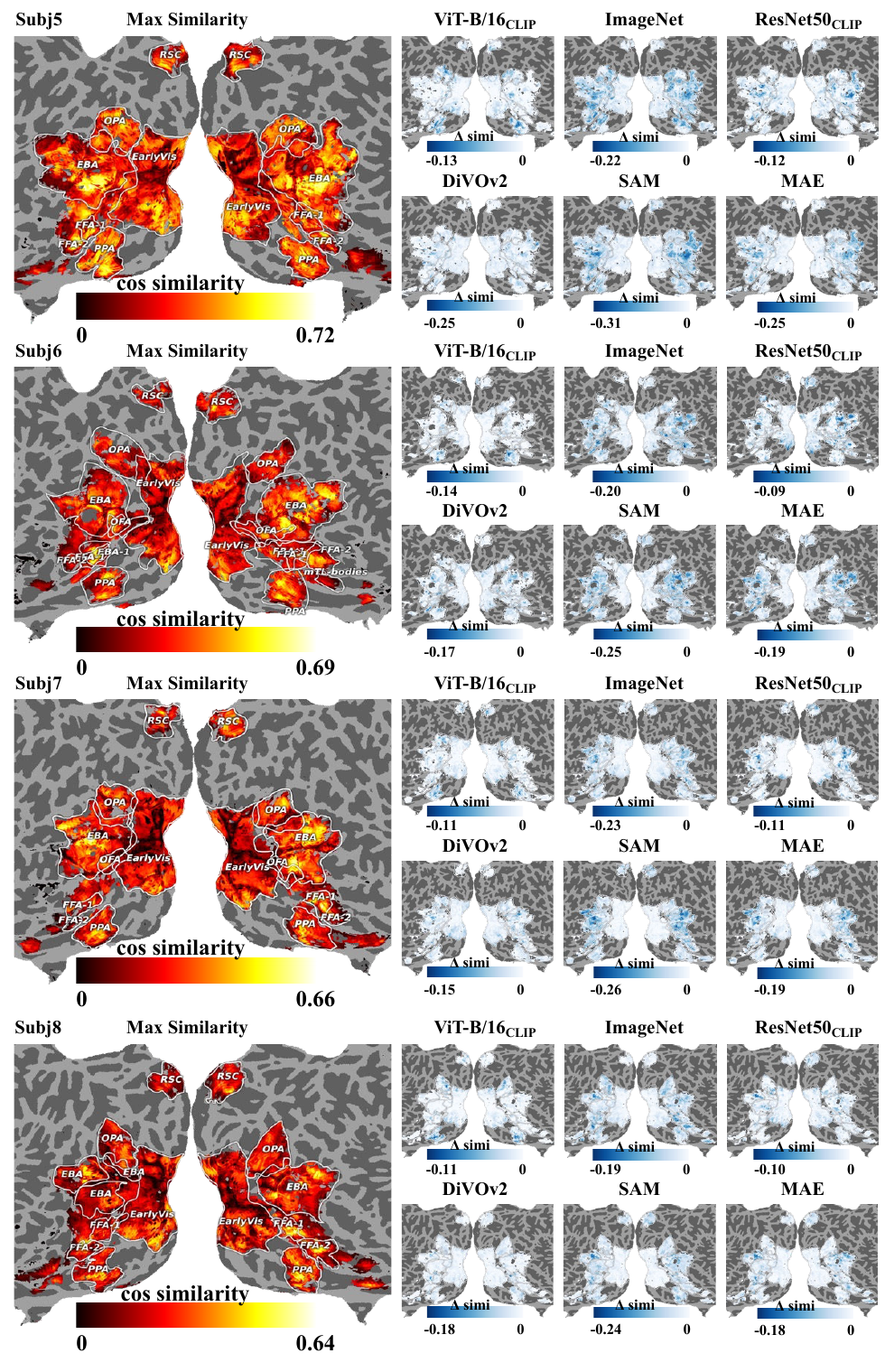}
    \caption{\textbf{Brain-SAEs Activation Similarity on S5-8.} Left: voxel-wise maximum similarity score across models. Right: difference in similarity scores between models and the voxel-wise maximum.}
    \label{AppendixFig2}
\end{figure}

For each subject (S1-8), we plot the voxel-wise maximum similarity score across models and the difference in similarity scores between models and the voxel-wise maximum. We visualize it in Figure~\ref{AppendixFig1} and Figure~\ref{AppendixFig2}.

\newpage

\newpage

\newpage
\section{Brain-Model Layer Alignment}
\begin{figure}[h]
    \centering
    \includegraphics[scale=0.8]{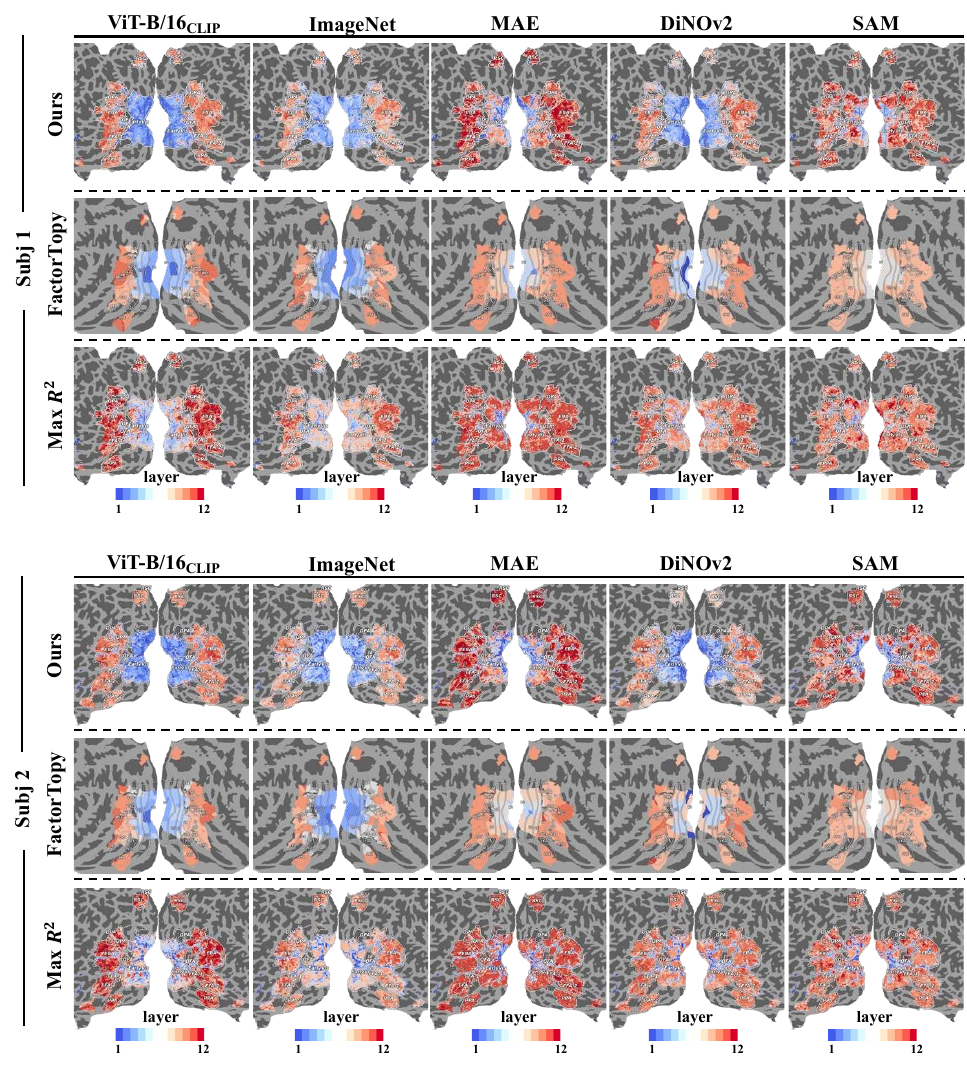}
    \caption{\textbf{Brain-Model Layer Alignment}. Layer alignment for S1 and S2 based on SAE-BrainMap, FactorTopy~\cite{yangBrainDecodesDeep2024} and Max $R^2$~\cite{wangBetterModelsHuman2023}, we visualize five models with attention structure and have 12 layers.}
    \label{AppendixFig4}
\end{figure}

\begin{figure}[h]
    \centering
    \includegraphics[scale=0.8]{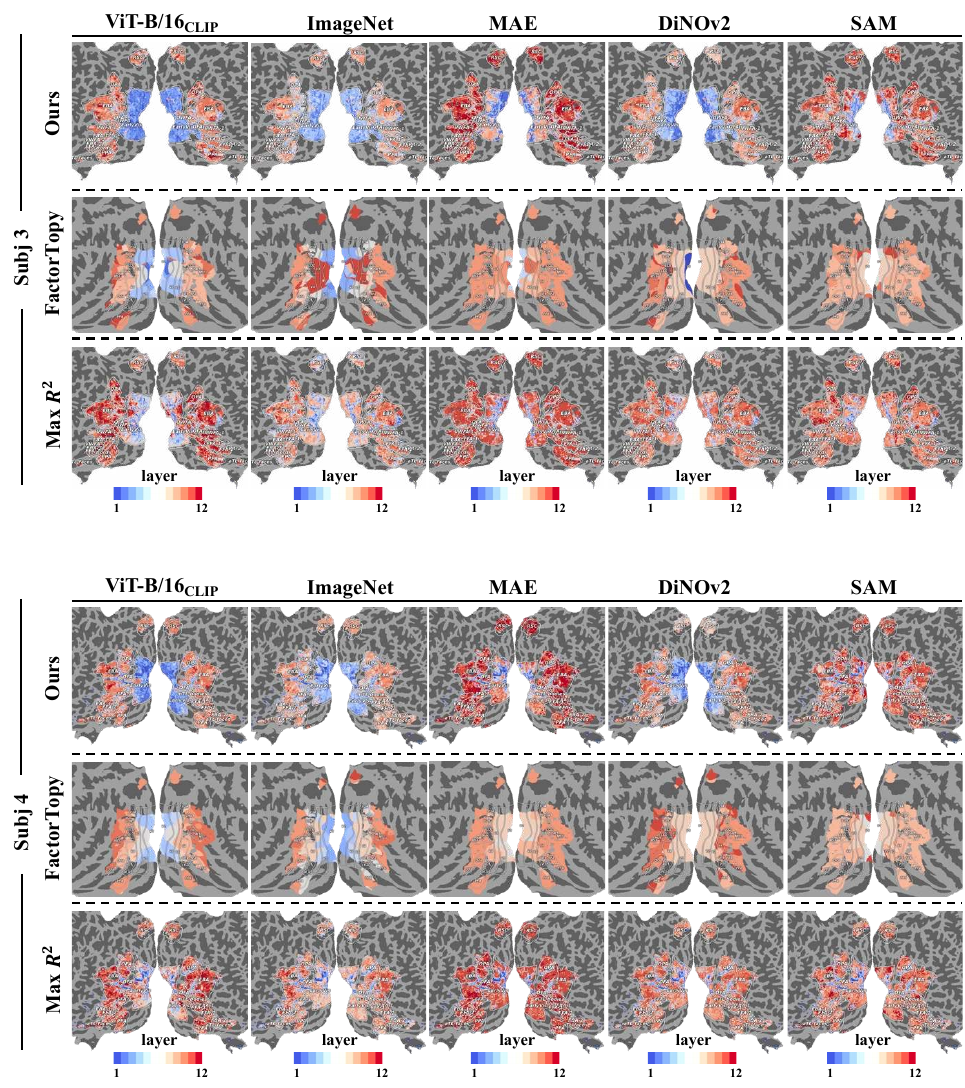}
    \caption{\textbf{Brain-Model Layer Alignment}. Layer alignment for S3 and S4 based on SAE-BrainMap, FactorTopy~\cite{yangBrainDecodesDeep2024} and Max $R^2$~\cite{wangBetterModelsHuman2023}, we visualize five models with attention structure and have 12 layers.}
    \label{AppendixFig5}
\end{figure}

\begin{figure}[h]
    \centering
    \includegraphics[scale=0.8]{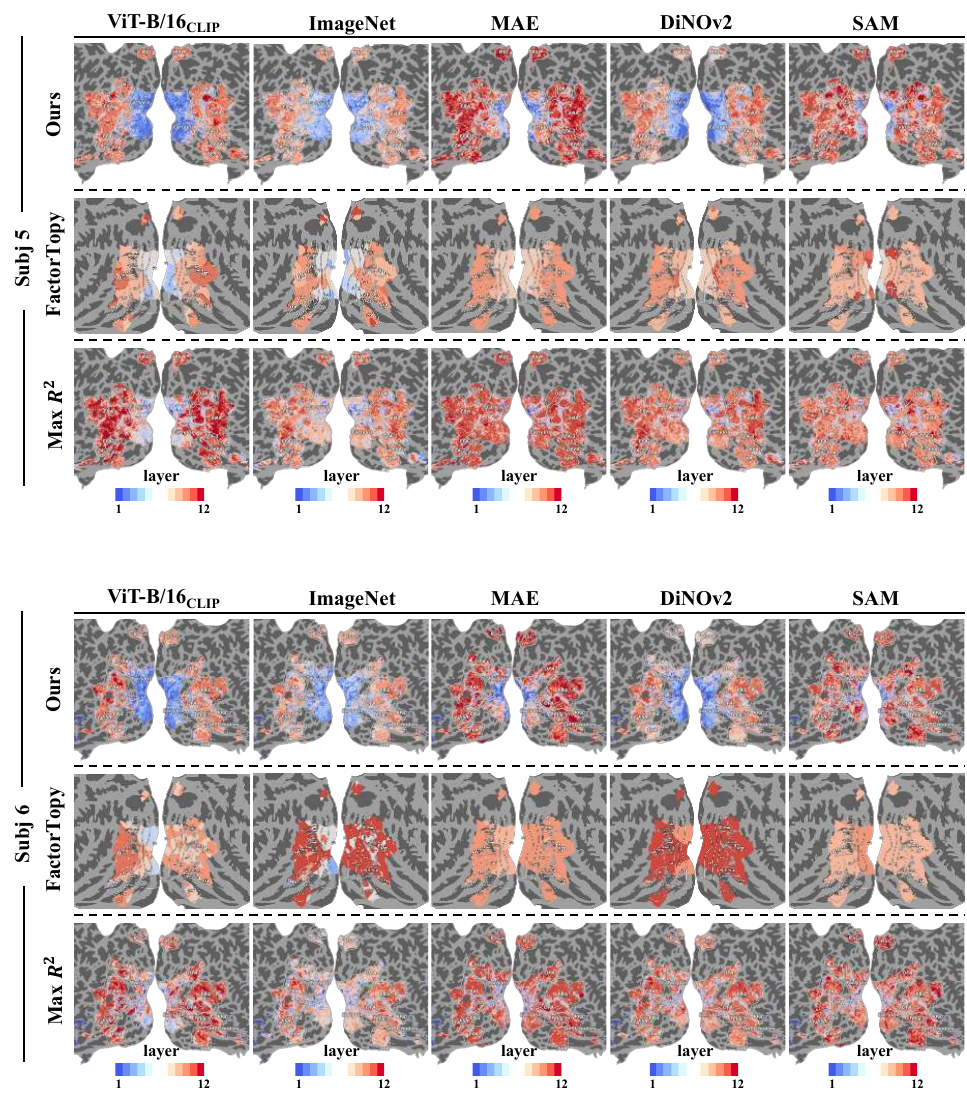}
    \caption{\textbf{Brain-Model Layer Alignment}. Layer alignment for S5 and S6 based on SAE-BrainMap, FactorTopy~\cite{yangBrainDecodesDeep2024} and Max $R^2$~\cite{wangBetterModelsHuman2023}, we visualize five models with attention structure and have 12 layers.}
    \label{AppendixFig6}
\end{figure}

\begin{figure}[h]
    \centering
    \includegraphics[scale=0.8]{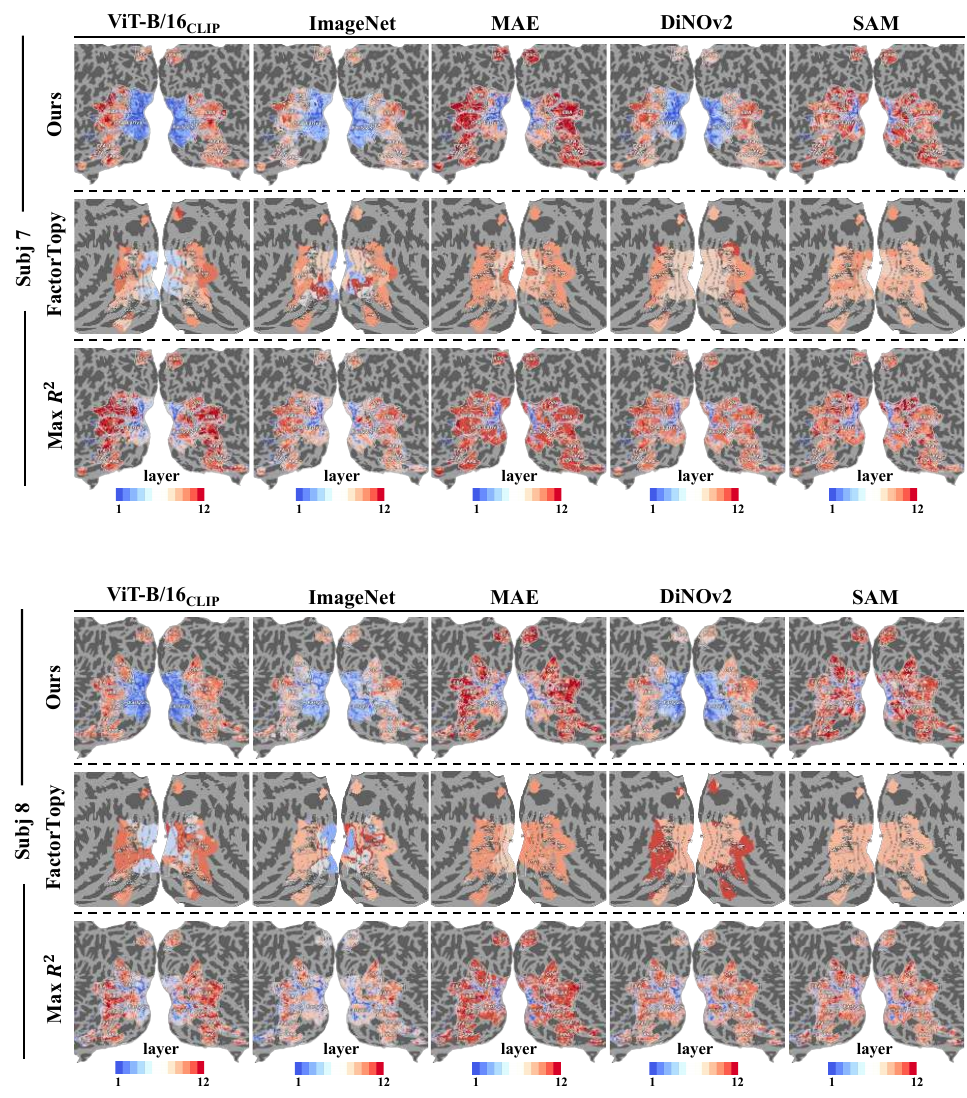}
    \caption{\textbf{Brain-Model Layer Alignment}. Layer alignment for S7 and S8 based on SAE-BrainMap, FactorTopy~\cite{yangBrainDecodesDeep2024} and Max $R^2$~\cite{wangBetterModelsHuman2023}, we visualize five models with attention structure and have 12 layers.}
    \label{AppendixFig7}
\end{figure}

We plot the layer-wise alignment between the brain and the model for each subject (S1-8). We visualize it in Figure~\ref{AppendixFig4},\ref{AppendixFig5},\ref{AppendixFig6},\ref{AppendixFig7}.

\newpage

\newpage

\newpage

\newpage
\section{Similarity Between Voxel Dictionary and Brain Encoder}
\begin{figure}[h]
    \centering
    \includegraphics[scale=0.25]{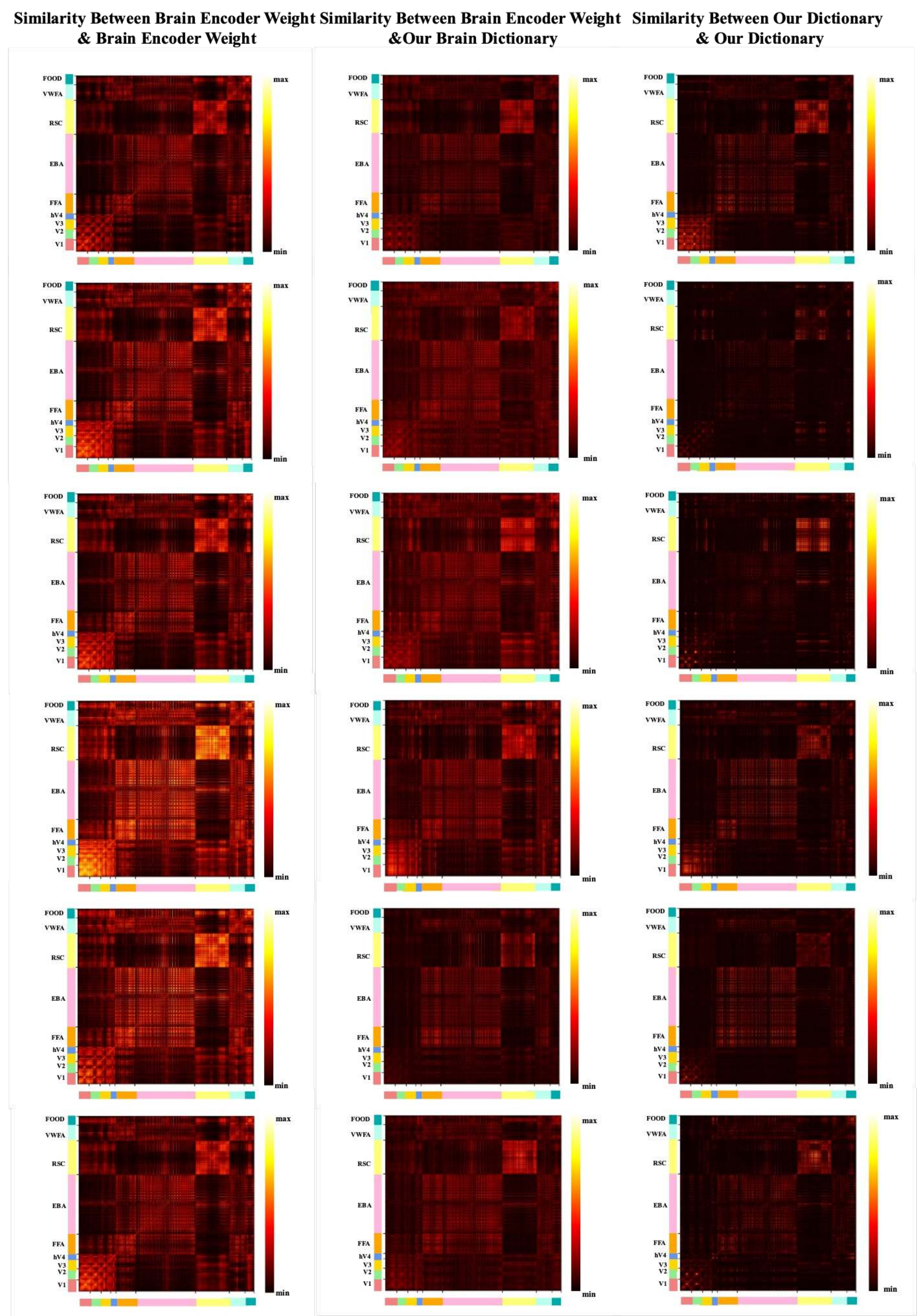}
    \caption{\textbf{Representation Similarity Matrix} between the brain encoder weights of the last layer of voxel dictionary, and their mutual similarity on Subject 5. The model from top to bottom is ResNet50\textsubscript{CLIP}, DiNOv2, ImageNet, MAE, SAM and ViT-B/16\textsubscript{CLIP}}
    \label{AppendixFig8}
\end{figure}

We utilize RSA to calculate the similarity between the Brain Encoder Weight and Voxel Dictionary. During the process, we consider the brain encoder weight as the representation of the brain voxel, as well as the voxel dictionary. We visualize the Subject 5's Similarity Matrix in Figure~\ref{AppendixFig8}.

\paragraph{Representational Similarity Analysis (RSA).}
For brain encoder weight $W \in \mathbb{R}^{v \times d}$ and voxel dictionary $D \in \mathbb{R}^{v \times d}$, where $v$ is the number of voxels and $d$ is the feature dimension, we first compute the Representational Dissimilarity Matrix (RDM), denoted as $\mathbf{D} \in \mathbb{R}^{v \times v}$, using pairwise correlation distances$
\mathbf{D}_{ij} = 1 - \rho(\mathbf{x}_i, \mathbf{x}_j)$, 
where $\rho(\cdot, \cdot)$ denotes the Pearson correlation between two row vectors $W_i$ and $D_j$. Given two RDMs, $\mathbf{D}_1$ and $\mathbf{D}_2$, we extract their upper triangular parts (excluding the diagonal), denoted as $\mathbf{d}_1$ and $\mathbf{d}_2$, respectively. The RSA score is then defined as the Spearman rank correlation between these two vectors,$
\text{RSA} = \text{Spearman}(\mathbf{d}_1, \mathbf{d}_2)
$.
This scalar RSA score reflects the degree of correspondence between Brain Encoder Weight and Voxel Dictionary.

\newpage
\section{Selectivity of SAEs Units}
\begin{figure}[h]
    \centering
    \includegraphics[scale=0.6]{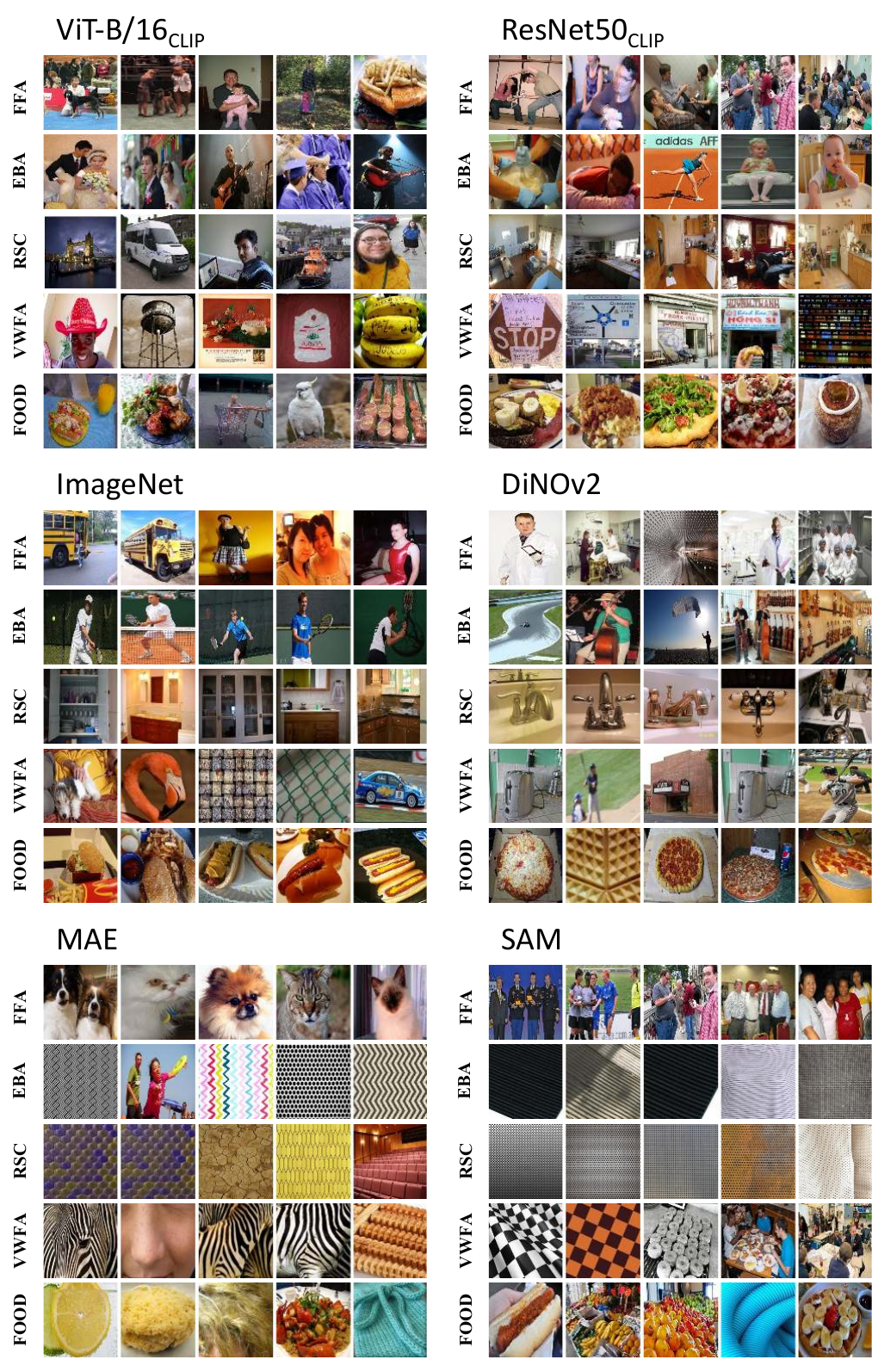}
    \caption{\textbf{Units Selectivity for S5} The top-5 images whose "CLS" token maximize activate the last layer's units that selected by the brain ROIs.}
    \label{AppendixFig9}
\end{figure}

We training the SAEs units with the "CLS" token for all models except SAM. We take the average of all patches in SAM's target layer output as the "CLS", which is the same with ~\cite{yangBrainDecodesDeep2024}. We visualize the selectivity of the units that shares the highest average cosine similarity with certain ROIs by visualize the top-5 images whose "CLS" token maximize activate the units, we visualize all the models of Subject 5 in Figure~\ref{AppendixFig9}. We found that not all the model's last layers units have the same selectivity with the brain ROIs.
MAE and SAM's last layer miss the information process of EBA, RSC and VWFA, replaced by several low level information. It is indicated that for these models, the last layer's functions need more low level information, which is in line with Figure~\ref{fig6}. The analysis dataset is the mixture of COCO~\cite{linMicrosoftCOCOCommon2015} Valid dataset and Broden Dataset~\cite{bauNetworkDissectionQuantifying2017}.

\newpage
\section{Subject 1-8 for selectivity of SAEs Units}

\begin{figure}[h]
    \centering
    \includegraphics[scale=0.6]{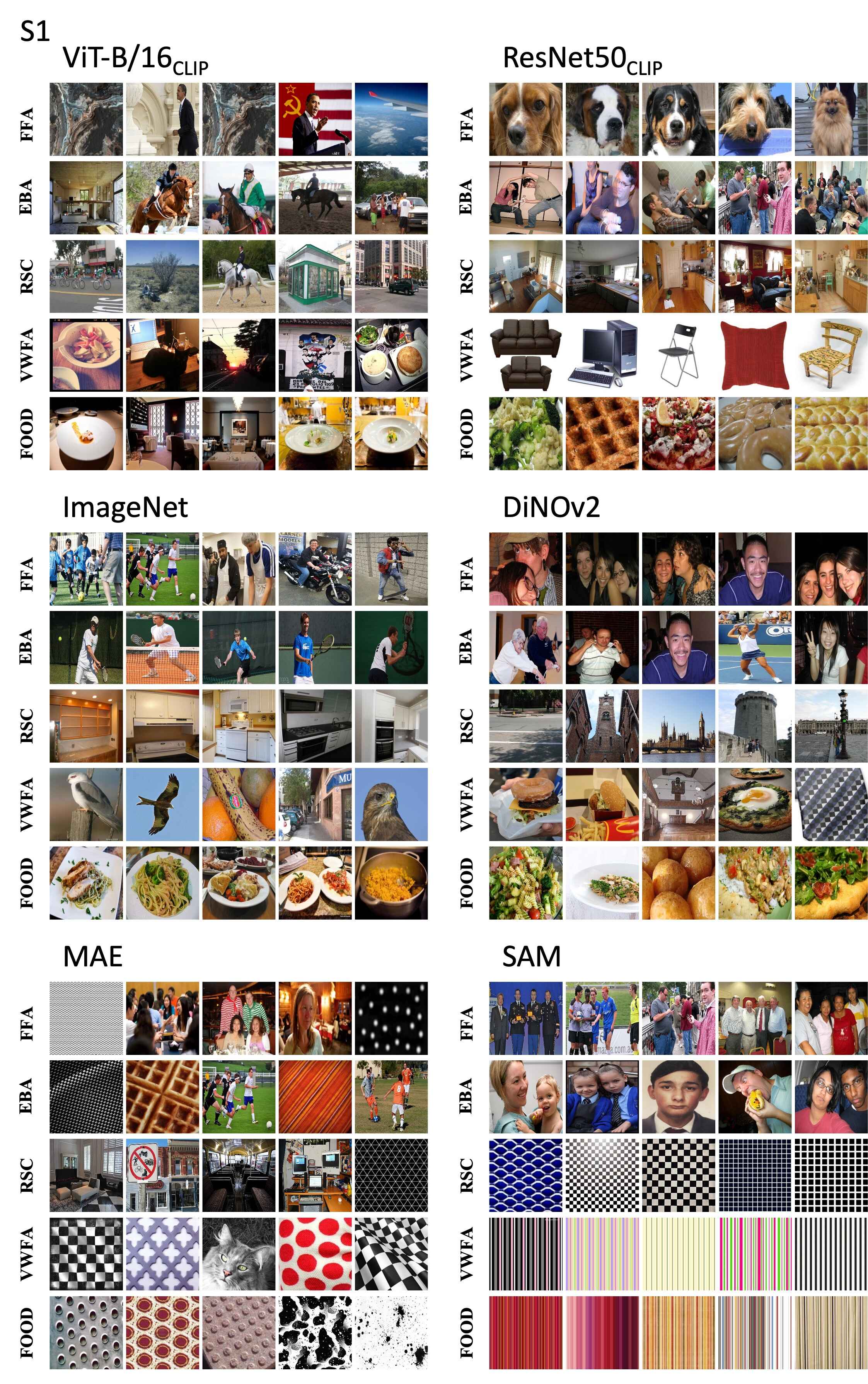}
    \caption{\textbf{Units Selectivity for S1} The top-5 images whose "CLS" token maximize activate the last layer's units that selected by the brain ROIs.}
    \label{AppendixFig10}
\end{figure}
\newpage

\begin{figure}[h]
    \centering
    \includegraphics[scale=0.6]{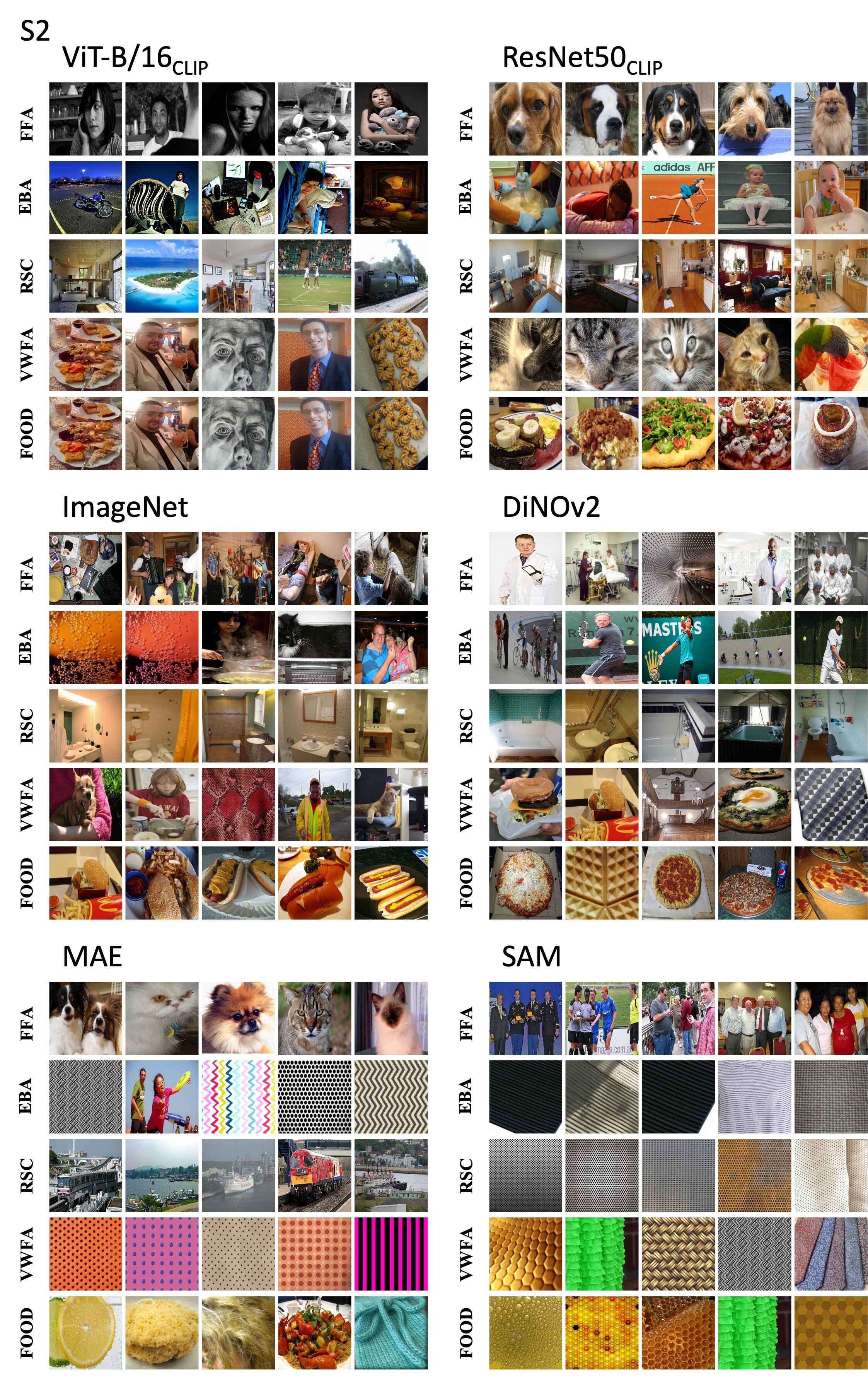}
    \caption{\textbf{Units Selectivity for S2} The top-5 images whose "CLS" token maximize activate the last layer's units that selected by the brain ROIs.}
    \label{AppendixFig11}
\end{figure}
\newpage

\begin{figure}[h]
    \centering
    \includegraphics[scale=0.6]{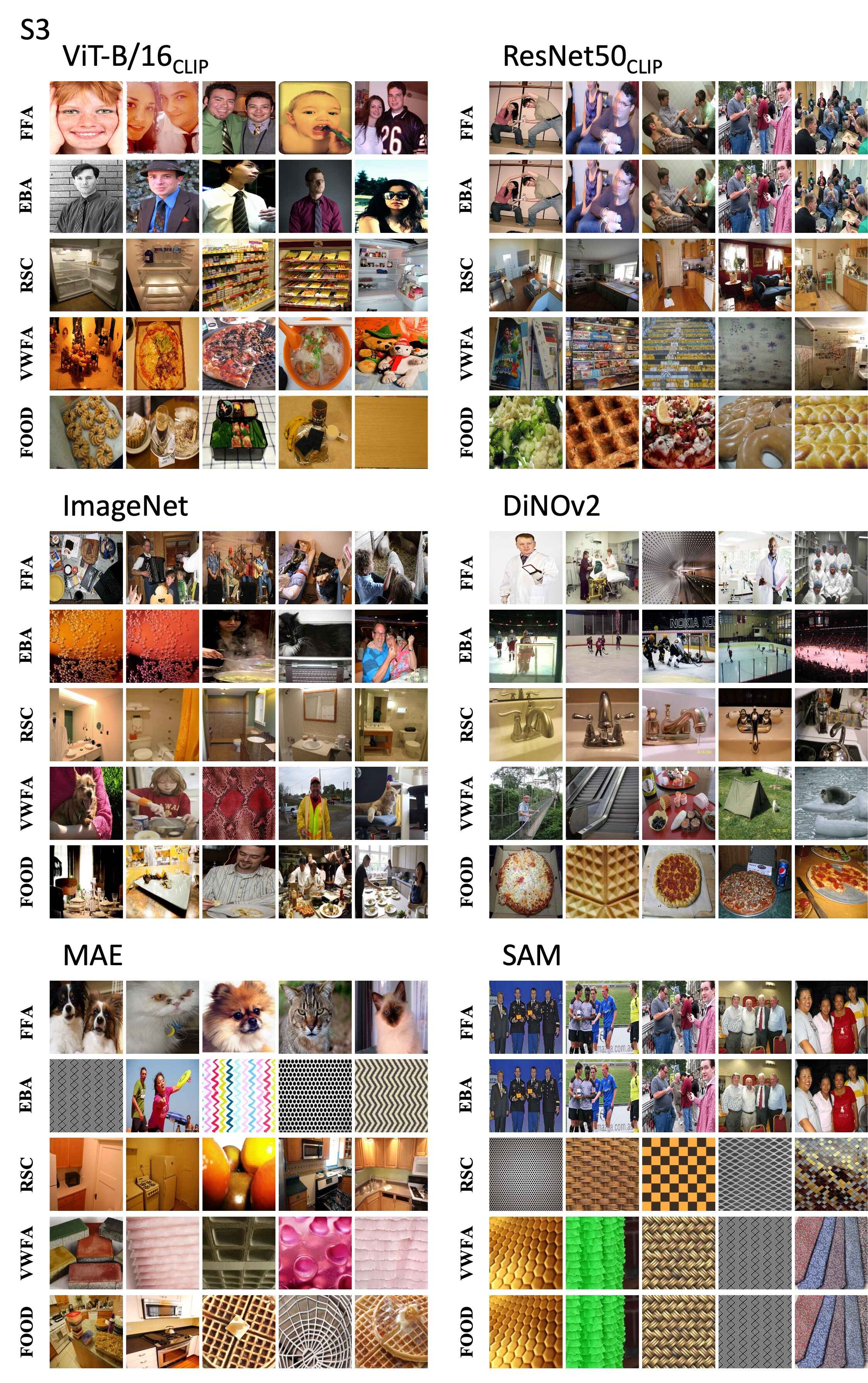}
    \caption{\textbf{Units Selectivity for S3} The top-5 images whose "CLS" token maximize activate the last layer's units that selected by the brain ROIs.}
    \label{AppendixFig12}
\end{figure}
\newpage

\begin{figure}[h]
    \centering
    \includegraphics[scale=0.6]{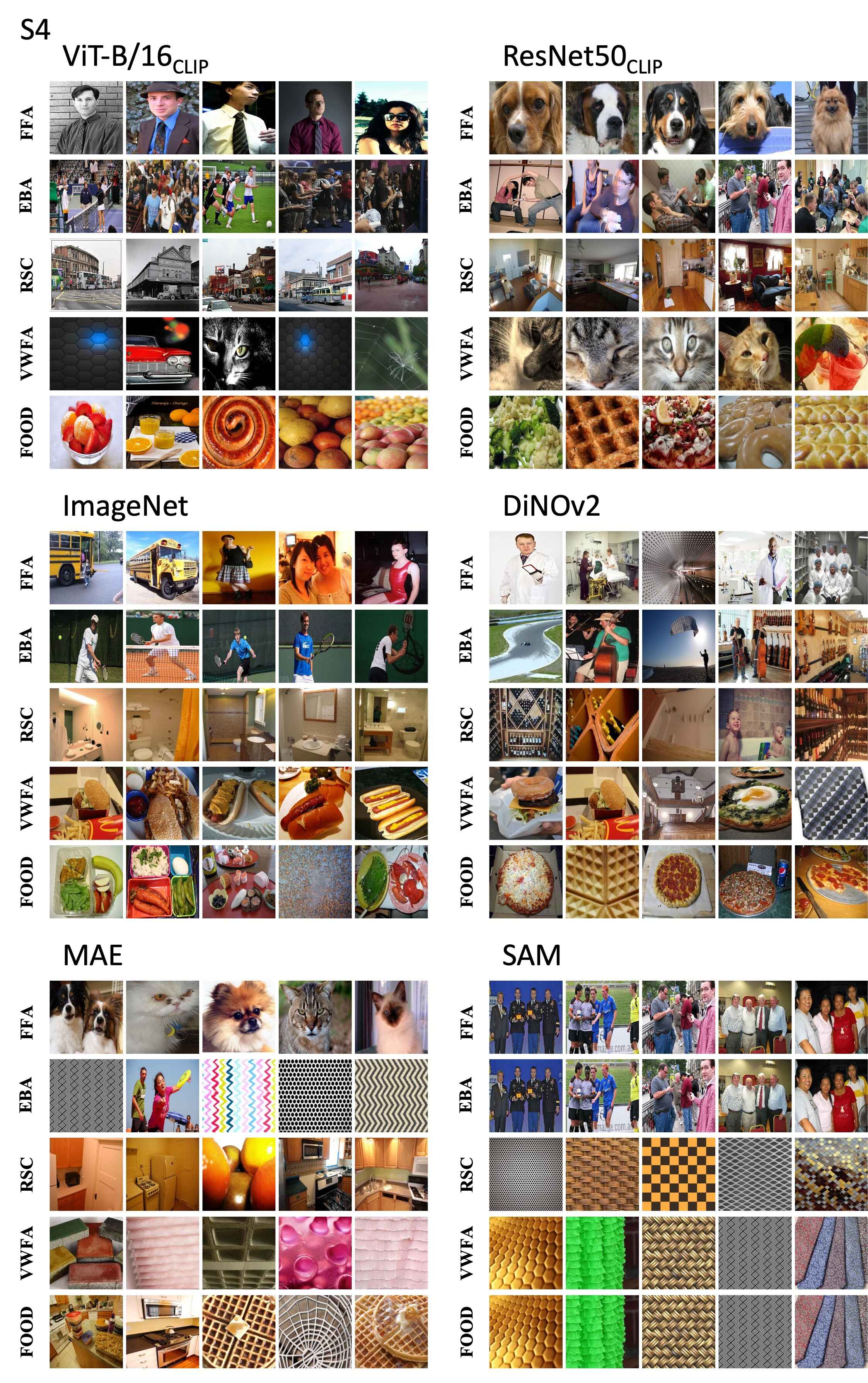}
    \caption{\textbf{Units Selectivity for S4} The top-5 images whose "CLS" token maximize activate the last layer's units that selected by the brain ROIs.}
    \label{AppendixFig13}
\end{figure}
\newpage

\begin{figure}[h]
    \centering
    \includegraphics[scale=0.6]{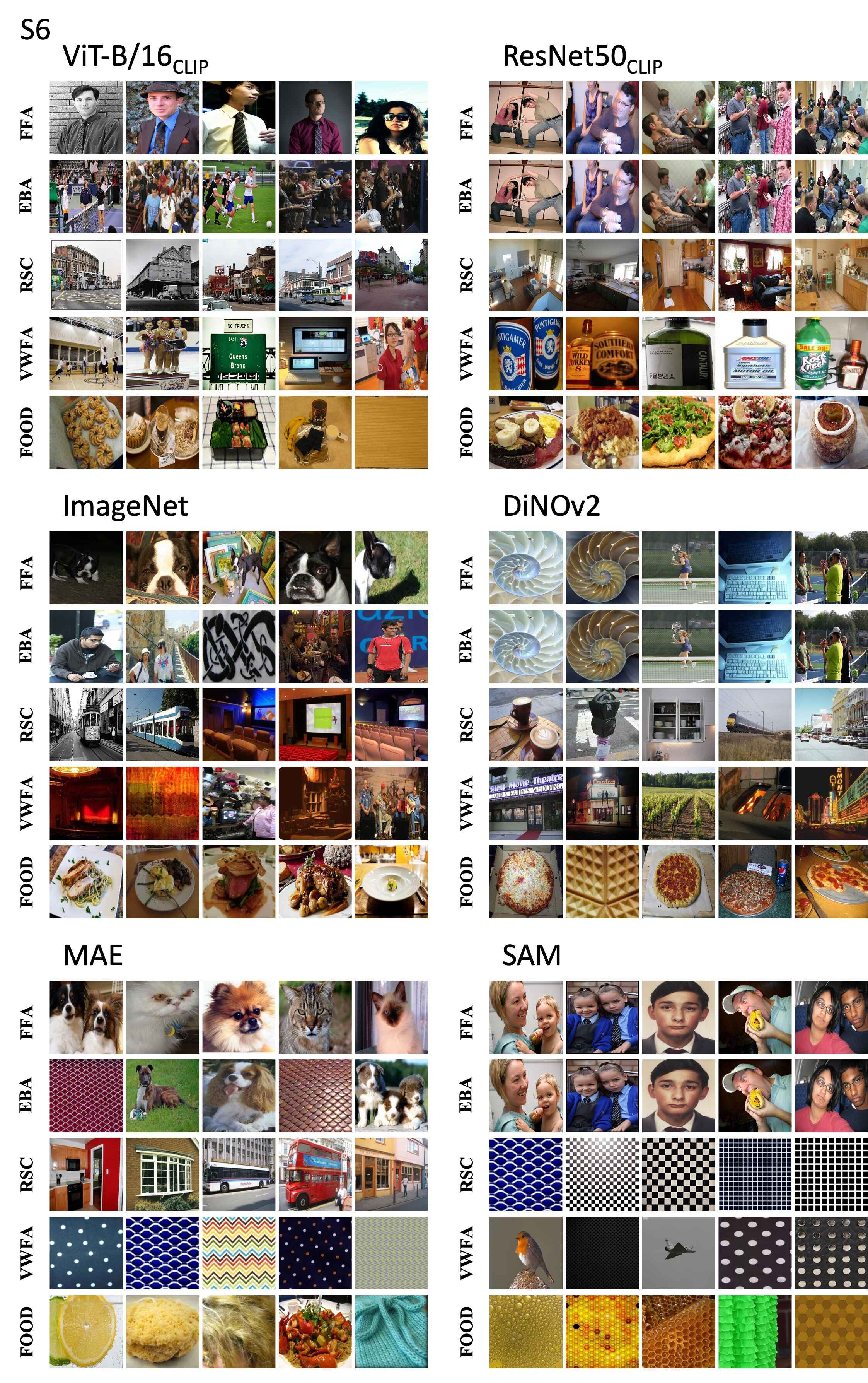}
    \caption{\textbf{Units Selectivity for S6} The top-5 images whose "CLS" token maximize activate the last layer's units that selected by the brain ROIs.}
    \label{AppendixFig14}
\end{figure}
\newpage

\begin{figure}[h]
    \centering
    \includegraphics[scale=0.6]{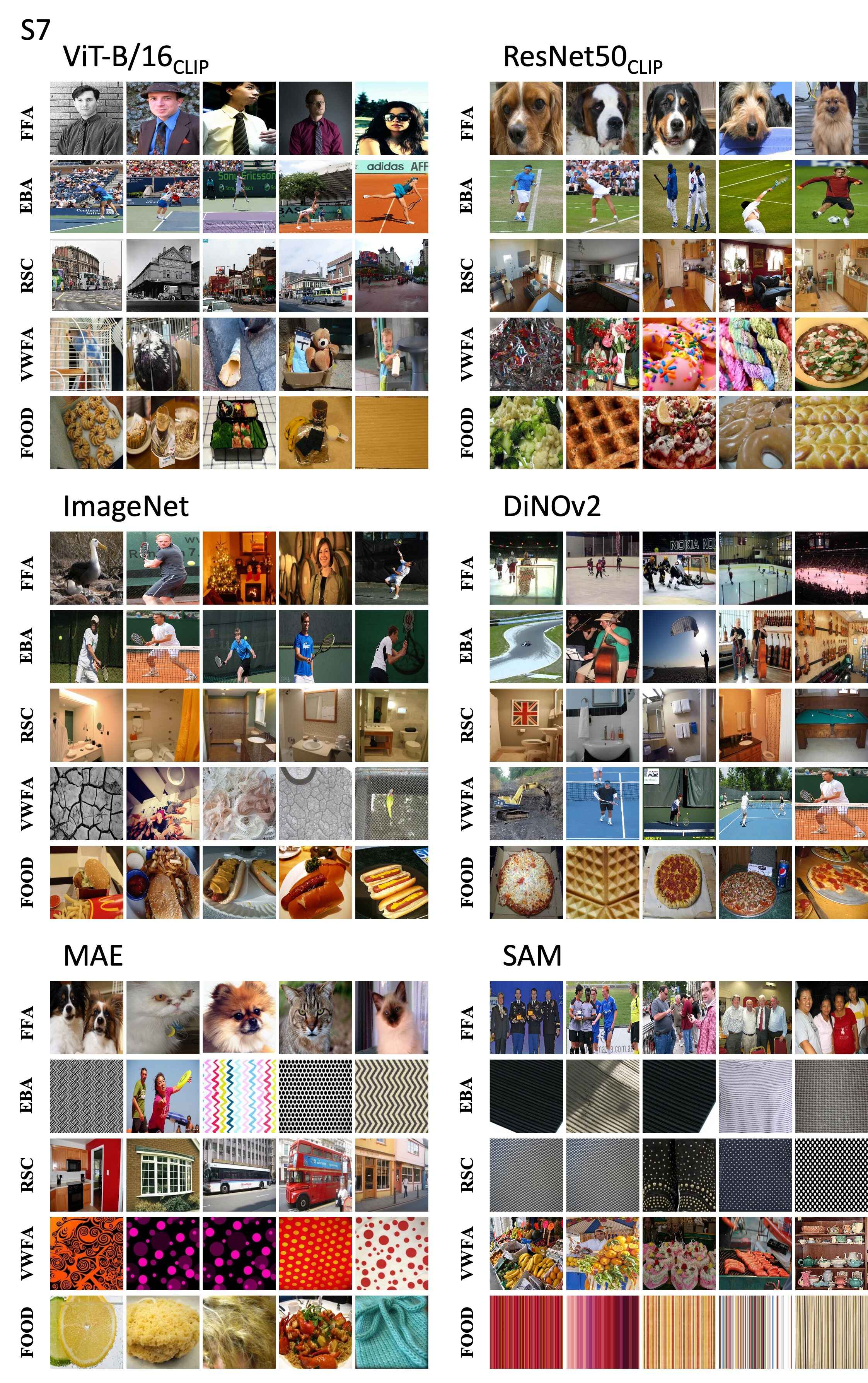}
    \caption{\textbf{Units Selectivity for S7} The top-5 images whose "CLS" token maximize activate the last layer's units that selected by the brain ROIs.}
    \label{AppendixFig15}
\end{figure}
\newpage

\begin{figure}[h]
    \centering
    \includegraphics[scale=0.6]{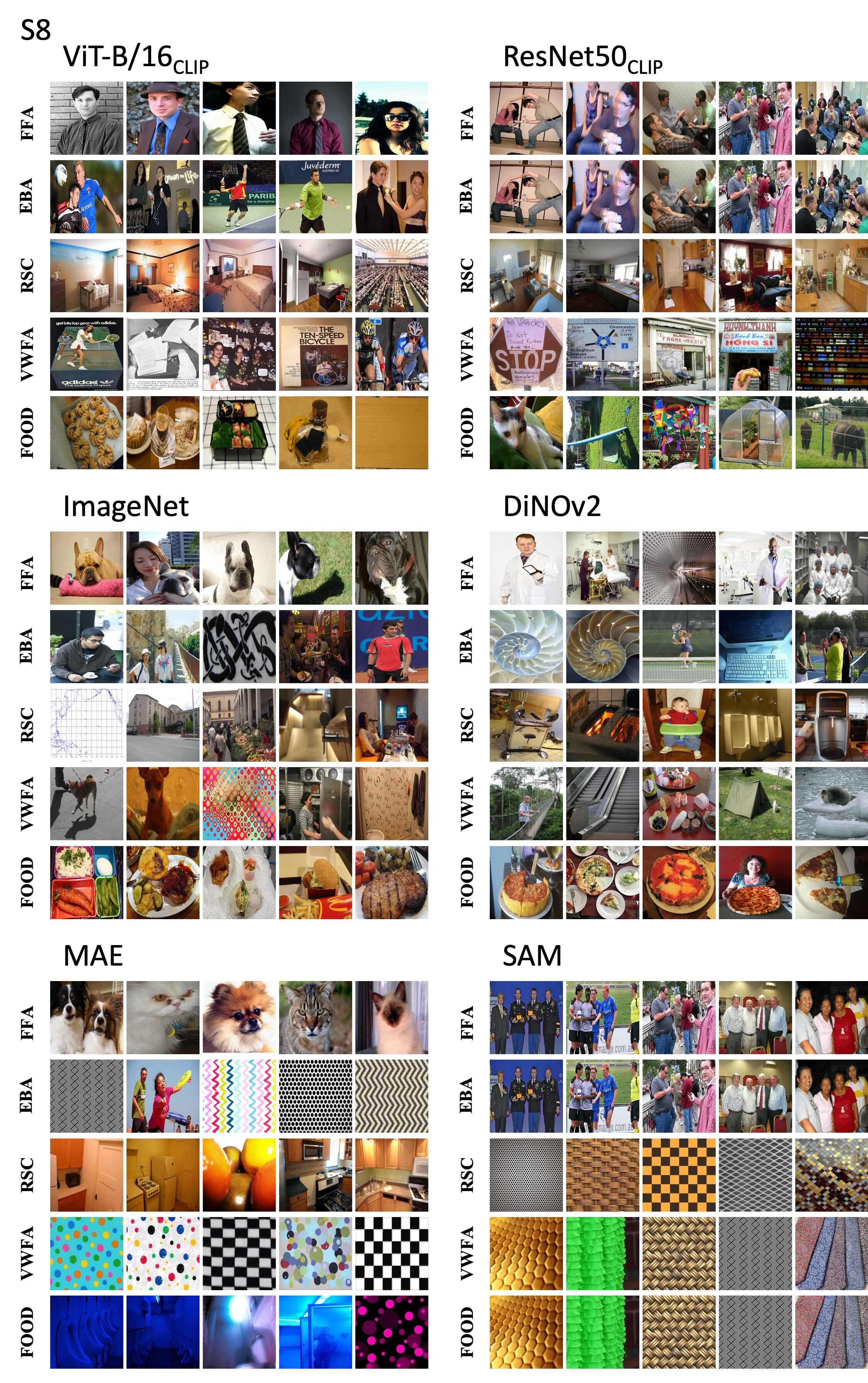}
    \caption{\textbf{Units Selectivity for S8} The top-5 images whose "CLS" token maximize activate the last layer's units that selected by the brain ROIs.}
    \label{AppendixFig16}
\end{figure}
\newpage

\section{Subject 1-8 for Similarity between voxel dictionary and brain encoder}

\begin{figure}[h]
    \centering
    \includegraphics[scale=0.3]{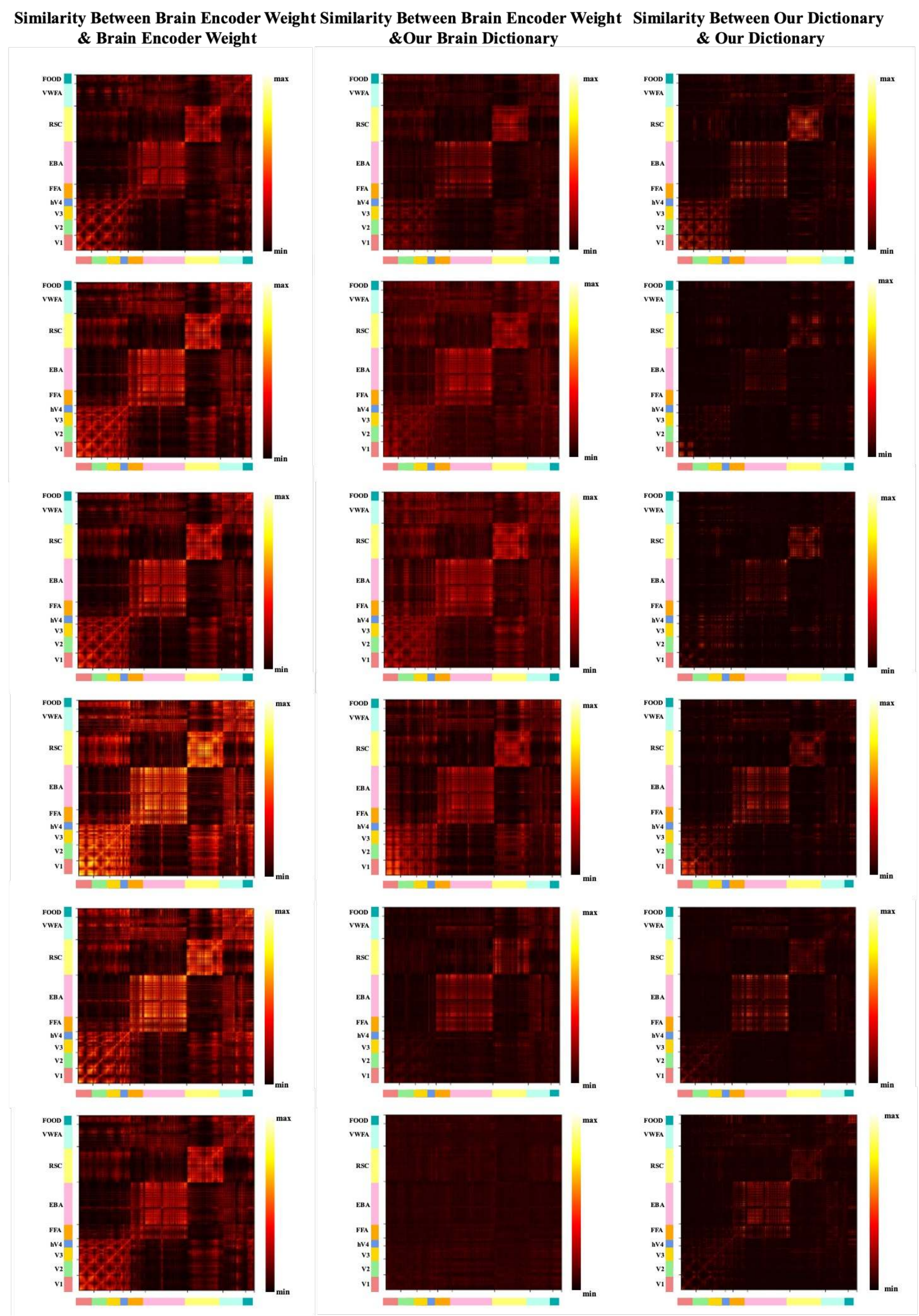}
    \caption{\textbf{Representation Similarity Matrix on S1} between the brain encoder weights of the last layer of voxel dictionary, and their mutual similarity. The model from top to bottom is ResNet50\textsubscript{CLIP}, DiNOv2, ImageNet, MAE, SAM and ViT-B/16\textsubscript{CLIP}}
    \label{AppendixFig17}
\end{figure}

\newpage

\begin{figure}[h]
    \centering
    \includegraphics[scale=0.6]{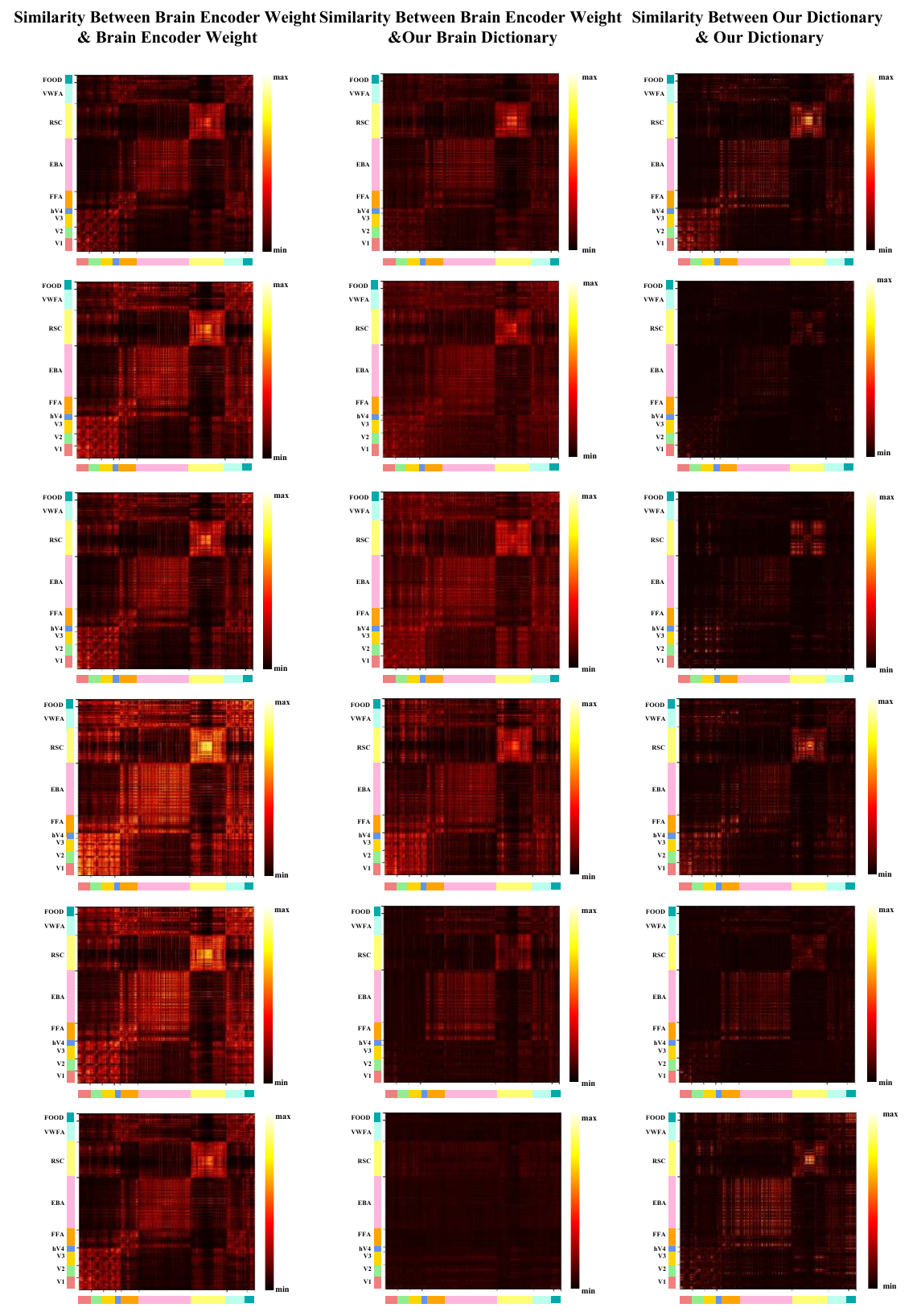}
    \caption{\textbf{Representation Similarity Matrix on S2} between the brain encoder weights of the last layer of voxel dictionary, and their mutual similarity. The model from top to bottom is ResNet50\textsubscript{CLIP}, DiNOv2, ImageNet, MAE, SAM and ViT-B/16\textsubscript{CLIP}}
    \label{AppendixFig18}
\end{figure}

\newpage

\begin{figure}[h]
    \centering
    \includegraphics[scale=0.3]{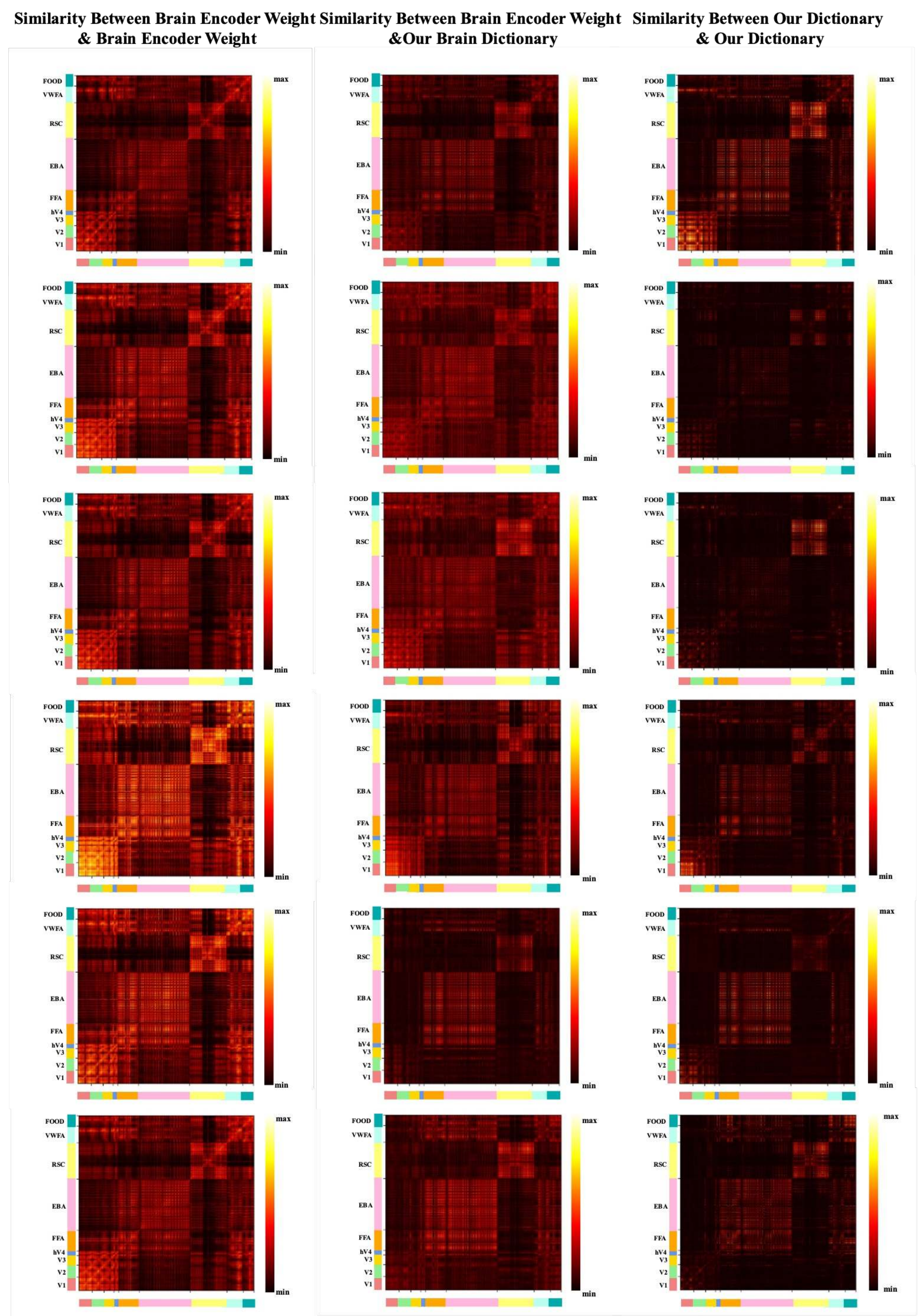}
    \caption{\textbf{Representation Similarity Matrix on S3} between the brain encoder weights of the last layer of voxel dictionary, and their mutual similarity. The model from top to bottom is ResNet50\textsubscript{CLIP}, DiNOv2, ImageNet, MAE, SAM and ViT-B/16\textsubscript{CLIP}}
    \label{AppendixFig19}
\end{figure}

\newpage

\begin{figure}[h]
    \centering
    \includegraphics[scale=0.25]{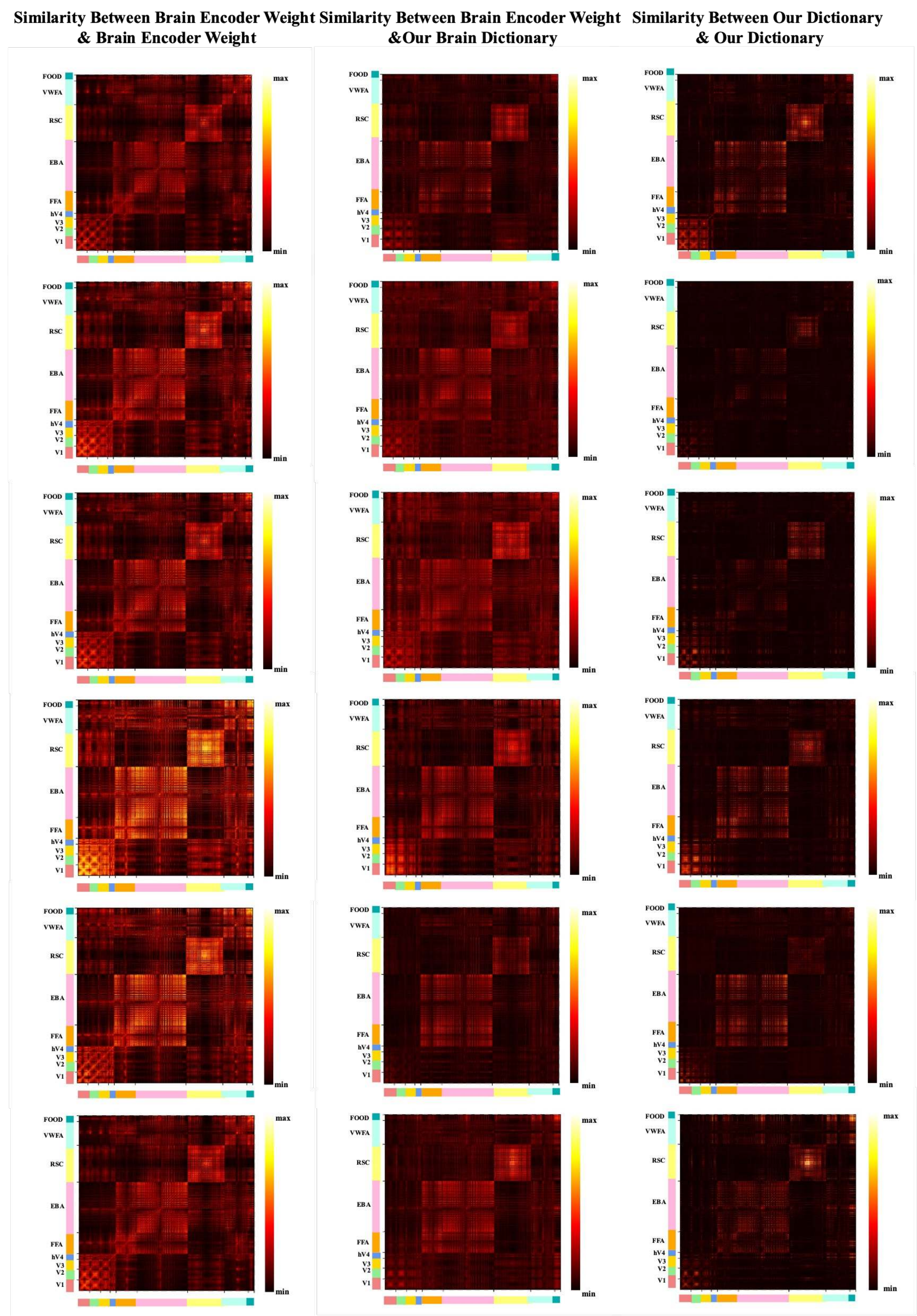}
    \caption{\textbf{Representation Similarity Matrix on S4} between the brain encoder weights of the last layer of voxel dictionary, and their mutual similarity. The model from top to bottom is ResNet50\textsubscript{CLIP}, DiNOv2, ImageNet, MAE, SAM and ViT-B/16\textsubscript{CLIP}}
    \label{AppendixFig20}
\end{figure}

\newpage

\begin{figure}[h]
    \centering
    \includegraphics[scale=0.25]{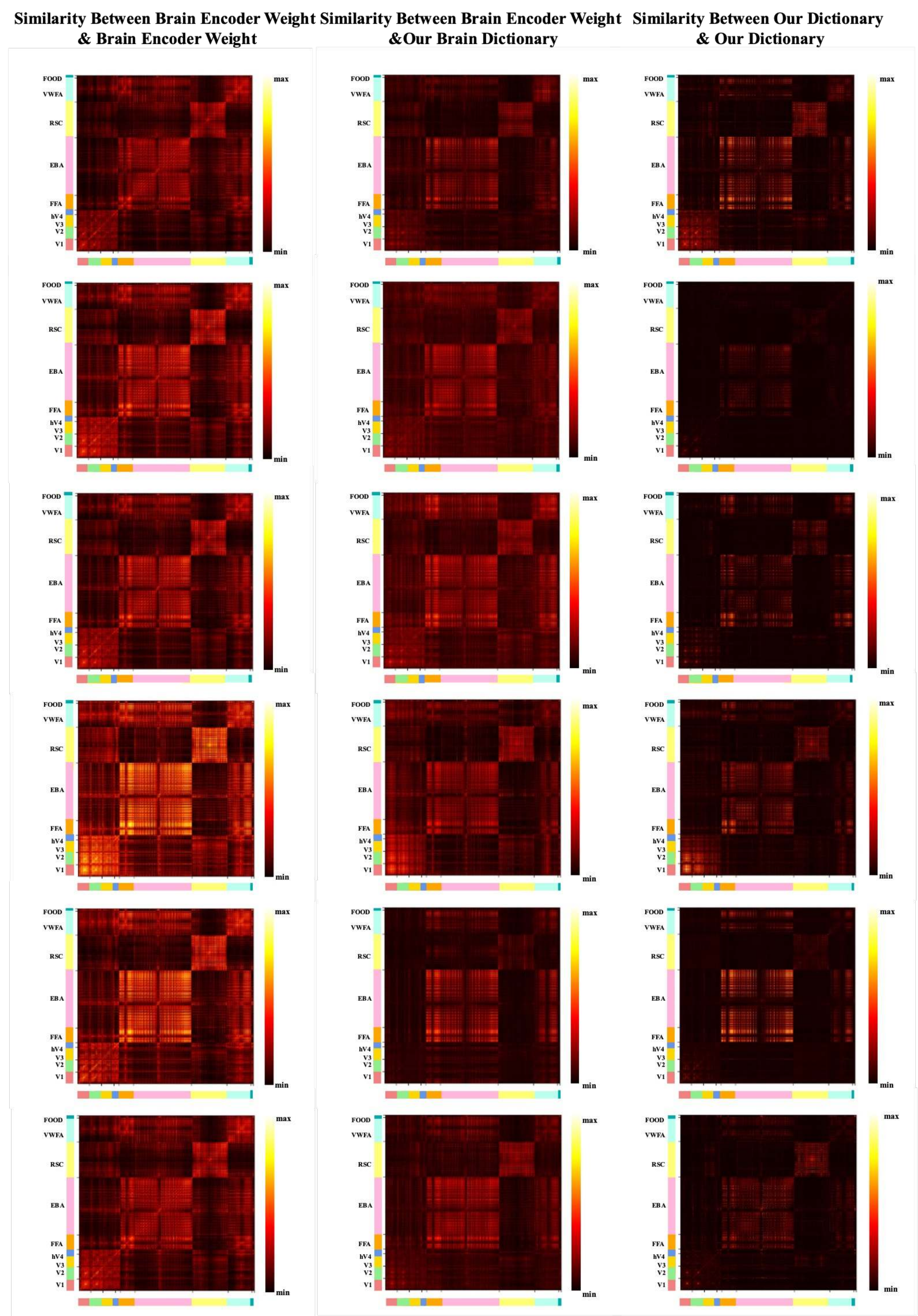}
    \caption{\textbf{Representation Similarity Matrix on S6} between the brain encoder weights of the last layer of voxel dictionary, and their mutual similarity. The model from top to bottom is ResNet50\textsubscript{CLIP}, DiNOv2, ImageNet, MAE, SAM and ViT-B/16\textsubscript{CLIP}}
    \label{AppendixFig21}
\end{figure}

\newpage

\begin{figure}[h]
    \centering
    \includegraphics[scale=0.2]{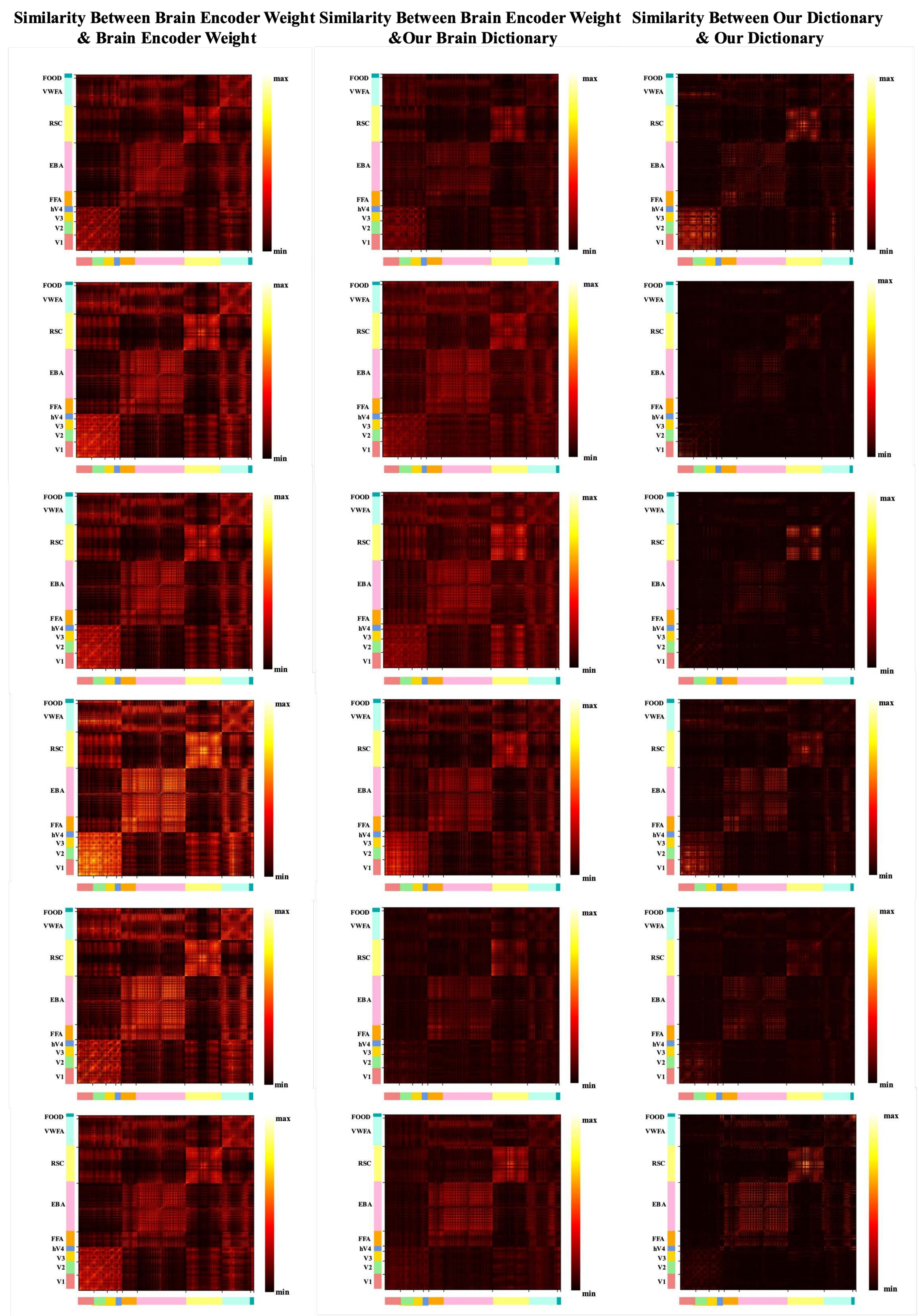}
    \caption{\textbf{Representation Similarity Matrix on S7} between the brain encoder weights of the last layer of voxel dictionary, and their mutual similarity. The model from top to bottom is ResNet50\textsubscript{CLIP}, DiNOv2, ImageNet, MAE, SAM and ViT-B/16\textsubscript{CLIP}}
    \label{AppendixFig22}
\end{figure}

\newpage

\begin{figure}[h]
    \centering
    \includegraphics[scale=0.2]{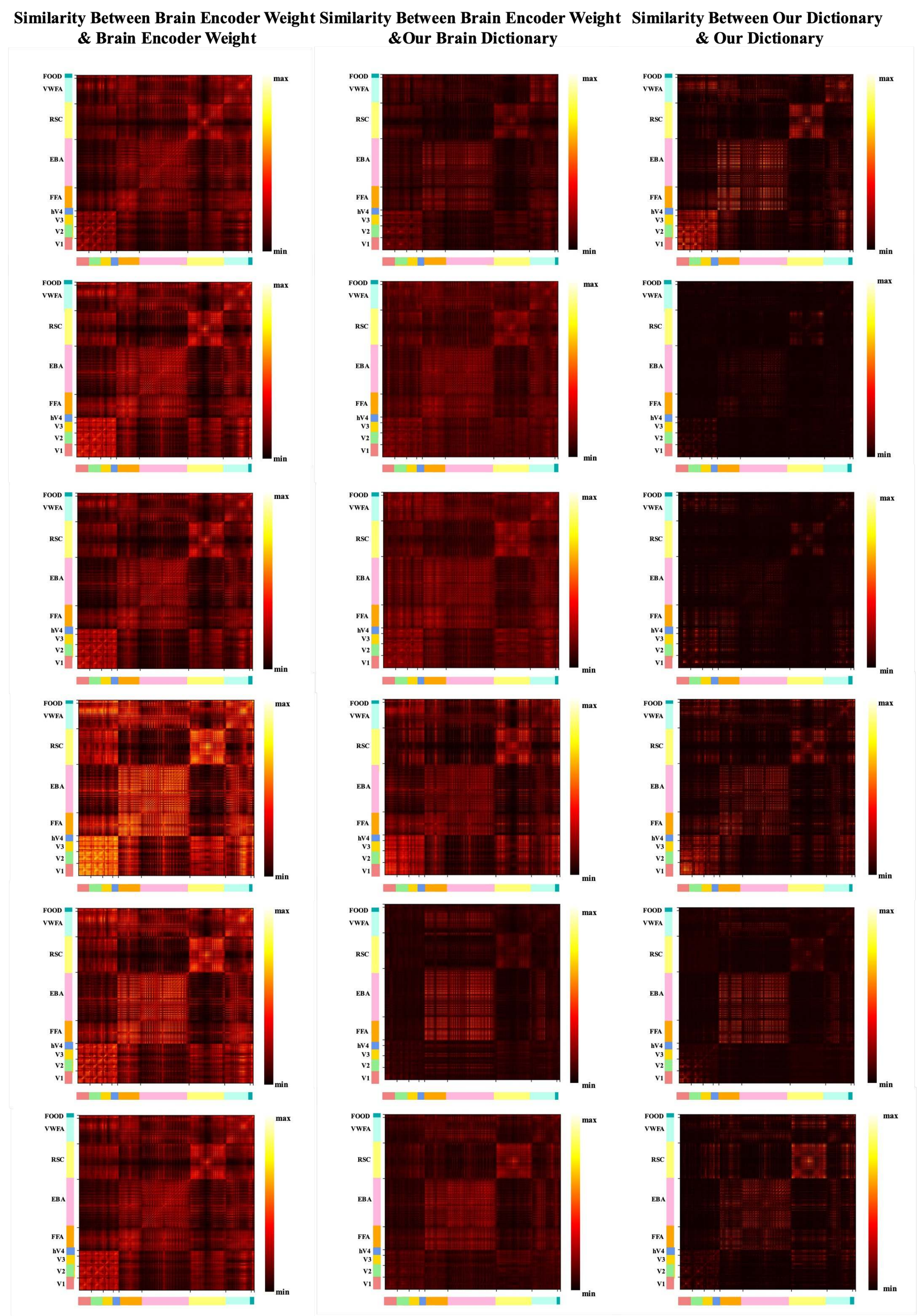}
    \caption{\textbf{Representation Similarity Matrix on S8} between the brain encoder weights of the last layer of voxel dictionary, and their mutual similarity. The model from top to bottom is ResNet50\textsubscript{CLIP}, DiNOv2, ImageNet, MAE, SAM and ViT-B/16\textsubscript{CLIP}}
    \label{AppendixFig23}
\end{figure}

\newpage
\section{Visual Information Procedure Analysis}
We visualize the visual information procedure for all subject (S1-S8) and models that with ViT structure and have 12 layers.

\begin{figure}[h]
    \centering
    \includegraphics[scale=0.6]{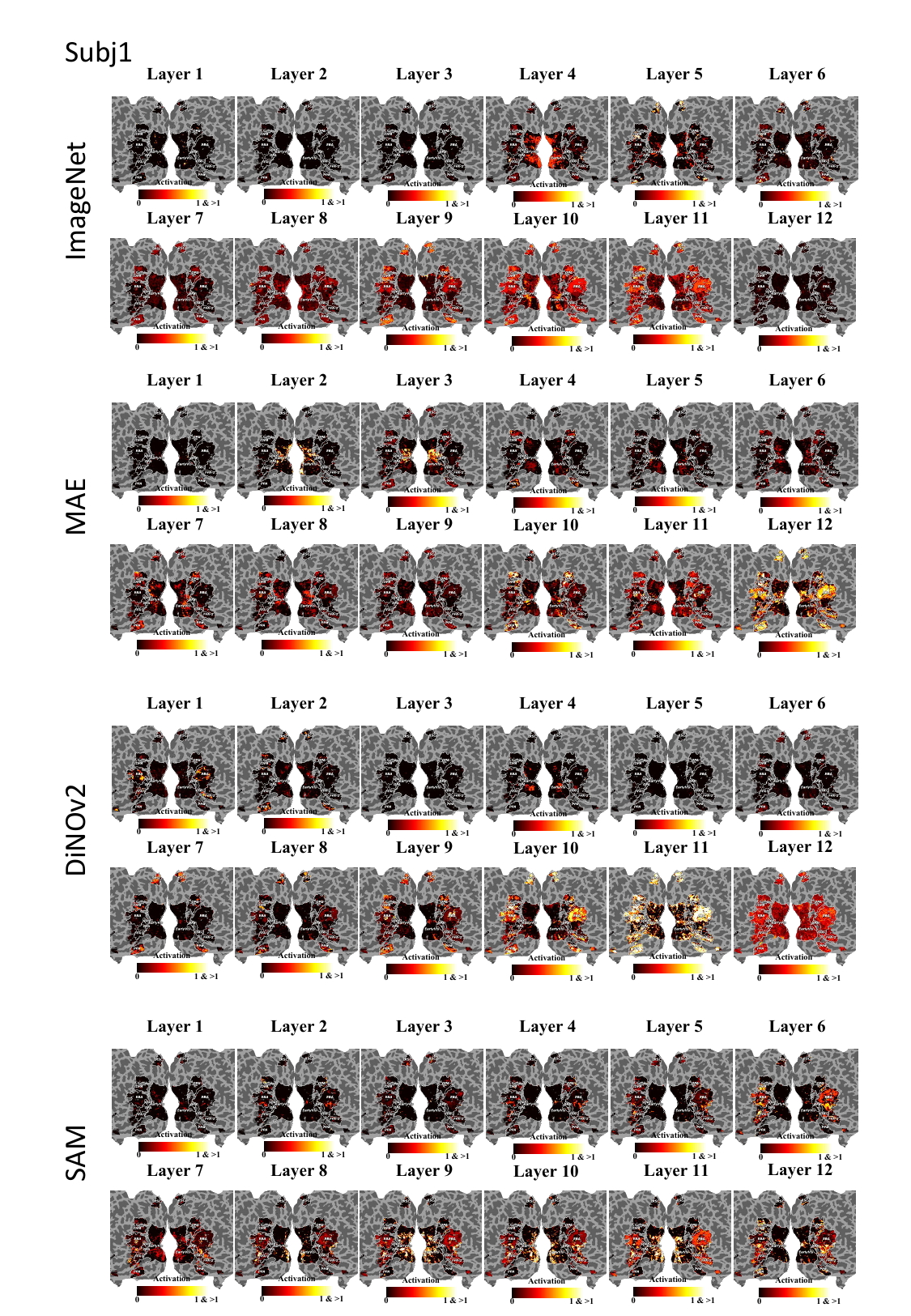}
    \caption{Layer wise Voxel Dictionary average activation for ImageNet, DiNOv2, MAE, SAM for S1.}
    \label{AppendixFig24}
\end{figure}

\newpage
\begin{figure}[h]
    \centering
    \includegraphics[scale=0.6]{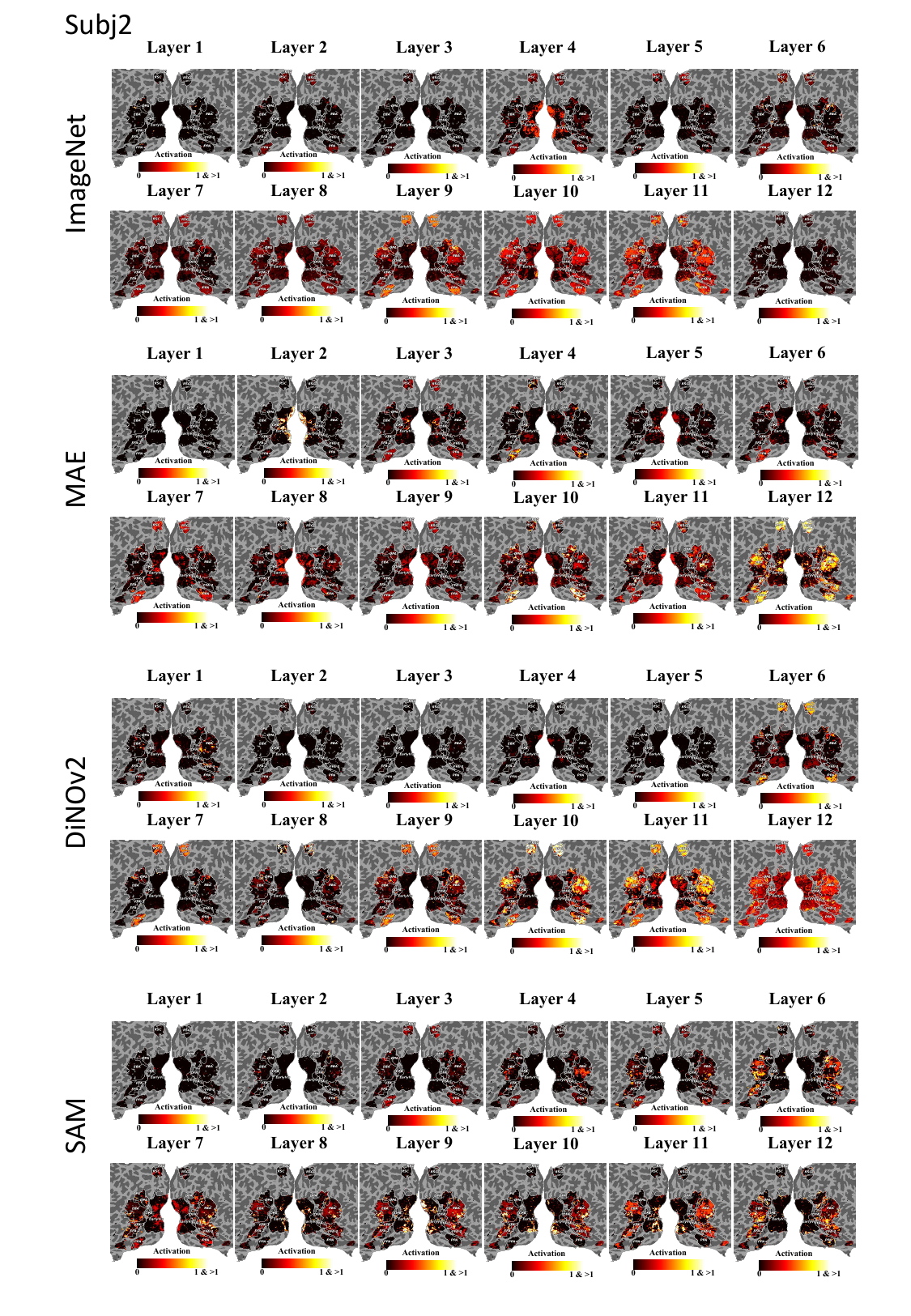}
    \caption{Layer wise Voxel Dictionary average activation for ImageNet, DiNOv2, MAE, SAM for S2.}
    \label{AppendixFig25}
\end{figure}

\newpage
\begin{figure}[h]
    \centering
    \includegraphics[scale=0.6]{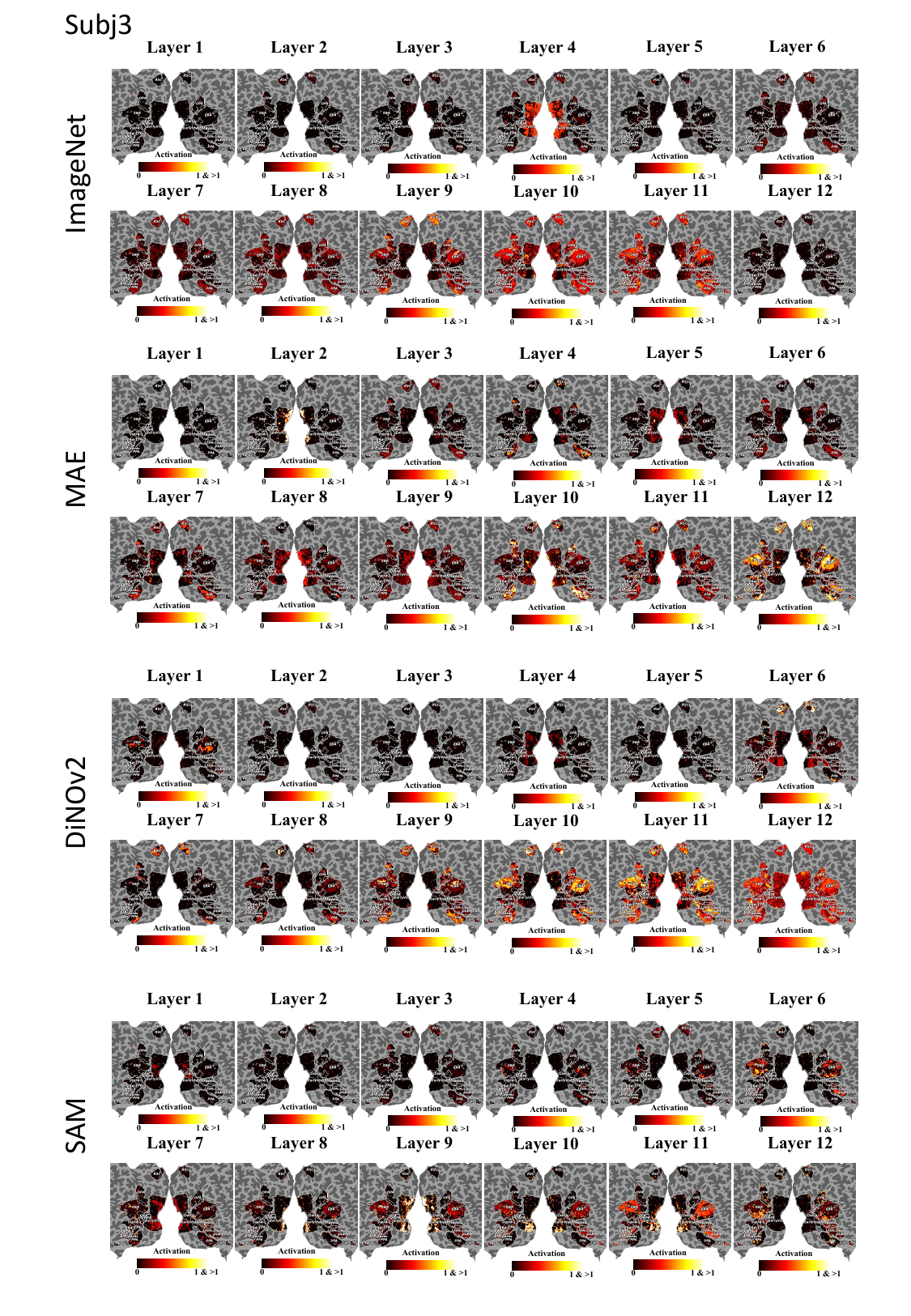}
    \caption{Layer wise Voxel Dictionary average activation for ImageNet, DiNOv2, MAE, SAM for S3.}
    \label{AppendixFig26}
\end{figure}

\newpage
\begin{figure}[h]
    \centering
    \includegraphics[scale=0.6]{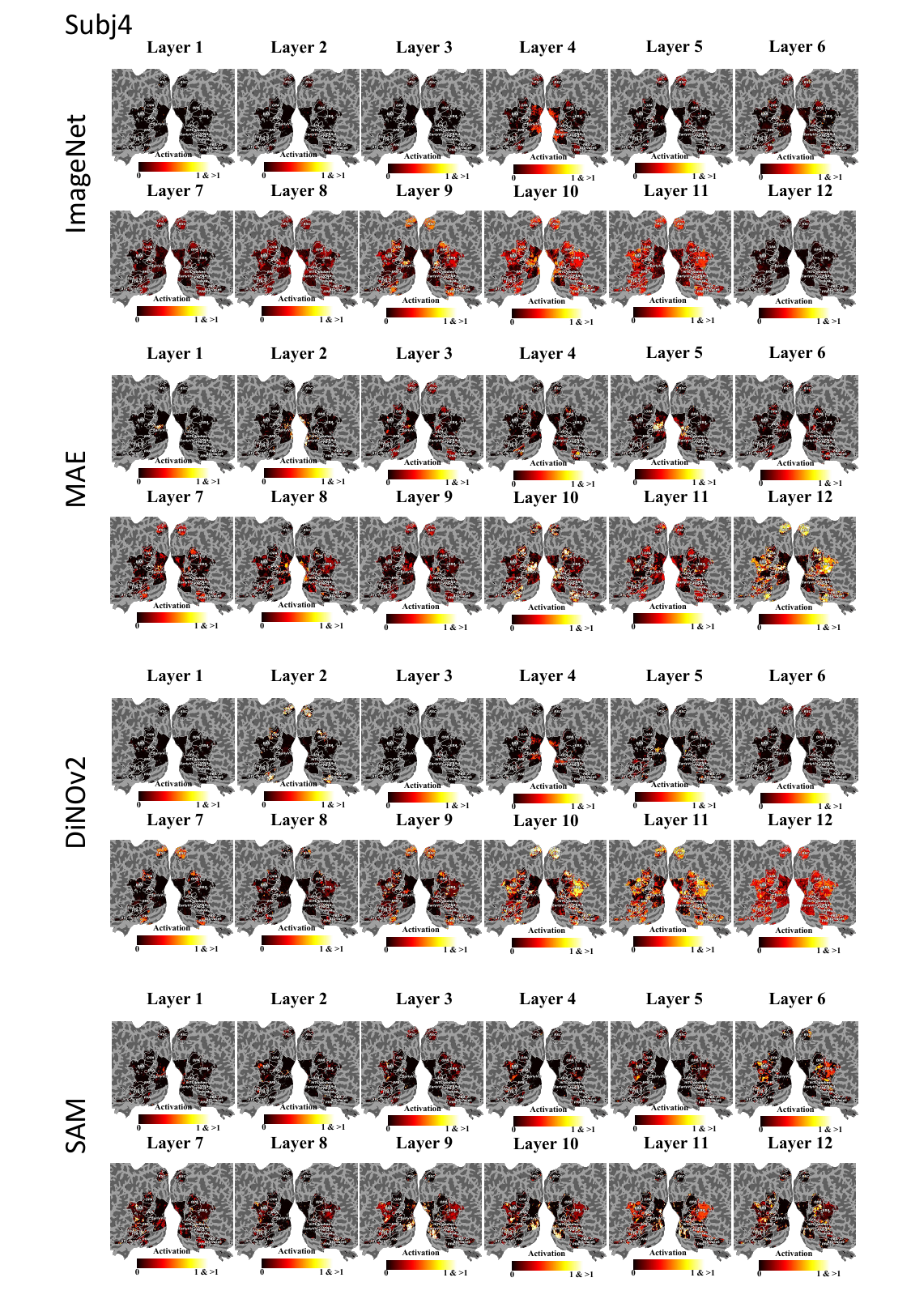}
    \caption{Layer wise Voxel Dictionary average activation for ImageNet, DiNOv2, MAE, SAM for S4.}
    \label{AppendixFig27}
\end{figure}

\newpage
\begin{figure}[h]
    \centering
    \includegraphics[scale=0.6]{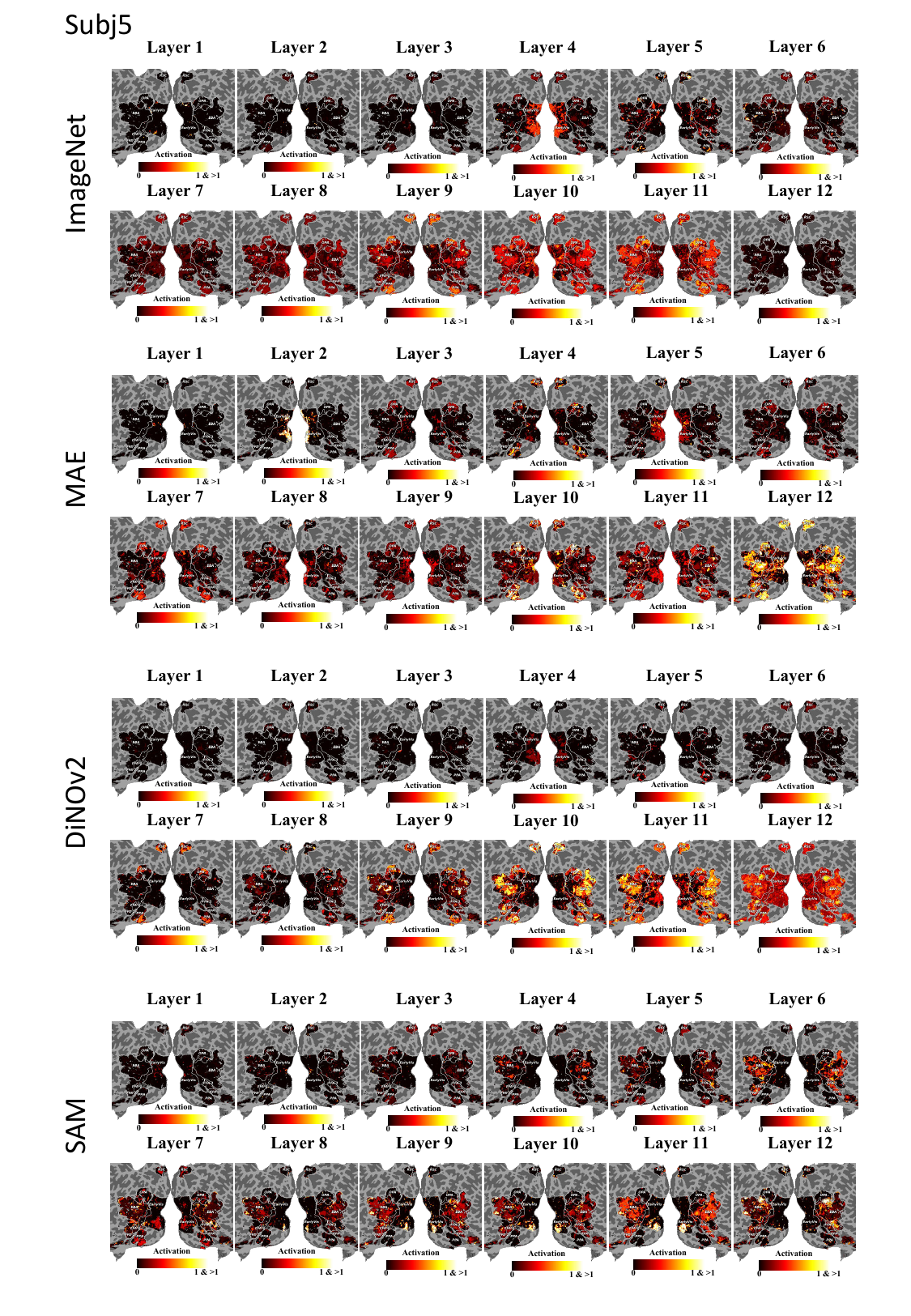}
    \caption{Layer wise Voxel Dictionary average activation for ImageNet, DiNOv2, MAE, SAM for S5.}
    \label{AppendixFig28}
\end{figure}

\newpage
\begin{figure}[h]
    \centering
    \includegraphics[scale=0.6]{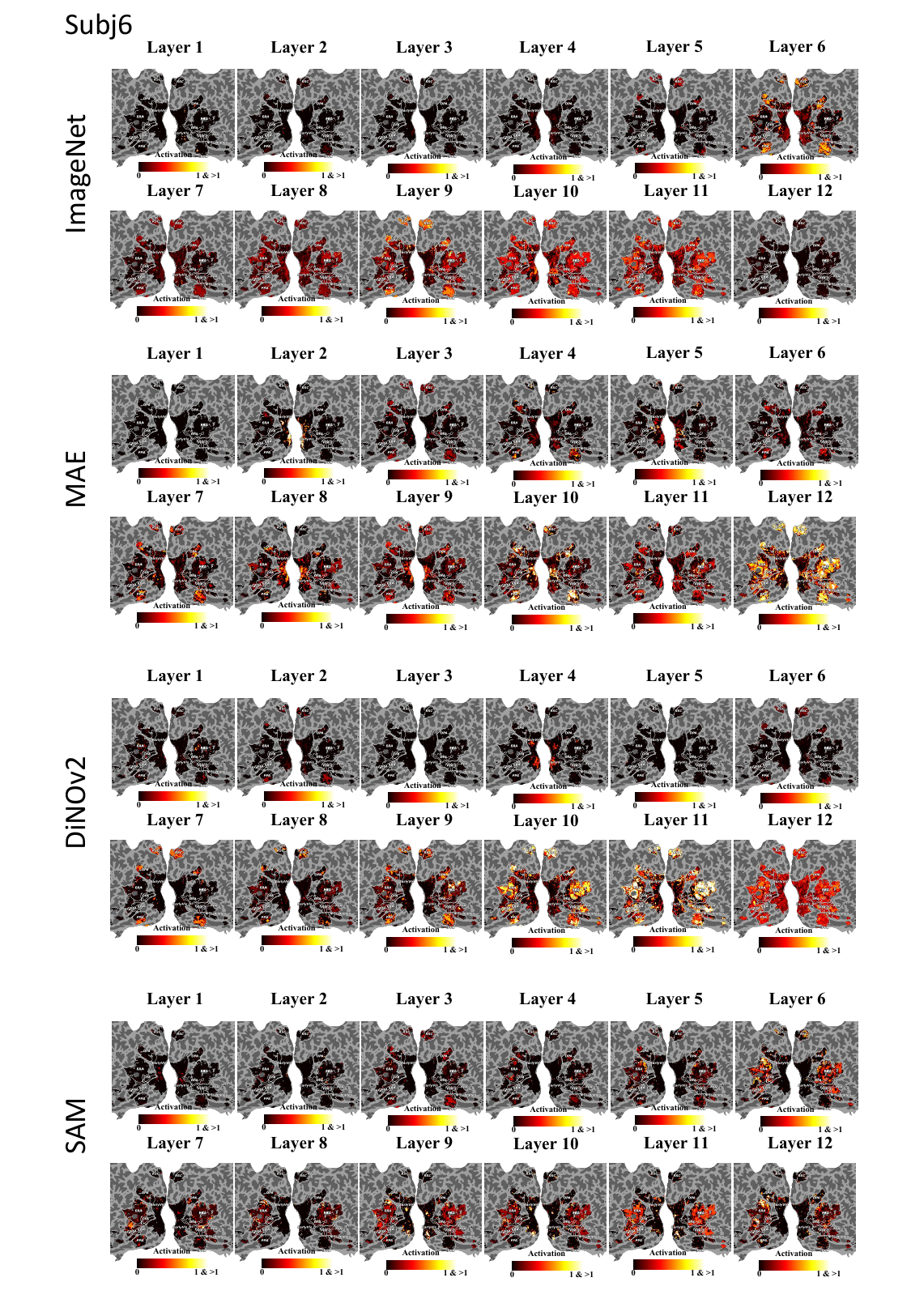}
    \caption{Layer wise Voxel Dictionary average activation for ImageNet, DiNOv2, MAE, SAM for S6.}
    \label{AppendixFig29}
\end{figure}

\newpage
\begin{figure}[h]
    \centering
    \includegraphics[scale=0.6]{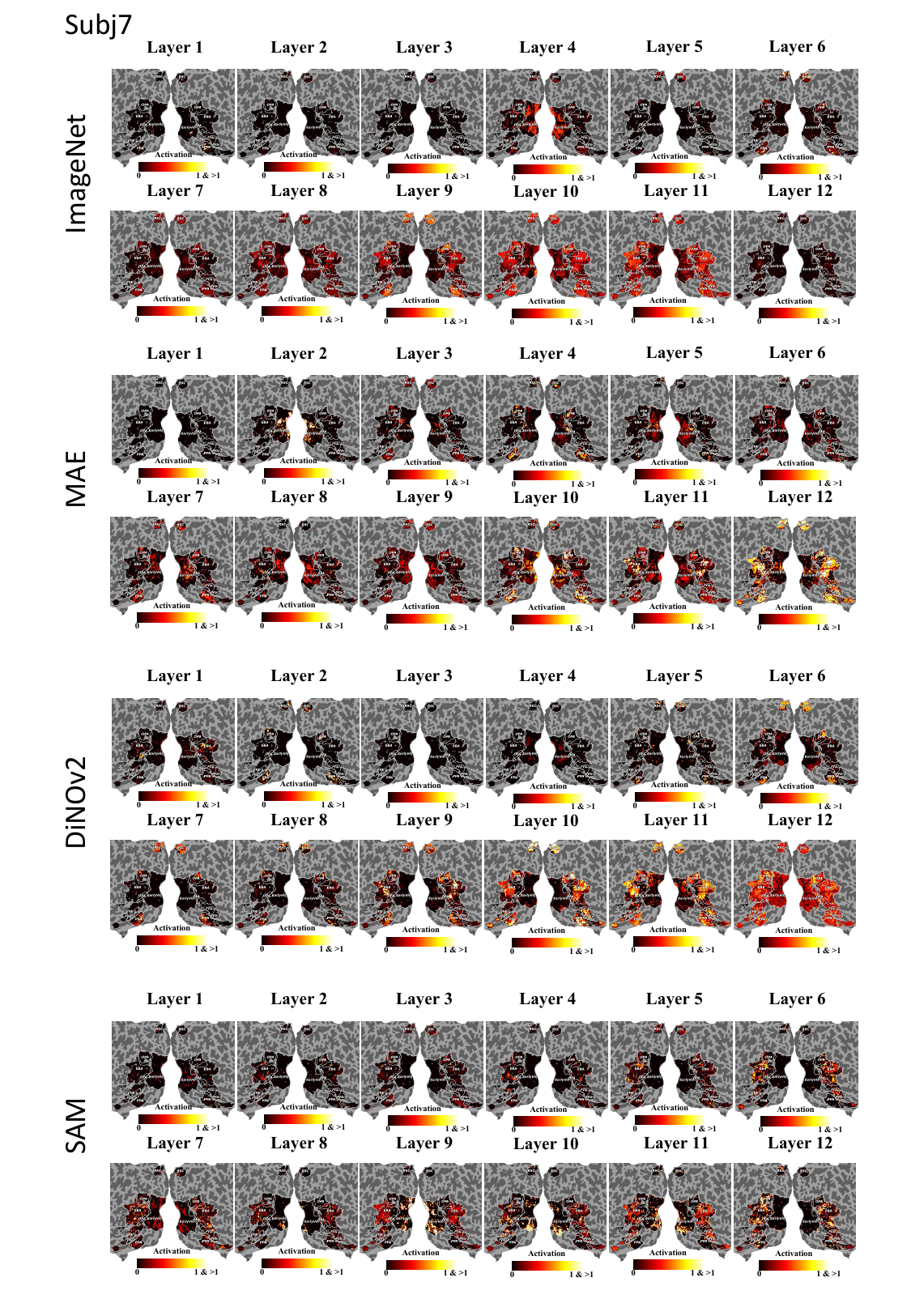}
    \caption{Layer wise Voxel Dictionary average activation for ImageNet, DiNOv2, MAE, SAM for S7.}
    \label{AppendixFig30}
\end{figure}

\newpage
\begin{figure}[h]
    \centering
    \includegraphics[scale=0.6]{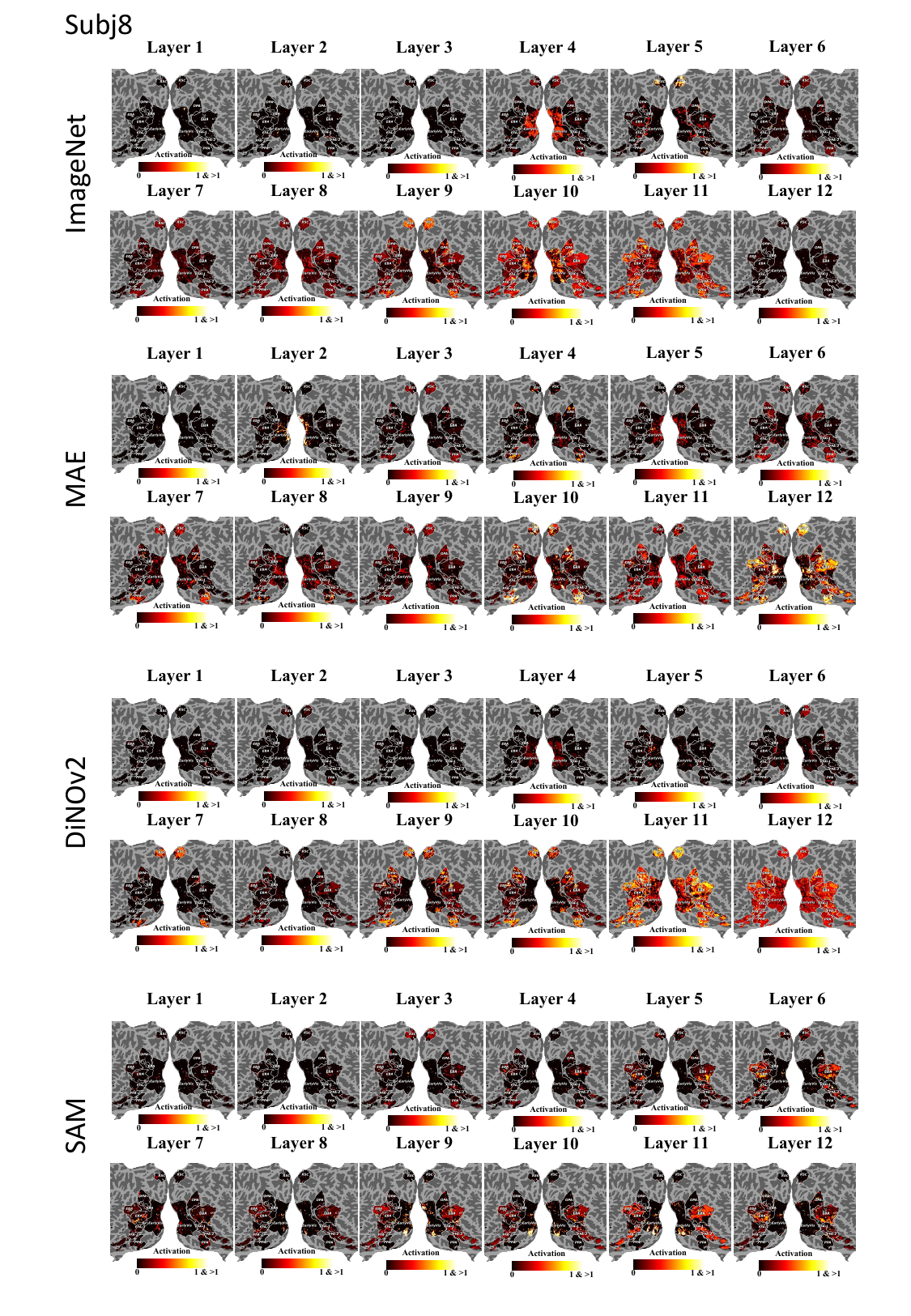}
    \caption{Layer wise Voxel Dictionary average activation for ImageNet, DiNOv2, MAE, SAM for S8.}
    \label{AppendixFig31}
\end{figure}

\newpage
\begin{figure}[h]
    \centering
    \includegraphics[scale=0.6]{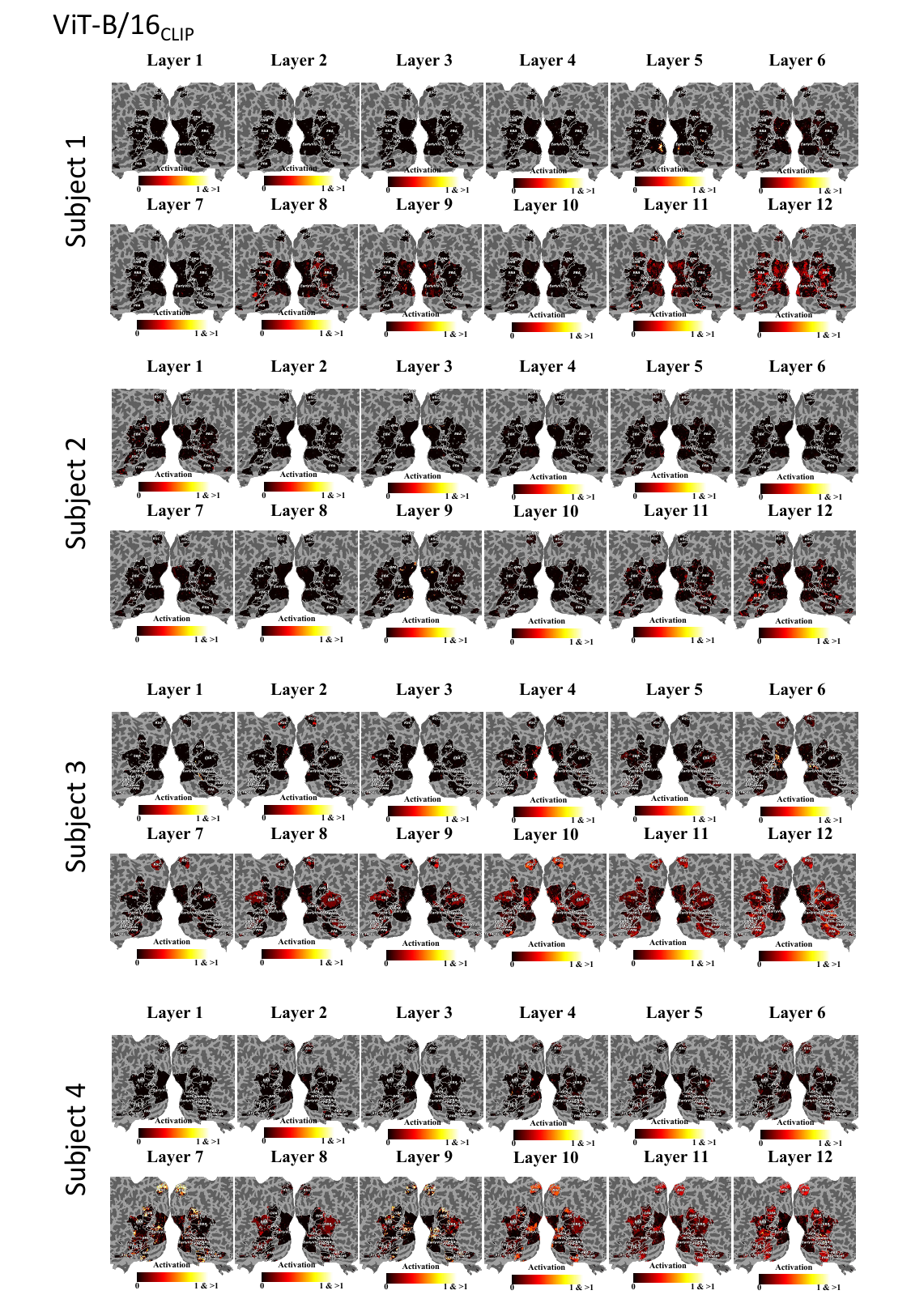}
    \caption{Layer wise Voxel Dictionary average activation for ViT-B/16\textsubscript{CLIP} for S1-4.}
    \label{AppendixFig32}
\end{figure}

\newpage
\begin{figure}[h]
    \centering
    \includegraphics[scale=0.6]{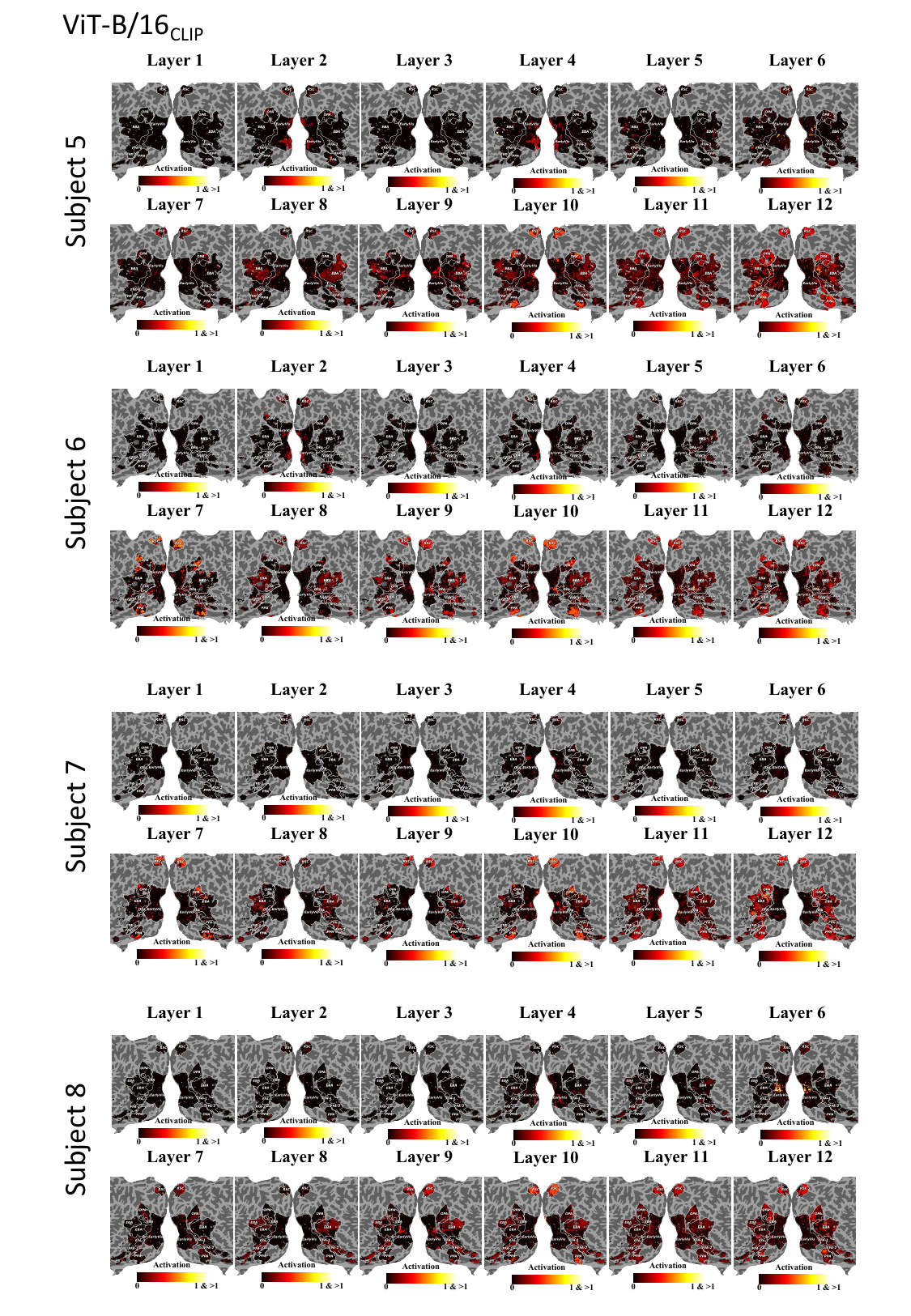}
    \caption{Layer wise Voxel Dictionary average activation for ViT-B/16\textsubscript{CLIP} for S5-8.}
    \label{AppendixFig33}
\end{figure}

\newpage
\section*{NeurIPS Paper Checklist}
\begin{enumerate}

\item {\bf Claims}
    \item[] Question: Do the main claims made in the abstract and introduction accurately reflect the paper's contributions and scope?
    \item[] Answer: \answerYes{} 
    \item[] Justification: We describe the contributions in the last paragraph of introduction~\ref{introduction}.

\item {\bf Limitations}
    \item[] Question: Does the paper discuss the limitations of the work performed by the authors?
    \item[] Answer: \answerYes{} 
    \item[] Justification: We discuss the limitations in Section~\ref{section 5}.

\item {\bf Theory assumptions and proofs}
    \item[] Question: For each theoretical result, does the paper provide the full set of assumptions and a complete (and correct) proof?
    \item[] Answer: \answerNA{} 
    \item[] Justification: The article dosen't include theoretical results.

    \item {\bf Experimental result reproducibility}
    \item[] Question: Does the paper fully disclose all the information needed to reproduce the main experimental results of the paper to the extent that it affects the main claims and/or conclusions of the paper (regardless of whether the code and data are provided or not)?
    \item[] Answer: \answerYes{} 
    \item[] Justification: We discribe the detail in section~\ref{setup} and Appendix~\ref{appendix2}.

\item {\bf Open access to data and code}
    \item[] Question: Does the paper provide open access to the data and code, with sufficient instructions to faithfully reproduce the main experimental results, as described in supplemental material?
    \item[] Answer: \answerYes{} 
    \item[] Justification: We provide code in supplemental material and mention all the dataset in section~\ref{setup}

\item {\bf Experimental setting/details}
    \item[] Question: Does the paper specify all the training and test details (e.g., data splits, hyperparameters, how they were chosen, type of optimizer, etc.) necessary to understand the results?
    \item[] Answer: \answerYes{} 
    \item[] Justification: Dataset splits are mentioned in section~\ref{setup} and section~\ref{section 4.3}.

\item {\bf Experiment statistical significance}
    \item[] Question: Does the paper report error bars suitably and correctly defined or other appropriate information about the statistical significance of the experiments?
    \item[] Answer: \answerYes{} 
    \item[] Justification: We illustrate the statistical information in Table~\ref{table1}, Table~\ref{table2}, and Table~\ref{table3}

\item {\bf Experiments compute resources}
    \item[] Question: For each experiment, does the paper provide sufficient information on the computer resources (type of compute workers, memory, time of execution) needed to reproduce the experiments?
    \item[] Answer: \answerYes{} 
    \item[] Justification: We provide computer resources in Appendix~\ref{appendix3}.
    
\item {\bf Code of ethics}
    \item[] Question: Does the research conducted in the paper conform, in every respect, with the NeurIPS Code of Ethics \url{https://neurips.cc/public/EthicsGuidelines}?
    \item[] Answer: \answerYes{} 
    \item[] Justification: We follow the NeurIPS Code of Ethics strictly.

\item {\bf Broader impacts}
    \item[] Question: Does the paper discuss both potential positive societal impacts and negative societal impacts of the work performed?
    \item[] Answer: \answerNA{} 
    \item[] Justification: We describe the alignment between deep learning model and brain, provide an insight rather than particular applications.
    
\item {\bf Safeguards}
    \item[] Question: Does the paper describe safeguards that have been put in place for responsible release of data or models that have a high risk for misuse (e.g., pretrained language models, image generators, or scraped datasets)?
    \item[] Answer: \answerNA{} 
    \item[] Justification: Our work poses no such risks.

\item {\bf Licenses for existing assets}
    \item[] Question: Are the creators or original owners of assets (e.g., code, data, models), used in the paper, properly credited and are the license and terms of use explicitly mentioned and properly respected?
    \item[] Answer: \answerYes{} 
    \item[] Justification: We respect the license and mention in Appendix~\ref{appendix3}.
    \item[] Guidelines:
    \begin{itemize}
        \item The answer NA means that the paper does not use existing assets.
        \item The authors should~ cite the original paper that produced the code package or dataset.
        \item The authors should state which version of the asset is used and, if possible, include a URL.
        \item The name of the license (e.g., CC-BY 4.0) should be included for each asset.
        \item For scraped data from a particular source (e.g., website), the copyright and terms of service of that source should be provided.
        \item If assets are released, the license, copyright information, and terms of use in the package should be provided. For popular datasets, \url{paperswithcode.com/datasets} has curated licenses for some datasets. Their licensing guide can help determine the license of a dataset.
        \item For existing datasets that are re-packaged, both the original license and the license of the derived asset (if it has changed) should be provided.
        \item If this information is not available online, the authors are encouraged to reach out to the asset's creators.
    \end{itemize}

\item {\bf New assets}
    \item[] Question: Are new assets introduced in the paper well documented and is the documentation provided alongside the assets?
    \item[] Answer: \answerNA{} 
    \item[] Justification: Our paper does not release new assets.

\item {\bf Crowdsourcing and research with human subjects}
    \item[] Question: For crowdsourcing experiments and research with human subjects, does the paper include the full text of instructions given to participants and screenshots, if applicable, as well as details about compensation (if any)? 
    \item[] Answer: \answerNA{} 
    \item[] Justification: Our experiment does not involve crowdsourcing nor research with human subjects.

\item {\bf Institutional review board (IRB) approvals or equivalent for research with human subjects}
    \item[] Question: Does the paper describe potential risks incurred by study participants, whether such risks were disclosed to the subjects, and whether Institutional Review Board (IRB) approvals (or an equivalent approval/review based on the requirements of your country or institution) were obtained?
    \item[] Answer: \answerYes{} 
    \item[] Justification: The methods were performed in accordance with relevant guidelines and regulations and approved by our university's IRB. All ethical regulations relevant to human research participants were followed.

\item {\bf Declaration of LLM usage}
    \item[] Question: Does the paper describe the usage of LLMs if it is an important, original, or non-standard component of the core methods in this research? Note that if the LLM is used only for writing, editing, or formatting purposes and does not impact the core methodology, scientific rigorousness, or originality of the research, declaration is not required.
    \item[] Answer: \answerNA{} 
    \item[] Justification: The core method in the research does not involve LLMs.

\end{enumerate}
\end{document}